\documentclass[journal]{IEEEtran}

\usepackage{cite}
\usepackage{graphicx}
\usepackage{amsmath}
\usepackage{algorithmic}
\usepackage{array}
\usepackage[caption=false,font=footnotesize]{subfig}
\usepackage{url}

\begin{document}

\title{A Survey of Near-Data Processing Architectures for Neural Networks}

\author{Mehdi Hassanpour, Marc Riera and Antonio Gonz\'{a}lez}

\maketitle

\begin{abstract}
Data-intensive workloads and applications, such as machine learning (ML), are fundamentally limited by traditional computing systems based on the von-Neumann architecture. As data movement operations and energy consumption become key bottlenecks in the design of computing systems, the interest in unconventional approaches such as Near-Data Processing (NDP), machine learning, and especially neural network (NN)-based accelerators has grown significantly. Emerging memory technologies, such as ReRAM and 3D-stacked, are promising for efficiently architecting NDP-based accelerators for NN due to their capabilities to work as both: High-density/low-energy storage and in/near-memory computation/search engine. In this paper, we present a survey of techniques for designing NDP architectures for NN. By classifying the techniques based on the memory technology employed, we underscore their similarities and differences. Finally, we discuss open challenges and future perspectives that need to be explored in order to improve and extend the adoption of NDP architectures for future computing platforms. This paper will be valuable for computer architects, chip designers and researchers in the area of machine learning.


\end{abstract}


\IEEEpeerreviewmaketitle

\section{Introduction}\label{s:intro}
\IEEEPARstart{T}{he} era of artificial intelligence and big data is introducing new workloads which operate on huge datasets. A clear example is found in machine learning (ML) and especially neural-network based techniques, which are commonly applied as a solution to find patterns and interpret large amounts of data. Despite the increasing popularity of ML algorithms, there are several challenges to efficiently implement and execute neural networks on conventional processing units based on the von-Neumann compute-centric architecture~\cite{vonneumann_arch} such as CPUs and GPUs. The main performance and energy bottleneck of traditional architectures is the memory hierarchy due to the huge number of inputs, weights, and partial outputs, resulting in numerous data movements which incur in higher energy consumption than the operations~\cite{pandiyan2014quantifying}\cite{kestor_energy_datamovement}. Thus, researchers are exploring novel hardware architectures to accelerate these algorithms by moving most of the computations "in/near-memory" and, hence, reduce the data movements as much as possible. In summary, modern applications based on machine learning are very data-intensive and demand a high level of parallelism and memory bandwidth~\cite{NDPBW}. To address these issues, some recent works have proposed near-data processing accelerators for neural networks with the promise to break the memory wall.

\begin{figure}[t!]
\centering
\subfloat[]{
    \includegraphics[width=0.45\columnwidth]{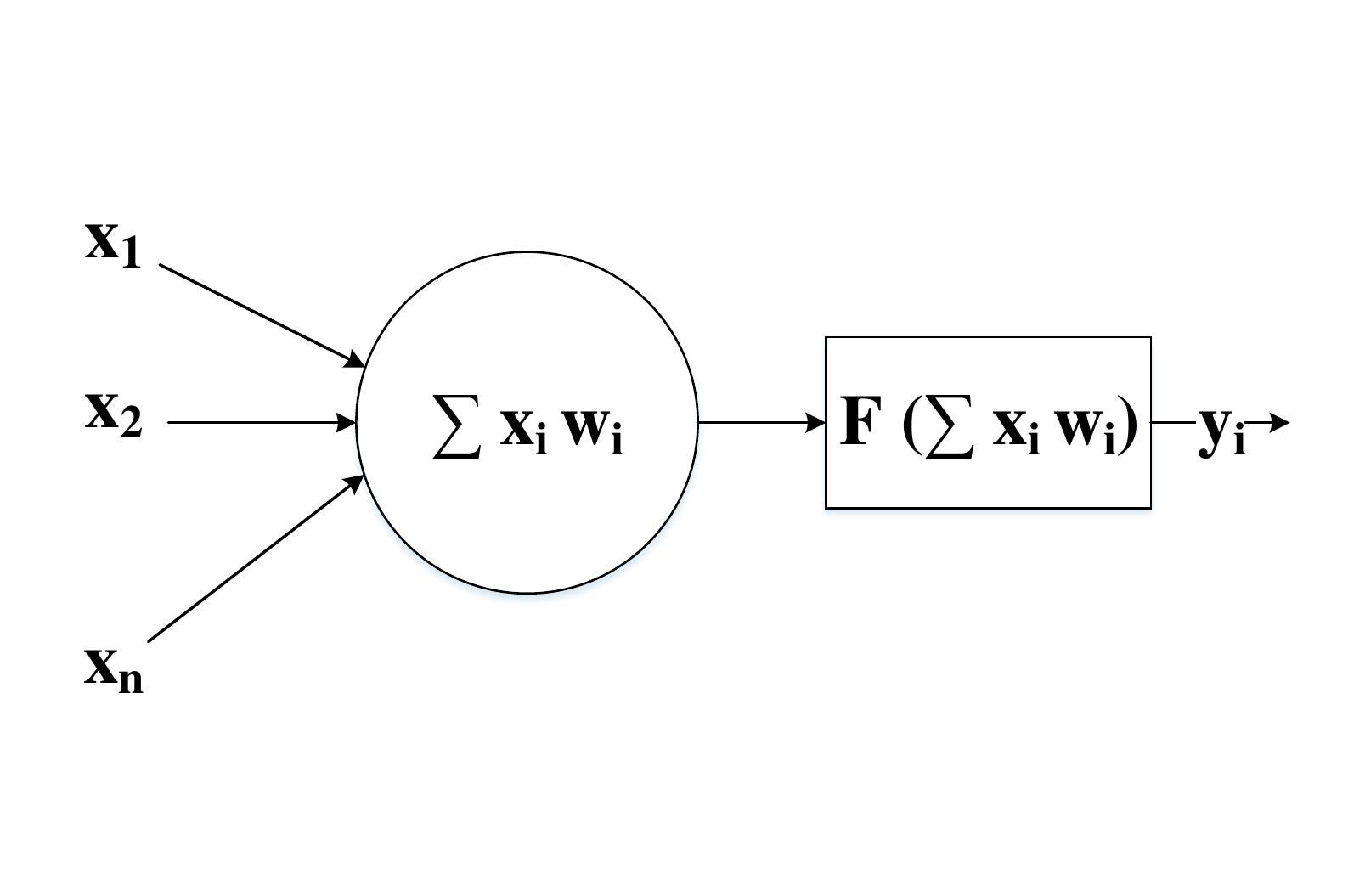}
    \label{fig:neuron}
}
\subfloat[]{
    \includegraphics[width=0.45\columnwidth]{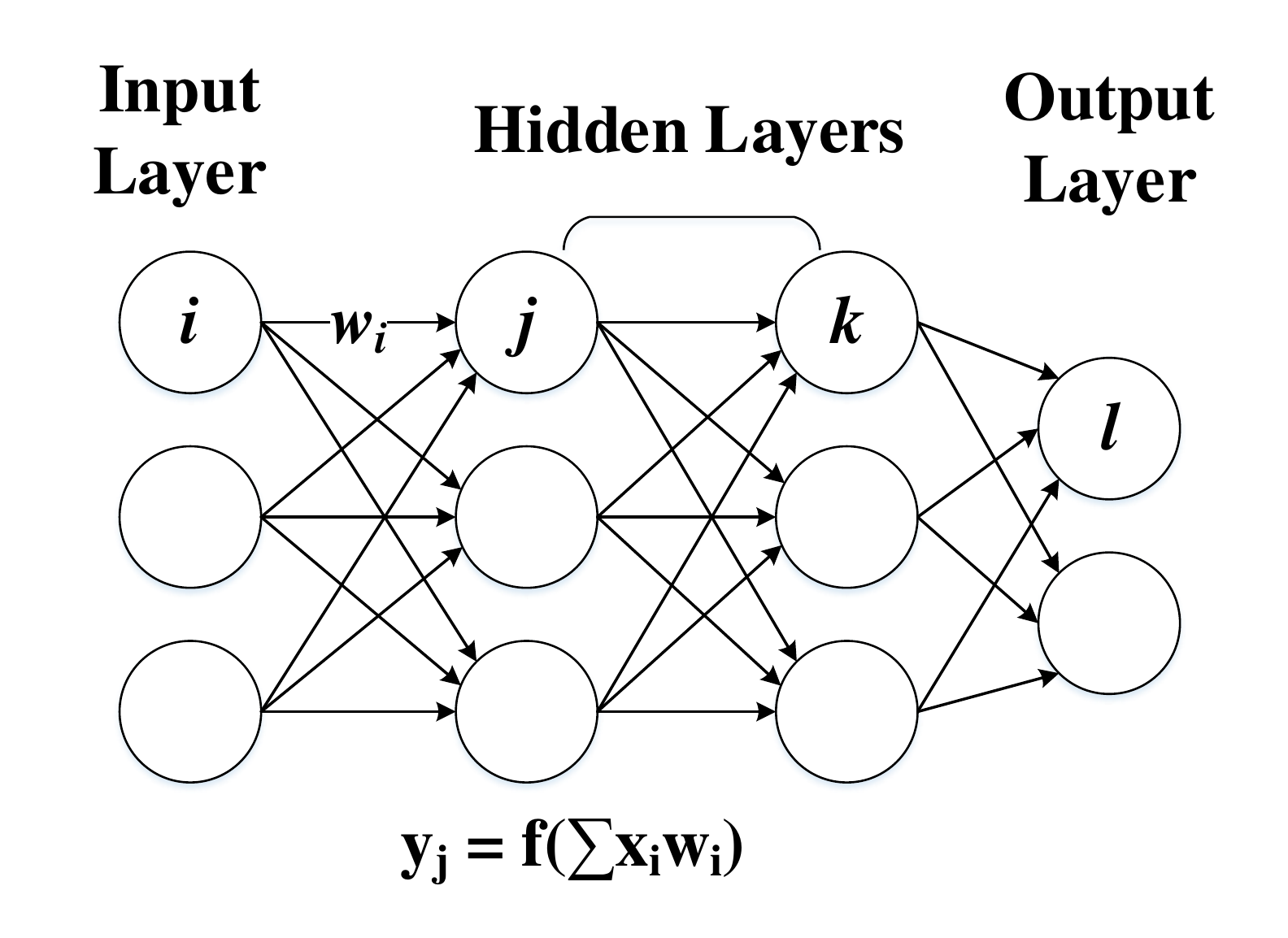}
    \label{fig:dnn}
}
\caption{(a) Computations of an artificial neuron and (b) DNN composed of three FC layers.}
\label{fig:an_dnn}
\end{figure}

Deep Neural Networks (DNN) have demonstrated to be the most effective machine learning solution for a broad range of classification and decision making problems such as speech recognition~\cite{speech_recognition_dnn}, image processing~\cite{image_processing_dnn} or machine translation~\cite{machine_translation_dnn}. Fig.~\ref{fig:neuron} and Fig.~\ref{fig:dnn} show an example of an artificial neuron and a DNN with its corresponding computations. DNNs are composed of a number of hidden layers between the input and output layers, forming feed-forward networks, which means that the outputs of one layer become the inputs of the next layer in the model, or recurrent networks, in which the output of a neuron can be the input of neurons in the same or previous layers. Each layer consists of a set of neurons interconnected with the neurons of adjacent layers according to the degree of relevance among them. The output of each neuron in a given layer is computed as an activation function of a weighted ($W$) sum of its inputs ($X$) as shown by Equation~(\ref{eq:neuron_fc}). The main computational cost of a DNN comes from the weighted sum of inputs of each neuron and layer since the activation functions can be mapped and implemented using look-up tables. Note that although the inputs of a layer depend on the outputs of the previous layer, there is still a high level of parallelism within each layer.

\begin{equation}
    y_j = f(\sum^{}_{i} W_i X_i + b)
    \label{eq:neuron_fc}
\end{equation}

Modern DNNs such as AlexNet~\cite{AlexNet}, GoogleNet~\cite{GoogleNet} are composed of hundreds of layers where a single execution can demand the evaluation of millions of model parameters (i.e. weights). In consequence, DNN models tend to be huge, their size ranges from tens to hundreds of megabytes, or even gigabytes, and computing the weighted sum of inputs for each neuron of a given layer requires a large number of data movements between the different levels of the memory hierarchy and the processing units. Furthermore, in order to be able to exploit the high level of parallelism of the DNN layers, a high memory bandwidth is required to provide the necessary data to feed multiple processing units. This enormous traffic in the memory hierarchy represents a great portion of energy consumption for any given device and, together with the high memory storage and memory bandwidth requirements, heavily constrains the efficiency of the compute-centric architectures.

\begin{figure}[t!]
\centering
\includegraphics[width=0.8\columnwidth]{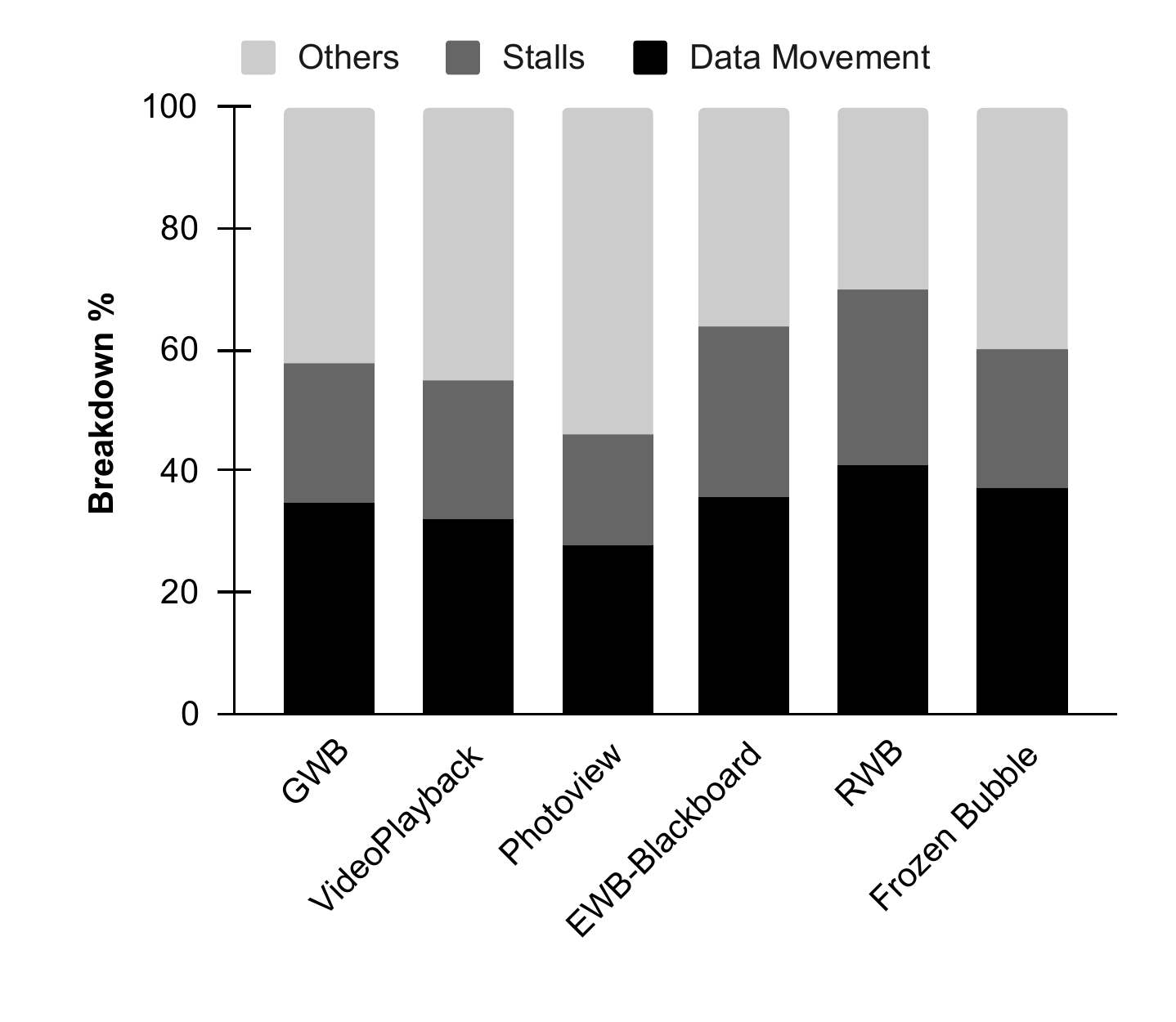}
\caption{Energy breakdown for emerging smartphone workloads on mobile platforms (adapted from~\cite{pandiyan2014quantifying}).}
\label{fig:mobile_energy_breakdown}
\end{figure}

The study in \cite{pandiyan2014quantifying} quantified the energy cost of data movement for emerging smartphone workloads on mobile platforms. To illustrate the high energy cost of the memory traffic, Fig.~\ref{fig:mobile_energy_breakdown} shows an example of the total energy breakdown for different mobile workloads, where \textit{Data Movement} considers the energy cost of moving data between different levels of the memory hierarchy, \textit{Stalls} indicate the energy consumption during stall cycles and the rest is represented as \textit{Others}. This analysis shows that on average $34.6\%$ of the total device's energy is spent on moving data from one level of the memory hierarchy to another. A similar study on a 28nm nVidia chip~\cite{NVIDIA28nmDallyLecture}, implies that the energy consumed for performing an operation is significantly lower than the amount spent for fetching operands. Specifically, the energy cost of fetching data from local memory (SRAM) and main memory (DRAM) is $26pJ$ and $16nJ$ respectively, compared to $1pJ$ and $20pJ$ for integer and floating point operations respectively. Consequently, some recent works have focused on reducing the data movements of the memory hierarchy by proposing accelerators based on memory-centric architectures.

\begin{figure}[t!]
\centering
\includegraphics[width=\columnwidth]{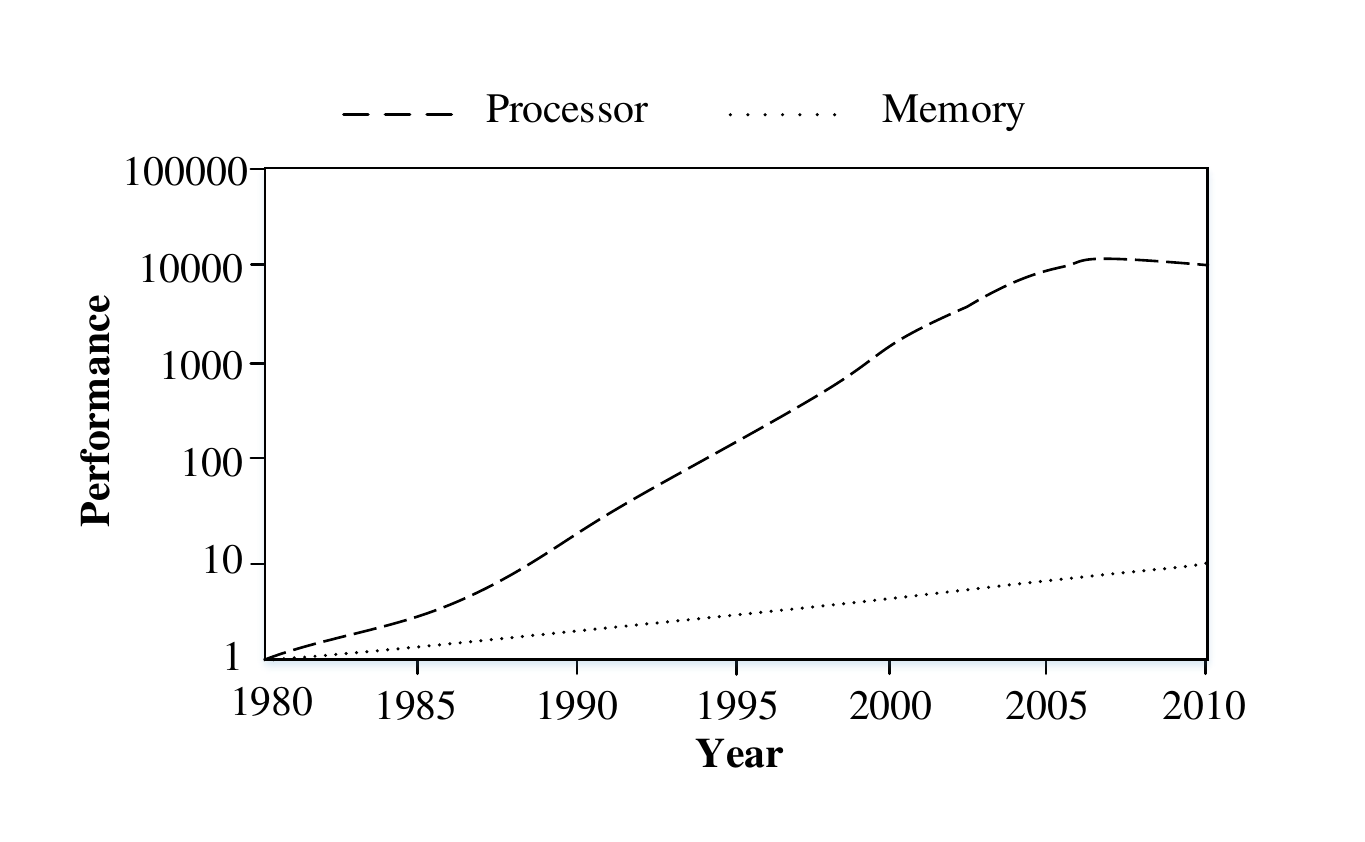}
\caption{Processor-Memory performance gap over time exemplifying the memory wall problem~\cite{HennessyPatterson12}.}
\label{fig:memory_wall}
\end{figure}

Most general-purpose computer systems and accelerators use a von-Neumann architecture, which separates the memory from the computing units. However, with the ever improving computational power of modern processing units, hardware communication fabrics struggle to advance at the same pace or even at a sufficient rate to efficiently support modern applications. It is well known that during the last 40 years the rate of improvement in microprocessor speed has exceeded the rate of improvement in DRAM memory speed. Both processor and memory performance have been improving exponentially, but the exponent for the processor is significantly higher than that of memory, in such a way that the difference between the two has also grown exponentially. Fig.~\ref{fig:memory_wall} illustrates the performance gap over time between the memory and the processor. Due to the end of Moore's law, the performance gap has become smaller in recent years, which has led to increased efforts to improve the memory technologies as well as the proposal of new memory-centric architectures. The ``Memory Wall"~\cite{wulf1995memorywall} problem has become the main bottleneck of data-centric applications since large amounts of data must be moved between the memory and the processing units, generating a large volume of communication traffic in bandwidth-limited systems~\cite{bandwidthwall}. DNNs are particularly susceptible to hitting the memory wall because of their large data demands and, hence, it has become one of the major challenges to be solved in order to execute ML algorithms efficiently.

\begin{figure}[t!]
\centering
\includegraphics[width=0.7\columnwidth]{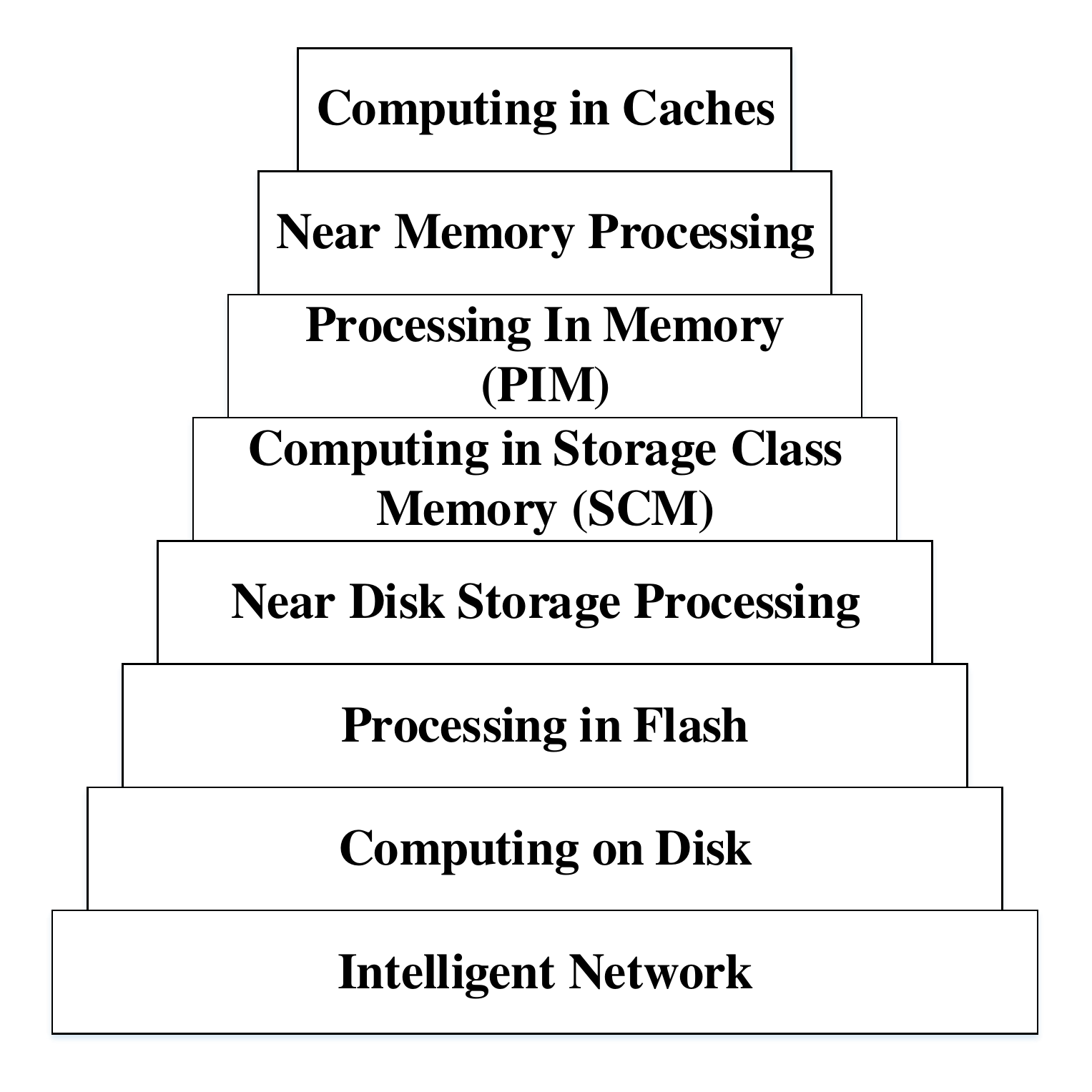}
\caption{Memory hierarchy enhanced with Near-Data Processing (NDP) capacity (adapted from~\cite{siegl2016data}).}
\label{fig:ndp_storage_hierarchy}
\end{figure}

In order to overcome the problems of compute-centric architectures and alleviate the cost of communicating data between memory and the processing units, many recently proposed architectures adopt a memory-centric design based on the so-called Near-Data Processing (NDP) paradigm. Fig.~\ref{fig:ndp_storage_hierarchy} shows how with the adoption of the NDP approach, storage and processing units will no longer be separated, and the memory hierarchy will shift towards a processing hierarchy or a hierarchical active-data storage~\cite{siegl2016data}.

NDP architectures can be further classified in two main categories. First, Near-Memory-Processing (NMP) consists on moving the processing units closer to the memory to mitigate the cost of the data movements. NMP has gained a lot of attention with the introduction of the 3D stacked memory technology, which allows the integration of logic and memory in the same chip by stacking multiple dies vertically. Popular state-of-the-art DNN accelerators such as DaDianNao~\cite{DaDianNao} and TPU~\cite{TPU} also follow the NMP approach by including bigger on-chip buffers to store inputs and weights, and reuse them across processing elements to limit the data transfers. On the other hand, Processing-In-Memory (PIM) consists in removing the necessity of moving data to the processing units by performing the computations inside the memory~\cite{ComputeDRAM}\cite{DRACC}\cite{DRISA}. The PIM approach is commonly implemented by exploiting the analog characteristics of emerging Non-Volatile Memories (NVM) such as ReRAM crossbars, but not only NVMs can be used to perform computations~\cite{[PIM_PCM]}\cite{PIM_STT_MRAM}, but also commodity memory technologies such as DRAMs and SRAMs have demonstrated the ability to perform logic operations with small changes to the memory array peripherals. Both NMP and PIM address the memory wall issue by spatially merging compute and storage units, drastically reducing the number of data transfers as well as their cost.

In this paper, we present a survey of near-data processing architectures for machine learning, and DNNs in particular. To achieve a balance between brevity and breadth, we only include state-of-the-art techniques implemented using commodity memories, 3D stacked memories based on DRAM, and ReRAM crossbars, although other emerging memories also provide NDP capabilities. We focus on describing the qualitative insights without generally including quantitative results, and paying special attention to the architectural and system-level techniques. This paper is expected to be useful for computer architecture researchers and practitioners.

The rest of the paper is organized as follows. Section~\ref{s:background} provides some background information on modern DNNs, commodity memories and emerging memory technologies. Section~\ref{s:ndp_sota} reviews the state-of-the-art NDP architectures for data-centric applications such as neural networks, and discusses the main advantages and disadvantages of the different proposals. Finally, Section~\ref{s:conclusions} concludes this paper with a discussion of future perspectives.

    

    

\section{Background}\label{s:background}
In the following subsections we review some terminology and concepts that may be helpful throughout this survey. First, we give a general description of deep neural networks (DNNs), including the main DNN categories and their different types of layers. Next, we review some fundamental concepts about conventional memory technologies such as DRAM and SRAM, which have proven to be capable of doing computations inside the memory arrays. Finally, we discuss new memory technologies such as ReRAM crossbars and 3D-stacked memories, which offer more opportunities, over commodity memory technologies, to implement a highly efficient DNN accelerator in terms of both performance and energy consumption.

\subsection{Deep Neural Networks (DNNs)}
Deep learning~\cite{Lecun_DEEP_LEARNING}\cite{Schmidhuber_DEEP_LEARNING} is a set of algorithms that is part of a larger family of machine learning methods based on artificial neural networks and data representation techniques that allow to learn and interpret features of large datasets. DNNs are a type of deep learning model that aims to mimic the human brain functionalities based on a very simple artificial neuron that performs a nonlinear function, a.k.a. activation function, on a weighted sum of the inputs. As discussed in Section~\ref{s:intro}, these artificial neurons are arranged into multiple layers. The term ``deep" refers to the use of a large number of layers, which results in more accurate models that capture complex patterns and concepts. DNNs are typically used as classifiers in order to identify which class the current input belongs to, so the last layer of the network often performs a softmax function that generates a probability distribution over a list of potential classes.

DNN's operation has two phases, training (or learning) and inference (or prediction), which refer to the DNN model construction and use respectively. The training procedure determines the weights and parameters of a DNN, adjusting them repeatedly until the DNN achieves the desired accuracy. During training, a large set of examples, with its corresponding labels indicating the correct classification, is used to execute the DNN's forward pass and measure the error against the correct labels. Then, the error is used in the DNN's backward pass to update the weights. On the other hand, inference uses the DNN model developed during the training phase to make predictions on unseen data. Although the DNN training is complex and computationally expensive, it is usually done only once per model, and then, the learned weights are used in inference as many times as it is required to classify new input data. On the other hand, DNN inference has strict latency and/or energy constraints and, hence, many research works focus on improving the execution of the inference phase. Nevertheless, a recent line of research with increasing popularity aims to propose efficient DNN accelerators for both phases.

DNNs can be classified in three main categories. Multi-Layer Perceptrons (MLP)~\cite{MLP} consist of multiple Fully-Connected (FC) layers in which every input neuron is connected, via synapses with particular weights, to every output neuron. Convolutional Neural Networks (CNN)~\cite{CNN} are composed of multiple convolutional layers to extract features, usually followed by one or several FC layers to perform the final classification. CNNs have proved to be particularly efficient for image and video processing. Finally, Recurrent Neural Networks (RNN)~\cite{RNN} consist of multiple layers of cells with feedback connections. RNN cells store information from past executions to improve the accuracy of future predictions, where cells consist of multiple single-layer FC networks commonly referred as gates. Each type of DNN is especially effective for a specific subset of cognitive applications. Moreover, for each application, each DNN has a different composition of layers with specific operations. The fully-connected (FC), convolutional and recurrent layers take up the bulk of the computations in most DNNs. Other types of layers performing pooling, normalization or activation functions are also common in modern DNNs. However, these other layers have no synaptic weights and represent a very low percentage of the DNN execution time. In consequence, state-of-the-art DNN accelerators focus on optimizing the execution of FC and convolutional layers.

\subsection{Conventional Memory Technologies}\label{s:background_cm}
Conventional memory technologies are widely used in the memory hierarchy of many current systems. DRAM is commonly employed for the main memory of most computer systems while SRAM is used to implement CPU caches and register files, as well as small buffers for different components. Both, DRAM and SRAM, are considered commodity memories, and a popular line of research tries to take advantage of the widespread use of these memories to perform near data processing by proposing minor changes to the memory chip circuitry, as we will describe later in Section~\ref{s:ndp_sota_cm}.

\begin{figure}[t!]
\centering
\includegraphics[width=0.9\columnwidth]{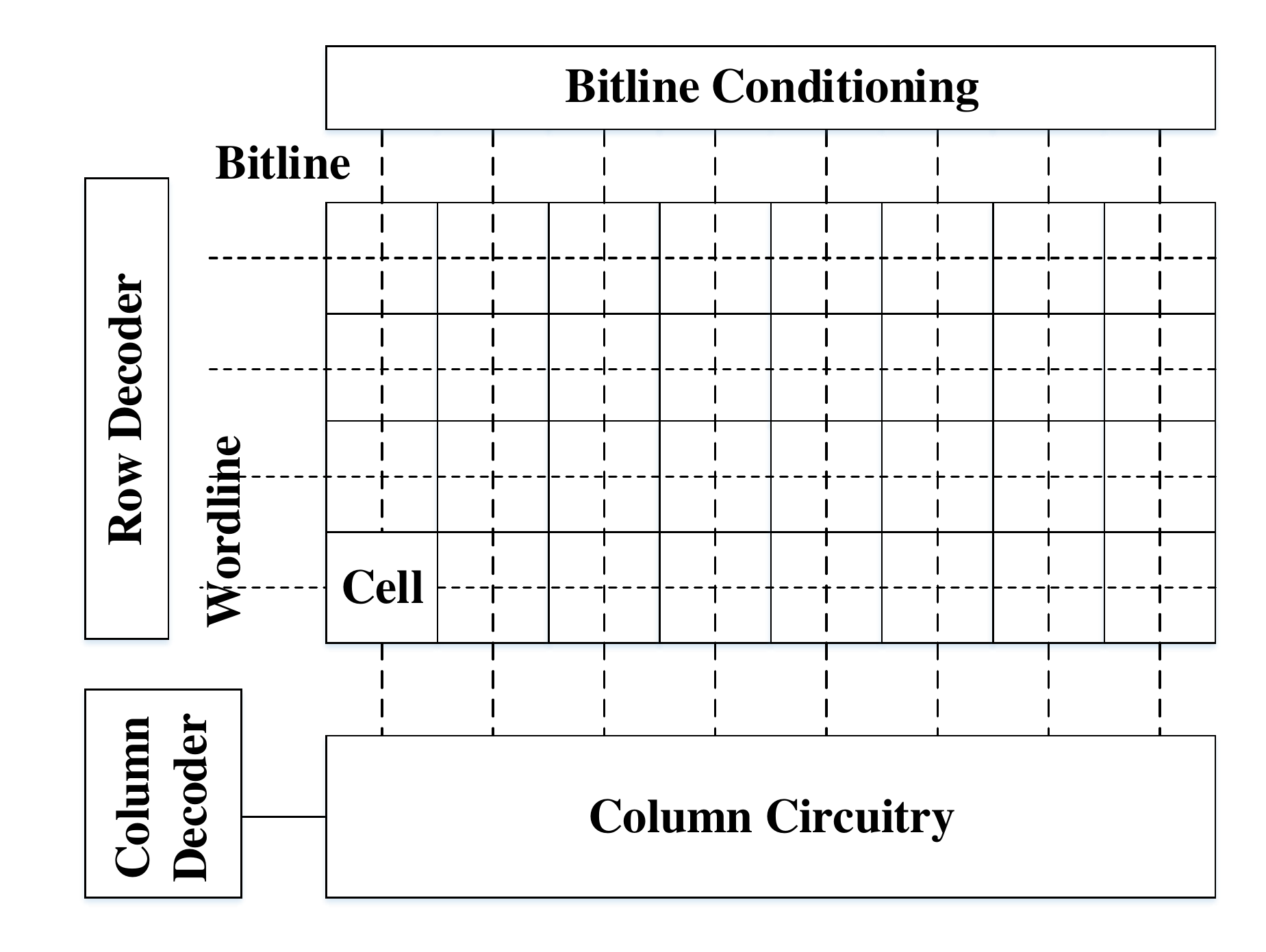}
\caption{General overview of conventional 2D memory array organization.}
\label{fig:2d_mem_array}
\end{figure}

A DRAM or SRAM chip is commonly divided into multiple memory banks connected to a global bus. Each bank consists of multiple memory arrays containing the DRAM/SRAM cells. Fig.~\ref{fig:2d_mem_array} shows an example of a typical 2D organization (i.e. memory array) of a DRAM or SRAM chip. The cells of a given row of the memory array are connected to a Word-Line (WL) while the cells in the same column share the same Bit-Line (BL). The read and write operations are performed by activating the WL of the corresponding row of the array, and exploiting the charge sharing effect of the bitlines. Among the peripherals of the memory array, the decoders are required to map and translate the addresses of the memory accesses to specific rows and columns of the array, while the sense amplifiers in the column circuitry detect small changes in the voltage of the bitlines and amplify the swing over a reference voltage to help to perform the read and write operations.

\begin{figure}[t!]
\centering
\subfloat[1T1C DRAM cell.]{
    \includegraphics[width=0.35\columnwidth]{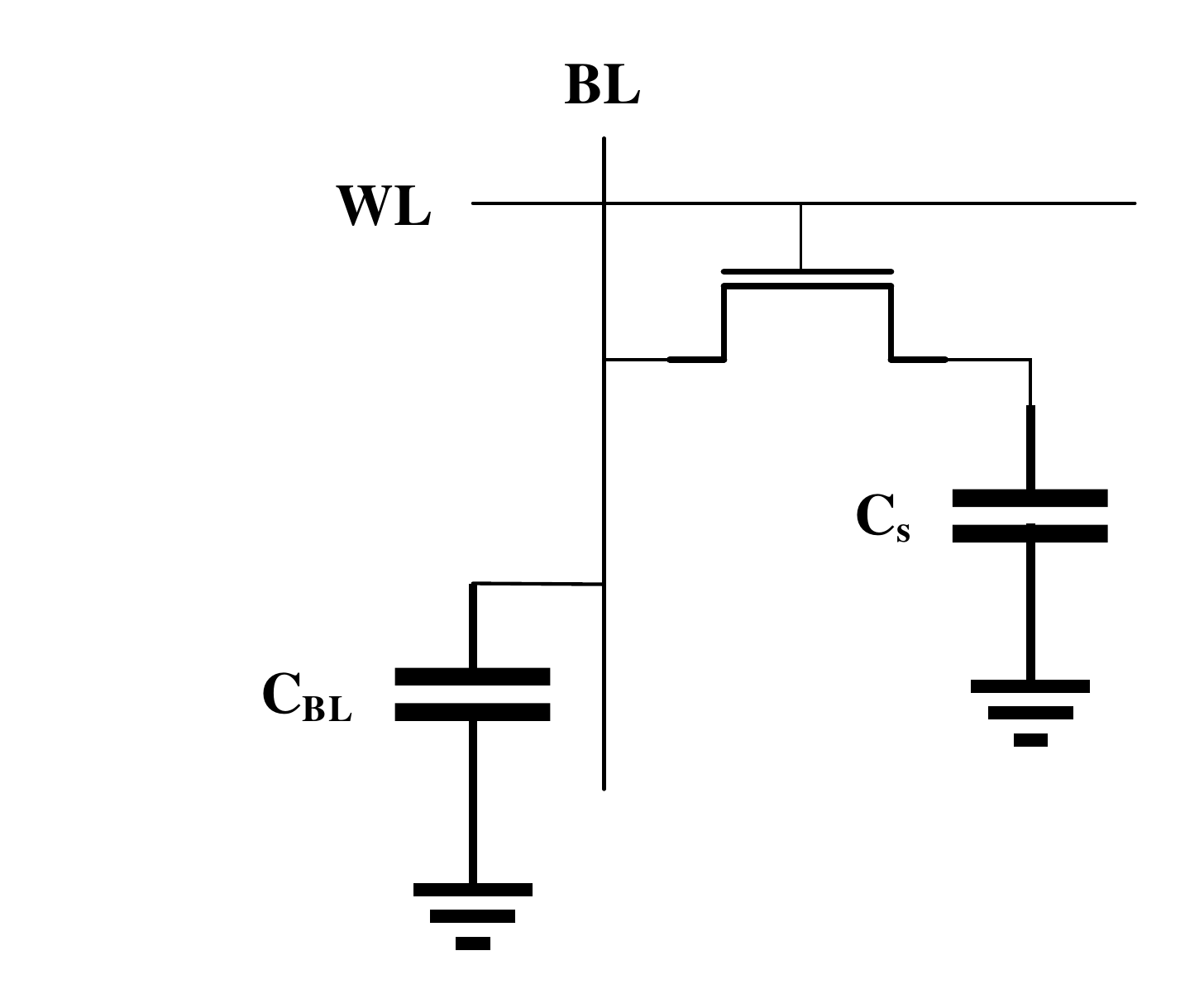}
    \label{fig:1t1c_dram_cell}
}
\subfloat[6T SRAM cell.]{
    \includegraphics[width=0.55\columnwidth]{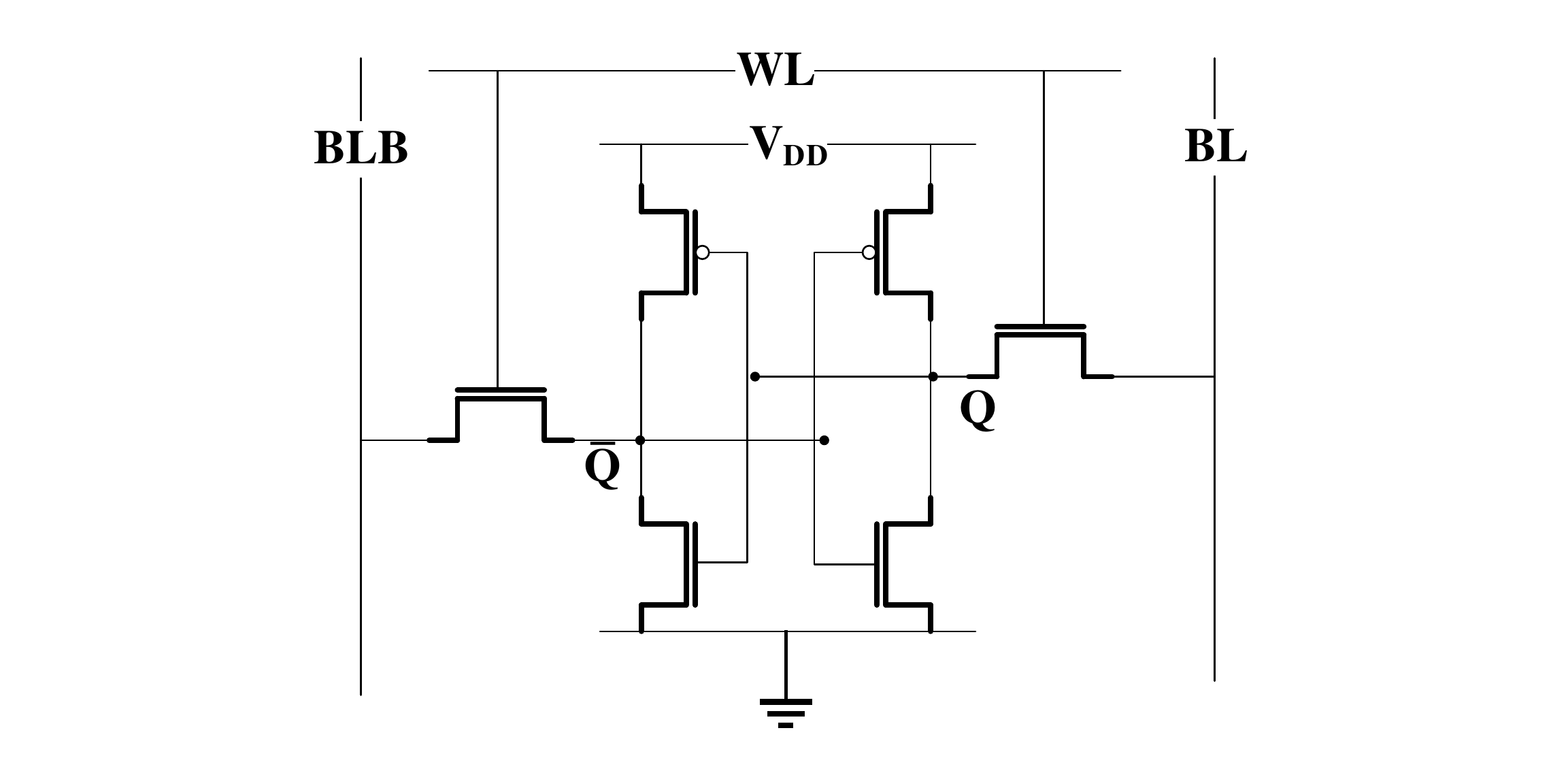}
    \label{fig:6t_sram_cell}
}
\caption{Examples of typical (a) DRAM and (b) SRAM cell designs.}
\label{fig:mem_cells}
\end{figure}

A DRAM cell stores one bit of information using a capacitor and an access transistor as shown in Fig.~\ref{fig:1t1c_dram_cell}, which is known as the 1T1C cell design. During the read operation, all the bitlines of the array are pre-charged equally to $V_{DD}/2$ using the peripheral drivers. Then, the corresponding WL of the row to be read is activated. Due to the charge sharing effect, the capacitance of the bitlines start to lose (if stored value is 0) or gain charge (if 1 is stored) changing the pre-charged voltages by a small amount, and destroying the values stored in the cells of the row. The small voltage swing in each of the bitlines is sensed and amplified to a stable state by the sense amplifiers and, hence, the value of each bitline is converted to a strong one (i.e. $V_{DD}$) or zero accordingly. Finally, since the WL is still active and connecting the capacitor of each cell of the row to the corresponding BL, the capacitors are fully charged (or discharged) restoring their original value. Similarly, to perform a write operation in a row of cells, the bitlines are pre-charged to $V_{DD}$ (to store 1) or 0 (to store 0), and  the corresponding WL is activated. Since the sense amplifiers hold the voltage of each bitline at a stable state, the capacitor of each cell will continue to charge (or discharge) until the pre-stored value is overwritten with the new value.

SRAM cells store bits of data and its negated version using four transistors in the form of cross-coupled inverters. In addition, two more transistors are employed to control the access to the bitlines. Fig.~\ref{fig:6t_sram_cell} shows a typical SRAM 6T cell design. On a read request, both bitlines, $BL$ and $\overline{BL}$, are pre-charged to $V_{DD}$. Then, when the WL of a cell is activated, one of the bitlines starts to discharge, that is, $BL$ if the stored value is zero or $\overline{BL}$ otherwise. The read operation has the potential to change the stored value and, hence, the size of the transistors must be chosen carefully so that the original value does not flip. To write a new value in a 6T cell, $BL$ and $\overline{BL}$ are driven to high and low to store a one, or the other way around to store a zero. Then, the WL is activated, and the bitlines overpower the cell with the new value.

DRAM cells are small since they are typically composed of one transistor and one capacitor, which provides higher density over SRAM memories. Despite the higher capacity of DRAMs, the 1T1C cells are not ideal due to the high leakage of the capacitor's charge, which might lead to losing the stored values after some time. In consequence, DRAM cells require additional circuitry to perform regular refreshes of the cell values, which has a negative impact on DRAM performance. On the other hand, SRAM is faster than DRAM and does not need any refresh circuitry. However, the 6T cells are large resulting in lower density and higher cost. Therefore, DRAM is preferred over SRAM for main memory due to its higher capacity and lower cost, while SRAM is employed for caches and small buffers due to its higher performance.

\subsection{Resistive Random Access Memory (ReRAM)}\label{s:background_reram}
In 1971, Leon Chua~\cite{chua1971memristor} theorized about the existence of a fourth passive electrical component that he referred to as the ``memristor'' (or memory resistor), complementing the quartet of fundamental electrical components together with the resistor, the inductor and the capacitor. In 2008, a team at HP Labs managed to develop the first physical memristor switch~\cite{strukov2008developingmemristor}, and later on started the prototyping of ReRAM memories using them.

\begin{figure}[t!]
\centering
\subfloat[]{
    \includegraphics[width=0.30\columnwidth]{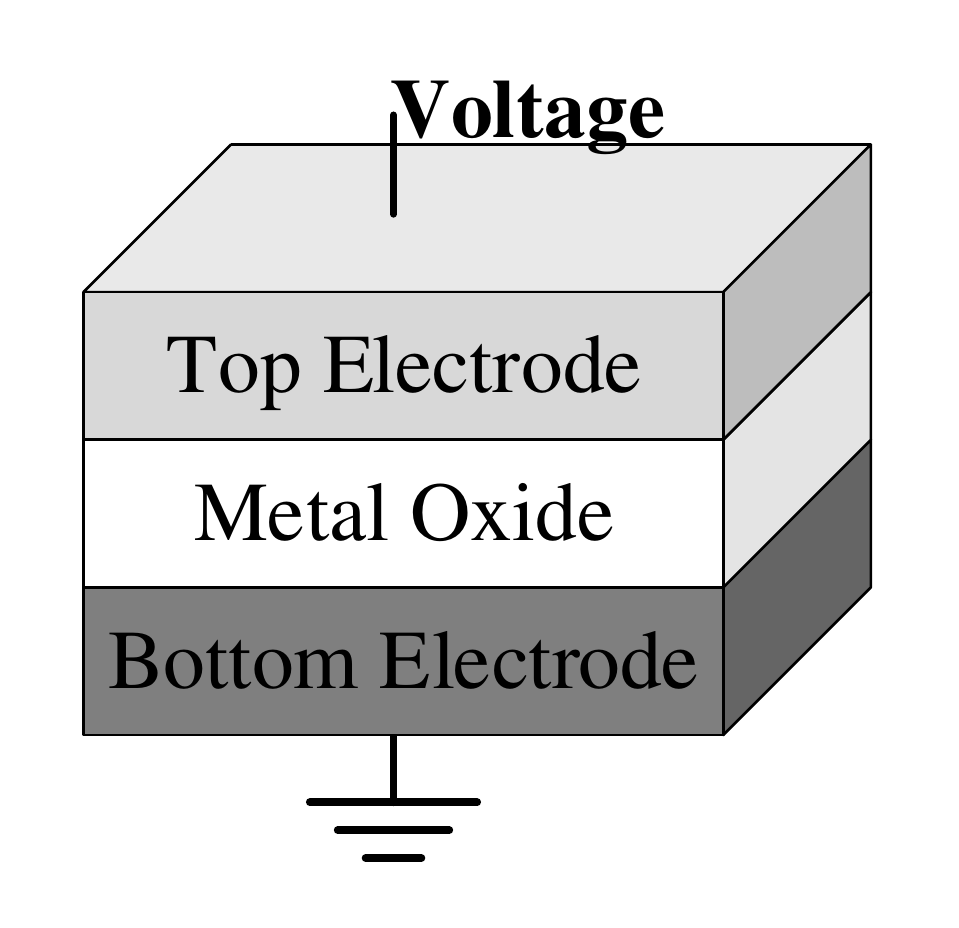}
    \label{fig:memristor_structure}
}
\subfloat[]{
    \includegraphics[width=0.30\columnwidth]{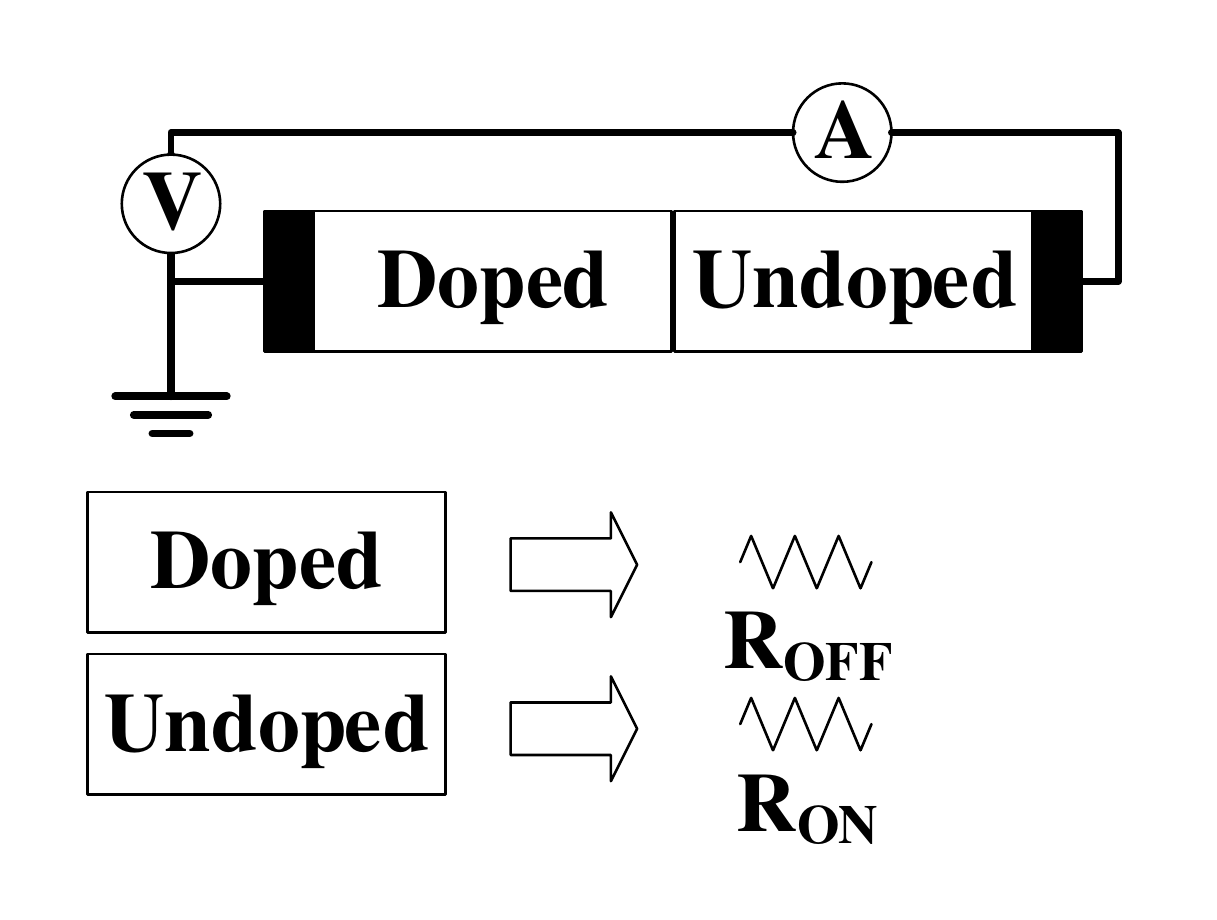}
    \label{fig:memristor_sw}
}
\subfloat[]{
    \includegraphics[width=0.30\columnwidth]{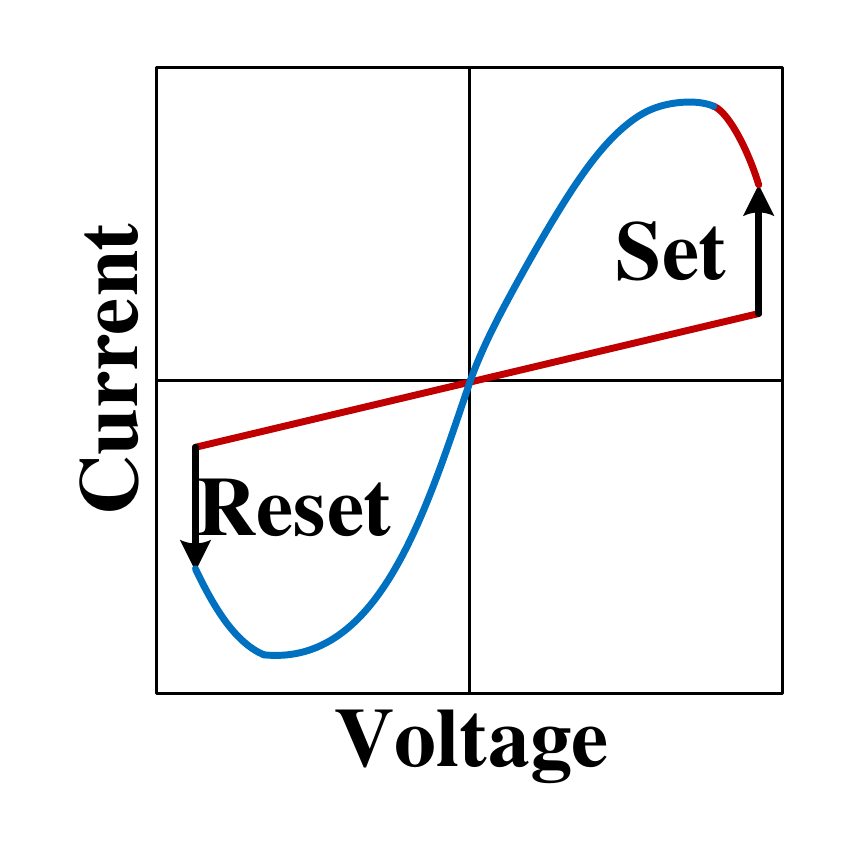}
    \label{fig:memristor_IV_curve}
}
\caption{(a) Structure of a ReRAM cell; (b) memristor switching behavior; (c) I-V curve of bipolar switching (adapted from~\cite{strukov2008developingmemristor} and~\cite{seal-lab-tech-report/reram-structure}).}
\label{fig:memristor}
\end{figure}

A memristor is a non-linear two-terminal electrical component with a layer of a resistive switching material sandwiched between the top and bottom electrodes as shown in Fig.~\ref{fig:memristor_structure}. The resistive switching material layer consists of two regions, "doped" and "undoped", with low ($R_{ON}$) and high ($R_{OFF}$) resistances respectively, which is equivalent to connecting two resistors in series as shown in Fig.~\ref{fig:memristor_sw}. The memristor's electrical resistance depends on the history of current that had previously flowed through the device, that is, the amount of electric charge flowing through the memristor device changes the length of the two regions based on the direction of the current and, as a result, the total resistance varies. The memristor device remembers its most recent resistance after turning off the power supply achieving the so-called non-volatility property~\cite{Akinaga_ReRAM}.

Fig.~\ref{fig:memristor_IV_curve} shows the I-V curve of a typical bipolar ReRAM cell. By applying an external voltage across a memristor, a ReRAM cell can be switched between a high resistance state (HRS) and a low resistance state (LRS), which are used to represent the logic “0” and “1”, respectively. Switching a cell from HRS (logic “0”) to LRS (logic “1”) is a SET operation, and the reverse process is a RESET operation. To SET the cell, a positive voltage that can generate sufficient write current is required, while to RESET the cell, a negative voltage with a proper magnitude is necessary. In addition, ReRAM cells can store more than one bit of information by employing various resistance levels, which can be realized by changing the resistance of a ReRAM cell gradually with finer write control. This multi-level cell (MLC) characteristic makes the ReRAM memories suitable for storing a large amount of DNN weights.

\begin{figure}[ht!]
\centering
\includegraphics[width=0.6\columnwidth]{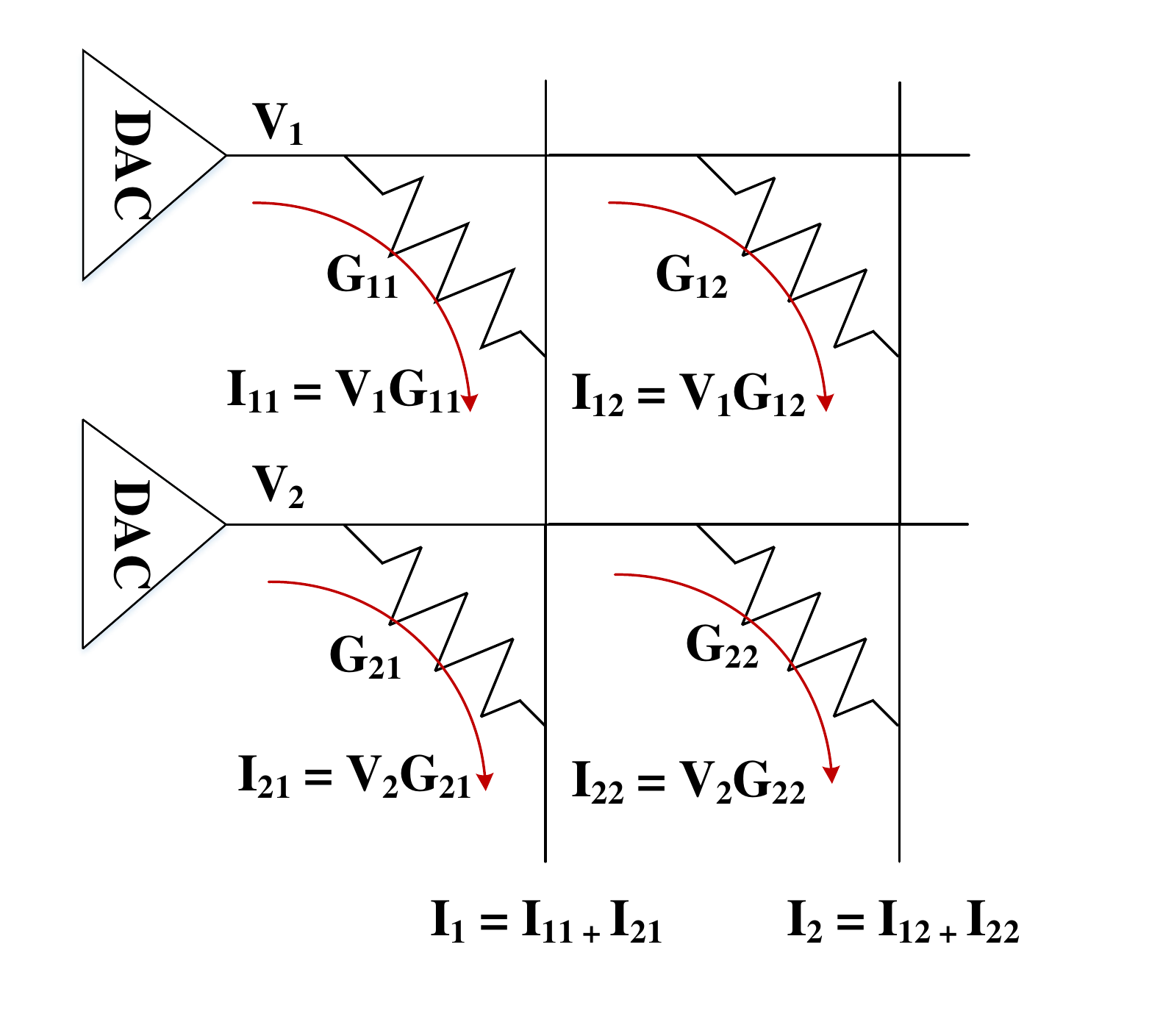}
\caption{Analog dot-product operation using ReRAM.}
\label{fig:reram_dot_product}
\end{figure}

Recent works have demonstrated the use of ReRAM devices to perform dot-product computations which, as described previously, are common in many DNN layers. Fig.~\ref{fig:reram_dot_product} shows an example of an analog MAC operation, and a MVM using a ReRAM crossbar. Similar to the conventional memory technologies, each BL of a ReRAM memory array connects to each WL through a ReRAM cell, which usually consists of a single memristor with an access transistor also known as the 1T1R cell design. Applying a voltage $V_i$ to a ReRAM cell with resistance $R_i$, results in a current of $V_i \times G_i$ passing from the cell to the bitline, as per Ohm's law, where $G_i = 1/R_i$. Then, from Kirchoff’s law, the total current from the bitline is the sum of currents generated through all the cells sharing the same bitline, that is, the total current (I) is the dot-product of input voltages at each row (V) and cell conductances (G) in a column as shown by Equation~\ref{eq:analog_mac}. This can be used to implement a DNN in a efficient manner. In this case, the synaptic weights of neurons are encoded as conductances of the ReRAM cells. Then, the total current of a BL is used to compute the output of a neuron in a given layer. Note that the dot-products are done massively in parallel performing a MVM in a single time step.

\begin{equation}
    I = \sum_{i}V_i \times G_i
    \label{eq:analog_mac}
\end{equation}

Implementing DNN accelerators using ReRAM devices is quite effective as ReRAM provides high capacity to store the weights, low latency to access them, and high performance PIM capabilities to perform the dot-products. In addition, since the leakage current of memristors is significantly lower than DRAM/SRAM cells, the energy efficiency can be highly improved. Furthermore, ReRAM devices can also be used for implementing bitwise operations and search operations which can be useful to implement different DNN layers. Section~\ref{s:ndp_sota_reram} describes state-of-the-art NDP accelerators for DNNs based on ReRAM architectures.

Despite the numerous benefits, the use of ReRAM devices also present several challenges~\cite{ReRAM_ACCELERATOR_SURVEY}:

\begin{itemize}
\item \textit{Challenges in Achieving High Performance}: Despite the low read latency of ReRAM, changing a memristor's resistance value requires a high and long enough voltage, which incurs in high write latency and energy consumption compared to SRAM. In consequence, the weights of a neural network are usually pre-stored in multiple ReRAM crossbars, and are reused across multiple executions to avoid writes as much as possible. Furthermore, ReRAM is not suitable for DNN training due to the high number of writes that are required to update the weights. In addition, the non-ideal characteristics and limitations of ReRAM, such as process variations, can reduce their performance even further.

\item \textit{Challenges in Analog/Digital Domain}: The dot-product computations within a ReRAM crossbar require input voltages in the wordlines and produce output currents in the bitlines, thus, operating in the analog domain. However, not all components in a DNN accelerator work in the analog domain, either due to the impossibility of implementing the operations of some layers in analog domain, or because some operations are executed much faster using digital components and cannot be accelerated with the use of ReRAM. Therefore, to perform computations using ReRAM, the digital inputs must be converted to analog voltages, and the analog outputs are converted back to digital, hence, requiring the use of DACs and ADCs and incurring in high power consumption and area overheads, which limits the scalability of ReRAM crossbars. For example, ISAAC's ADCs account for 58\% of power and 31\% of area for a given tile of the accelerator~\cite{ISAAC}. In addition, the precision of the analog computations, and the overall accuracy of the DNN model, may be affected by the signal degradation of using DACs/ADCs, external noise signals, or non-zero wire resistance, among other sources of errors.

\item \textit{Reliability Challenges of ReRAM}: Most ReRAM reliability issues come from the high defect rate and process variations (PV) that happen during fabrication. The faults are considered \textit{soft} when the resistance of the memristor can still be changed but the result is different from the expected value, or \textit{hard} when the memristor gets stuck at a given resistance state. The hard faults may happen not only due to PV but also due to the ReRAM's limited write endurance, which is another important obstacle for its general applicability. To mitigate these issues, redundancy-based techniques can be used, however, they incur in significant complexity and area overheads.
\end{itemize}

\subsection{3D-Stacked Memory}\label{s:background_3d}
3D stacking aids the design of future computing systems by allowing for wider buses and faster on-chip memories, and by enabling the composition of heterogeneous dies, built using different process technologies, within a single package. High density 3D memory is a promising emerging technology for the memory system of DNN accelerators. It consists of stacking multiple memory dies on top of each other, which increases the memory capacity and bandwidth compared to 2D memory, and also reduces the access latency due to the shorter on-chip wiring interconnection~\cite{davis2005prosandconsof3d}. These aspects can lead to an overall improvement in performance and power efficiency. Compared with the conventional 2D DRAM, 3D memory provides an order of magnitude higher bandwidth (160 to 250 GBps) with up to 5x better energy efficiency and, hence, 3D memory is an excellent option for meeting the high throughput, low energy requirements of scalable DNN accelerators.

The 3D memory dies are commonly based on DRAM, but the integration of other memory technologies is being actively researched with very promising results. For example, the 3D XPoint developed by Intel and Micron is a commercial Non-Volatile Memory (NVM) based on ReRAM. On the other hand, recent advances in low-capacitance through-silicon vias (TSVs) technology have enabled 3D memory that includes a few DRAM dies on top of a logic chip. Although there are numerous implementations of 3D-stacked memory technologies, the Hybrid Memory Cube (HMC) is the preferred choice for most DNN accelerator proposals~\cite{neurocube/ISCA.2016.41}~\cite{DEEPTRAIN}. There are several works that place multiple arrays of NN processing elements on the logic layer of a HMC to improve performance and power efficiency. Section~\ref{s:ndp_sota_3d} describes state-of-the-art NDP accelerators for DNNs based on DRAM HMC architectures.

\begin{figure}[t!]
\centering
\includegraphics[width=0.60\columnwidth]{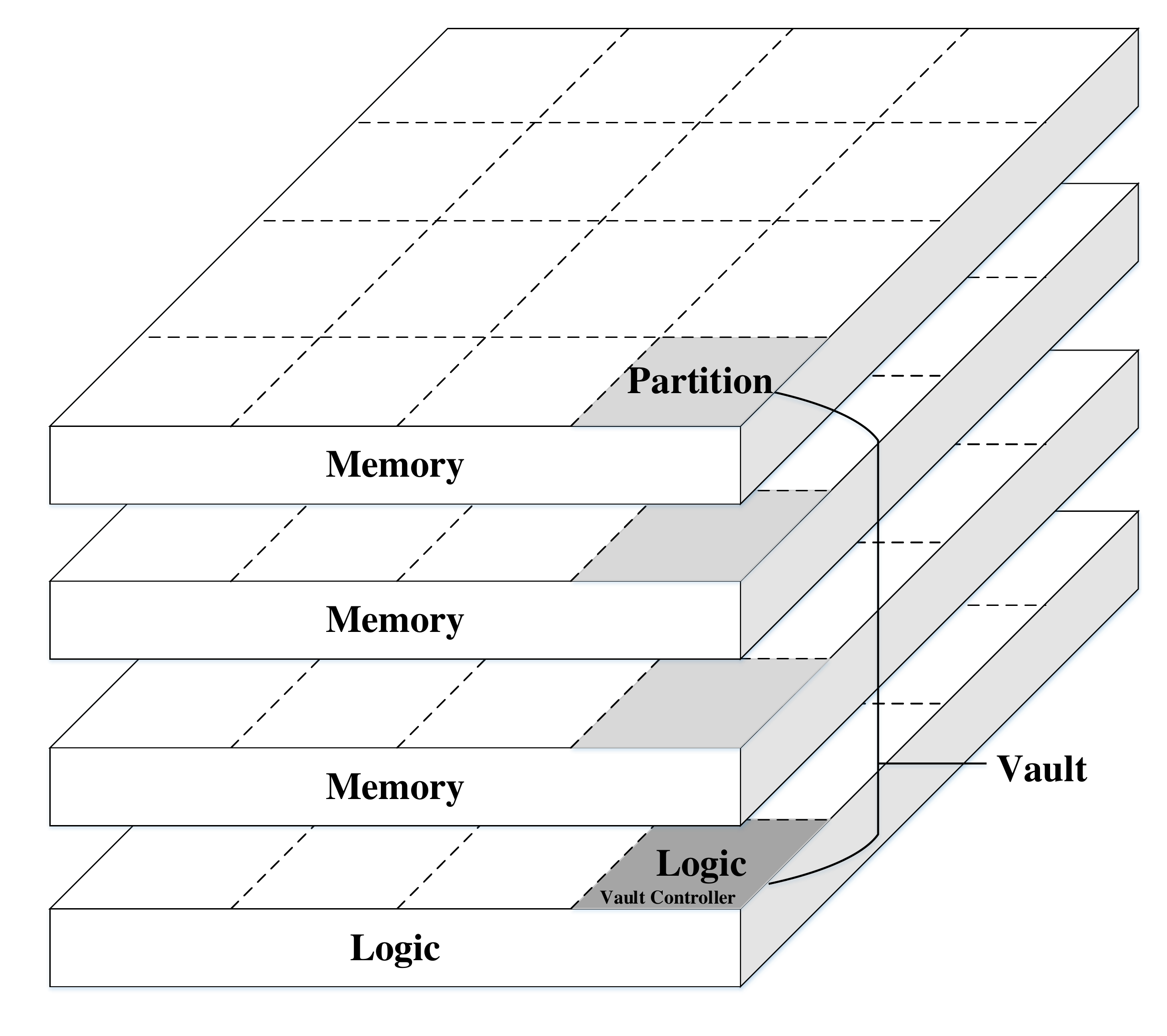}
\caption{High-level architecture of Hybrid Memory Cube (HMC) (adapted from~\cite{hybrid-memory-cube}).}
\label{fig:3d_stack_memory}
\end{figure}

The Hybrid Memory Cube (HMC) is designed for high performance data-centric applications. Fig.~\ref{fig:3d_stack_memory} shows a high-level diagram of a HMC, which is usually composed of multiple vertically stacked DRAM dies with a single logic layer at the bottom. The different layers are interconnected using thousands of TSVs to attain the much desired high memory bandwidth, and each DRAM die is divided into multiple partitions in a 2D grid where the corresponding partitions on the vertical direction form a single vault. In addition, the large number of TSVs can be organized into multiple independently-operated channels that exploit memory level parallelism, that is, one channel per vault where each vault has an independent vault controller on the logic die and, thus, multiple partitions in the DRAM die can be accessed simultaneously. The HMC provides highly parallel access to the memory which is well suited to the highly parallel architecture of the DNN accelerators~\cite{HMC_Analysis}. The logic and memory dies can be fabricated in different process technology nodes. However, the area of the logic die relative to the memory dies is constrained by the package, and power dissipation is limited by much tighter thermal constraints.

To fully exploit the benefits of 3D-stacked memory, there are several challenges to address. First, given the characteristics of 3D memory, it may be worth revisiting the design of on-chip buffers in the logic die, since the lower latency and energy cost of the accesses to main memory allow for smaller and faster on-chip buffers with different use cases. Second, 3D integration technology also provides opportunities to rethink where computations are executed, potentially moving some operations closer to the actual memory locations. Third, HMC changes the memory and compute hardware creating a highly parallel system with multiple vertical channels and, thus, which opens the door to new approaches for dataflow scheduling and efficient partitioning of DNN computations. Last but not least, the thermal issues for 3D stacking as well as higher manufacturing complexity lead to lower yield and problematic testability. Therefore, stacking multiple dies can increase the operation temperature resulting in performance degradation and, hence, proper cooling methods and better manufacturing solutions are required to extend the adoption of this technology.

To summarize, 3D-stacking is not just a solution for the memory wall but rather a key-enabler in single chip packaging, with low to medium risk involved, to accomplish a reduction in system-level power as well as form factor. HMC integrates fast logic layer and dense memory layers and allows for embedding fast processing units close to a large and high-bandwidth memory system. Furthermore, HMC offers a unique novel abstract bus interface to communicate with the CPU host, and high parallelism via the so-called independent vaults, making it a very promising solution for DNN accelerators.


\section{Near-Data Processing Architectures}\label{s:ndp_sota}
Recent advances and discoveries in new memory technologies, together with the increasing use of cognitive applications based on neural networks, have led to a growing interest in the area of Near-Data Processing (NDP) architectures for data-centric applications and specially DNNs. In the following subsections, we discuss and categorize multiple state-of-the-art approaches based on the memory technology employed. Section~\ref{s:ndp_sota_cm} describes NDP techniques based on commodity memory technologies such as DRAM and SRAM not only to improve the execution of neural networks but also for bulk bitwise operations. In Section~\ref{s:ndp_sota_3d}, we review two DNN accelerators based on 3D-stacked DRAM memory with HMCs. Finally, Section~\ref{s:ndp_sota_reram} gives a general overview of the NDP architectures of several accelerators for DNNs based on ReRAM.

\subsection{Commodity Memory based NDP Architectures}\label{s:ndp_sota_cm}
\subsubsection{Ambit}
Seshadri et al. proposed Ambit~\cite{seshadri2017ambit}, an in-memory accelerator to perform bulk bitwise operations using commodity DRAM technology. Many important applications can benefit of bitwise operations on large bit vectors, but its throughput is limited by the memory bandwidth available to the processing units. In current general purpose systems, such as CPU or GPGPU, a bulk bitwise operation requires a large number of data movements in the memory hierarchy, which result in high latency, memory traffic, and energy consumption. To overcome these bottlenecks, Ambit exploits the analog operation of DRAM technology to perform bitwise operations completely inside the DRAM memory arrays, thereby exploiting the full internal DRAM bandwidth and the memory-level parallelism across multiple arrays. Ambit consists of two components, \textbf{Ambit-AND-OR} and \textbf{Ambit-NOT}.

\begin{figure}[t!]
\centering
\includegraphics[width=\columnwidth]{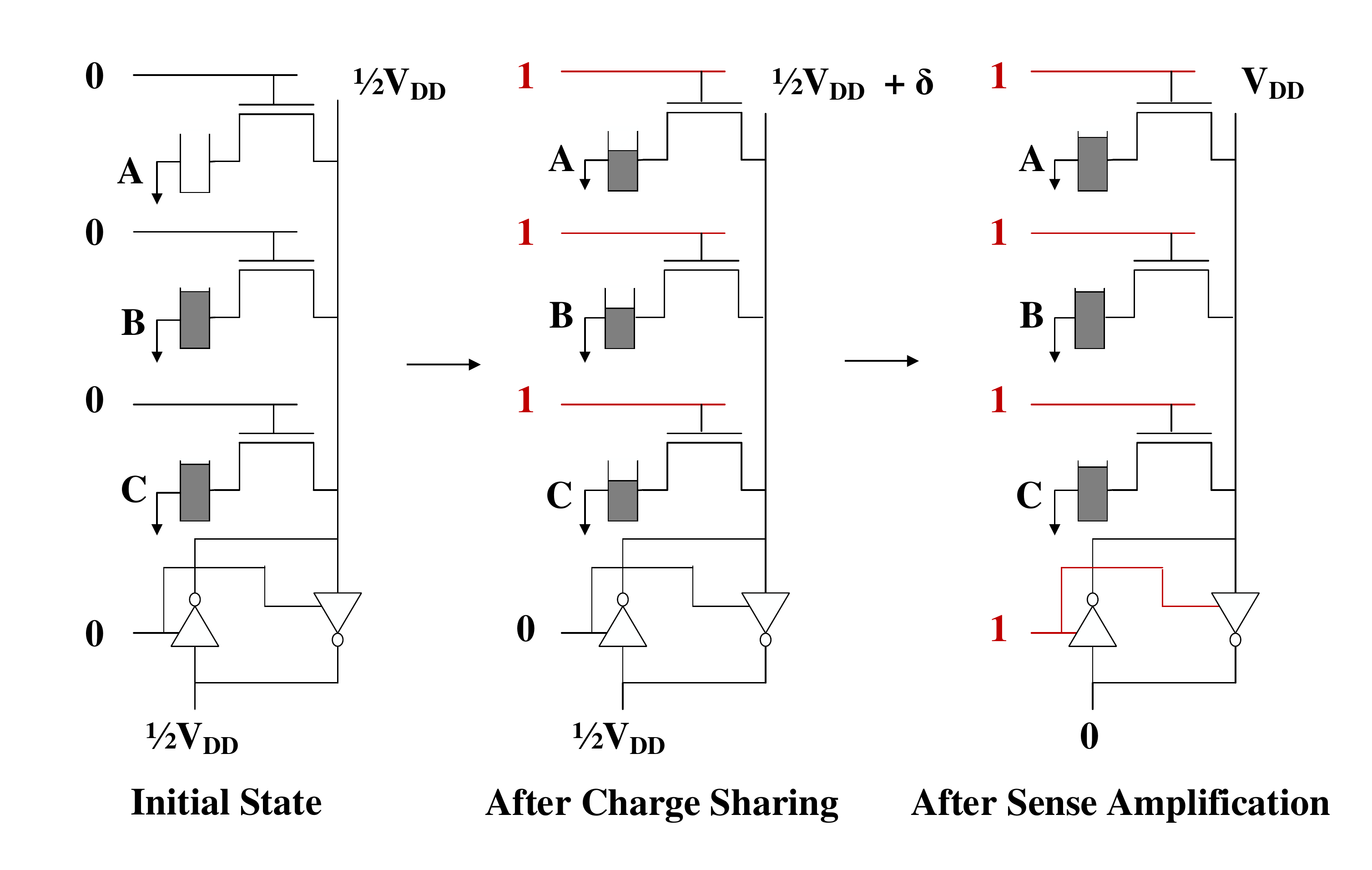}
\caption{Example of a Triple-Row Activation (TRA) (adapted from~\cite{seshadri2017ambit}).}
\label{fig:tra}
\end{figure}

Ambit-AND-OR is based on the key idea of Triple-Row Activation (TRA), which consists of a simultaneous activation of three DRAM rows that share the same set of sense amplifiers, enabling the system to perform bitwise AND and OR operations. TRA simultaneously connects a sense amplifier with three DRAM cells on the same bitline, where the final state of the bitline is expected to be VDD if at least two of the three cells are initially fully charged, or 0, if at least two of the three cells are initially fully empty. Due to the charge sharing principle and the use of sense amplifiers in the column peripherals, TRA results in a majority function across the cells in the three rows.

Fig.~\ref{fig:tra} shows an example of how TRA works when two of the three cells sharing the same bitline are initially in the charged state, that is, cells $B$ and $C$ are storing a logic 1 while $A$ stores a 0. First, the bitline is pre-charged to $VDD/2$ in the initial state. Then, when the wordlines of all the three cells are raised simultaneously, charge sharing results in a positive deviation on the bitline. Therefore, after sense amplification, the sense amplifier drives the bitline to VDD as discussed in Section~\ref{s:background_cm}, and as a result, fully charges all the three cells overwriting the original values. By controlling the initial value of one of the three cells, i.e. $C$ in the example, TRA is used to perform a bitwise AND (C = 0) or OR (C = 1) of the other two cells, $A$ and $B$, as shown by Equation~\ref{eq:tra}, where $A$, $B$ and $C$ represent the logical values of the three cells performing the bitwise majority function.

\begin{equation}
    AB + BC + CA = C(A + B) + \bar{C}(AB)
    \label{eq:tra}
\end{equation}

A naive mechanism to support TRA may be very costly and unreliable due to five potential issues regarding its implementation. First, when simultaneously activating three cells, the deviation on the bitline may be smaller than when activating only one cell, which may cause the sense amplifiers to detect wrong values. Second, the transistors and bitlines may not behave ideally due to process variations, affecting the reliability of TRA and the correctness of its results. The authors tackle these two issues by performing rigorous SPICE simulations, and proving that TRA is reliable and works correctly under several circumstances. Third, TRA overwrites the data of all the three source cells with the final result value, thereby destroying their original values. Fourth, DRAM cells leak charge over time, which may lead to unexpected TRA functionality if the cells involved have leaked significantly. Fifth, conventional DRAM is not able to simultaneously activate three arbitrary rows by default and, hence, TRA requires multiple row decoders and a wider address bus.

To overcome the last three issues, they present a practical implementation based on three ideas. First, they restrict the TRA to a designated set of rows, and employ RowClone~\cite{seshadri2013rowclone} to perform the required copy/initialization operations efficiently inside the DRAM arrays before and after completing each bitwise operation, avoiding the overwrite (issue 3) and leakage (issue 4) problems. Second, they reserve and define a fixed subset of DRAM row addresses and map them so that a single address is used to perform a TRA on a predefined set of designated rows. Finally, they split the row decoder in two parts: one small part handles all the activations of the designated rows, and the other part handles the activation of regular data rows, reducing the complexity of the row decoder (issue 5).

\begin{figure*}[t!]
\centering
\includegraphics[width=.95\textwidth]{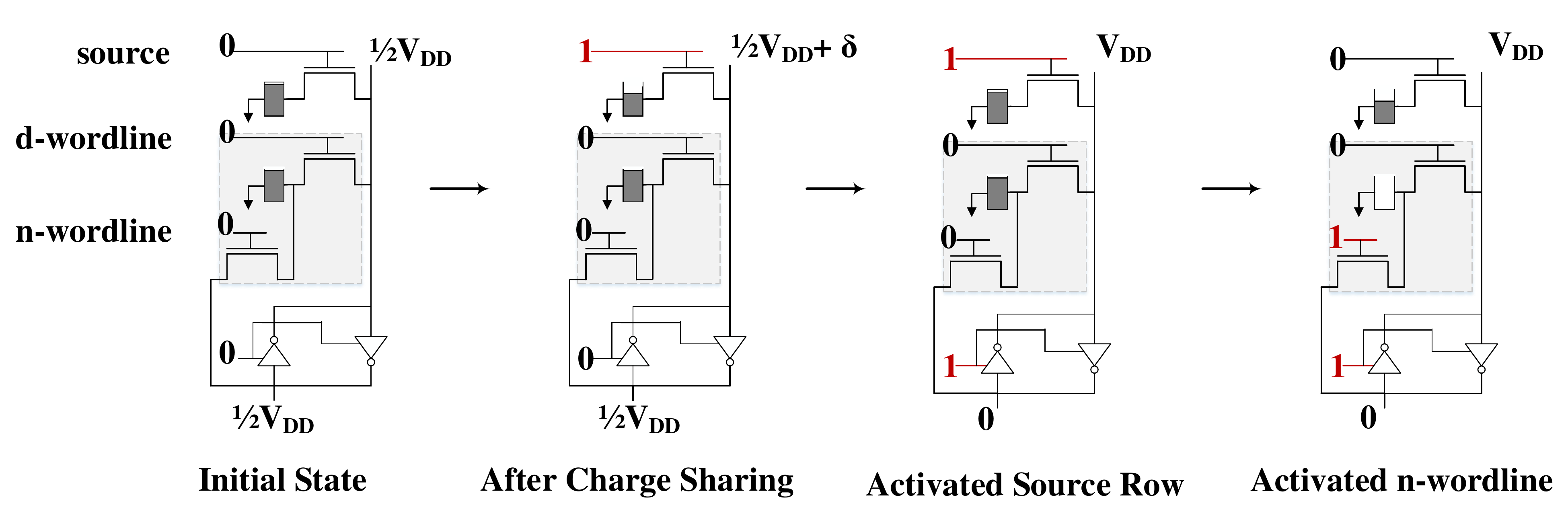}
\caption{Bitwise NOT using a dual-contact cell (DCC) connected to a sense amplifier (adapted from~\cite{seshadri2017ambit}).}
\label{fig:ambit_not_dcc}
\end{figure*}

On the other hand, Ambit-NOT exploits the fact that at the end of the sense amplification process, due to the two inverters present inside the sense amplifiers, the voltage level of the $\overline{bitline}$ represents the negated logical value of the cell. Their key idea is to perform bulk bitwise NOT operations in DRAM by transferring the data on the $\overline{bitline}$ to a cell that can also be connected to the bitline. For this purpose, a small subset of rows with dual-contact cells is introduced as well as modest changes to the sense amplifiers. A dual-contact cell (DCC) is a DRAM cell with two transistors (i.e. a 2T1C cell design). In a DCC, one transistor, controlled by the d-wordline (or data wordline), connects the cell capacitor to the bitline, and the other transistor, controlled by the n-wordline (or negation wordline), connects the cell capacitor to the $\overline{bitline}$. The negated value of a source cell can be transferred on to the DCC connected to the same bitline by performing two consecutive single row activations. Fig.~\ref{fig:ambit_not_dcc} shows an example of a bitwise NOT using a DCC. First, the source cell wordline is activated, driving the bitline to the data value corresponding to the source cell and the $\overline{bitline}$ to the negated value. Then, the mechanism activates the n-wordline, enabling the transistor that connects the DCC to the $\overline{bitline}$, and overwriting the DCC capacitor value with the negated value of the source cell.

\emph{Discussion}. Ambit-AND-OR and Ambit-NOT allow Ambit to perform any bulk bitwise operation efficiently inside DRAM. Ambit largely exploits the existing DRAM structure and interface, and hence incurs in a low cost implementation on top of commodity DRAM designs. In addition, Ambit can be easily integrated with any DRAM technology including 3D-stacked HMC, since the underlying DRAM microarchitecture is the same. Ambit provides great improvements in performance and energy reduction, that is, up to 32x/35x averaged across seven bulk bitwise operations, which can be leveraged by different applications such as neural networks.

\subsubsection{Neural Cache}
Following a parallel line of research, Eckert et al. proposed \emph{Neural Cache}~\cite{eckert2018neuralcache}, a bit-serial in-cache accelerator for DNNs using commodity SRAM technology. The \emph{Neural Cache} architecture re-purposes cache structures to transform them into massively parallel compute units capable of running DNN inference. The authors propose to extend the column peripherals of SRAM arrays of a commodity Last Level Cache (LLC) with extra logic to perform arithmetic computations directly inside the memory. Similar to Ambit, the \emph{Neural Cache} architecture enables in-memory processing inside the SRAM arrays based on the observation that by activating two wordlines of two cells sharing the same bitlines, bitwise AND and NOR operations can be performed between the values of the two cells after the sense amplification process of $BL$ and $BLB$. Although this idea was first proposed, alongside the copy and zero initialization operations, in the paper \emph{Compute Caches}~\cite{aga2017computecache}, \emph{Neural Cache} further extends it by adding support for more complex operations such as addition, multiplication and reduction, which are required for most DNN computations.

\begin{figure}[t!]
\centering
\includegraphics[width=.80\columnwidth]{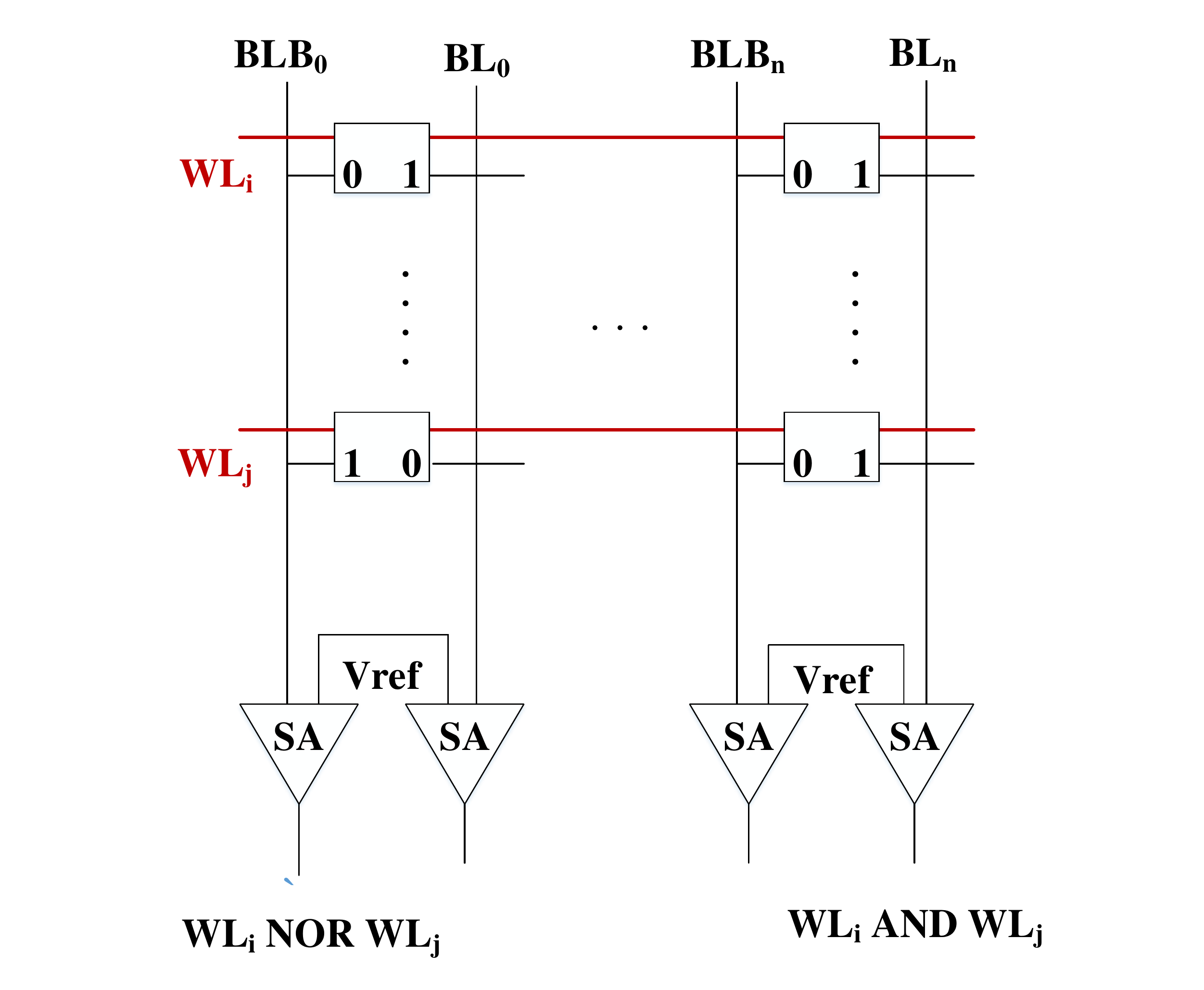}
\caption{SRAM circuit for bitwise NOR/AND operations (adapted from~\cite{aga2017computecache}). An AND operation is performed by sensing BL. A NOR operation can be performed by sensing bitline bar (BLB).}
\label{fig:sram_inplace_and_nor}
\end{figure}

In contrast to the DRAM TRA that is used in Ambit, the activation of multiple rows in SRAM memory arrays does not modify the values of the cells and, thus, the copy of both, source operands in a designated space and results back to the destiny, is not required. \emph{Neural Cache} prevents data corruption due to multi-row access by lowering the wordline voltage to bias against the write of the SRAM cells. The robustness comes at the cost of increased delay during compute operations, but does not affect conventional array read/write accesses. In order to perform bitwise AND and NOR operations between two rows of SRAM cells, \emph{Neural Cache} initially precharges all the bitlines of a given SRAM array to ‘1’. Then, the wordlines corresponding to the two rows of cells with which to operate are activated. An AND operation is performed by sensing the bitline (BL). If both the activated bits in a column have a ‘1’, then the BL stays high and it is sensed as a ‘1’. If any one of the bits were ‘0’ it will lower the BL voltage below $V_{ref}$ and will be sensed as a ‘0’. A NOR operation can be performed by sensing the bitline bar (BLB). Fig.~\ref{fig:sram_inplace_and_nor} shows an example of the SRAM circuit for in-place operations.

\begin{figure}[t!]
\centering
\includegraphics[width=1.00\columnwidth]{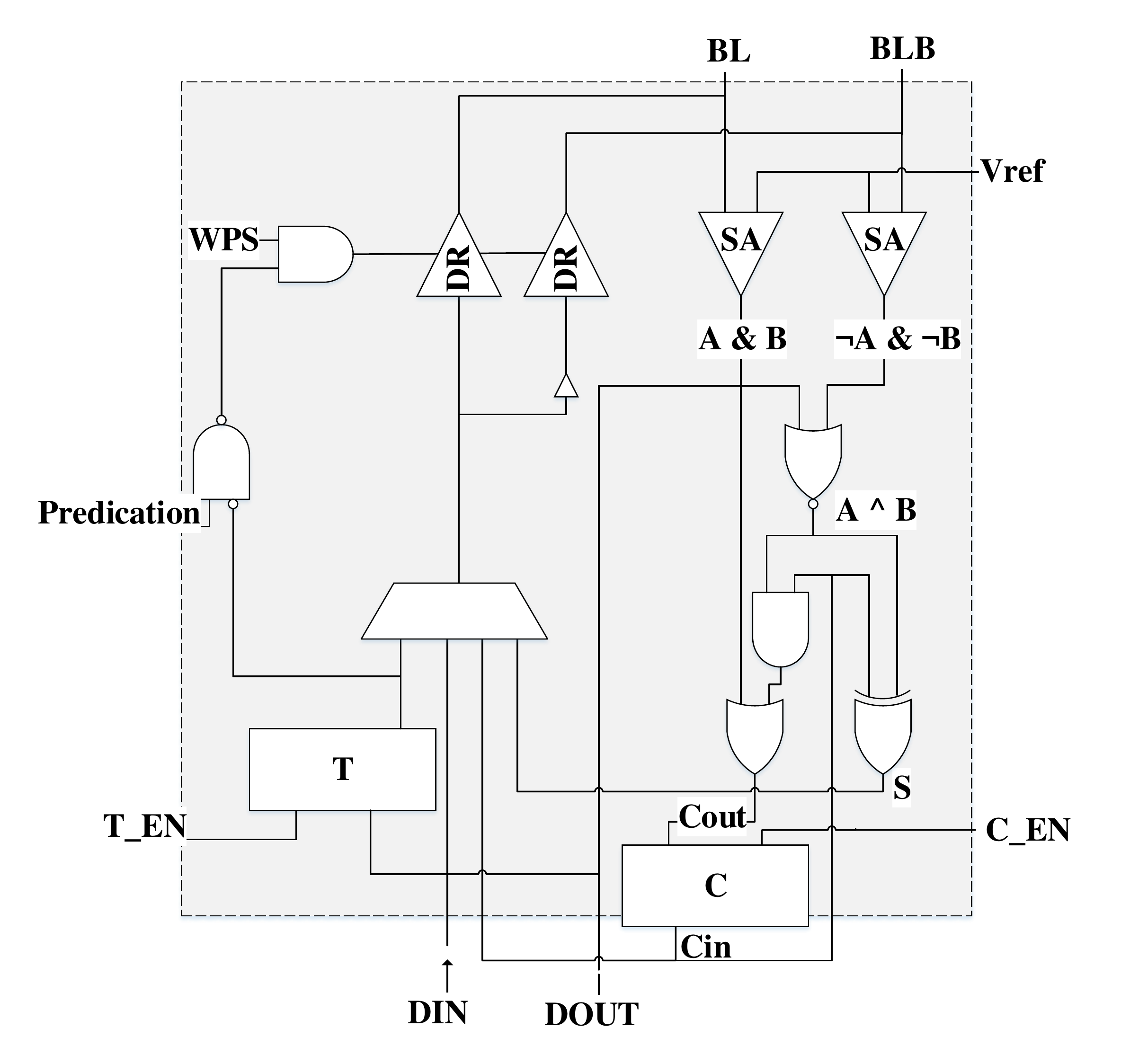}
\caption{Bitline peripherals in \emph{Neural Cache} (adapted from~\cite{eckert2018neuralcache}).}
\label{fig:sram_bl_peripherals}
\end{figure}

In conventional architectures, arrays are generated, stored, accessed, and processed element-by-element in the vertical direction along the bitlines, which is well known as the bit-parallel or regular data layout. \emph{Compute Cache} supports several simple bitwise operations that are bit-parallel and do not require communication between bitlines. However, \emph{Neural Cache} requires support for more complex operations, and the main challenge is to facilitate the interaction between bitlines. For example, the addition operation requires carry propagation between bitlines.

The authors propose a bit-serial implementation with a transposed data layout to address the above challenge. The key idea of the bit-serial arithmetic is to process one bit of multiple data elements every cycle. Although bit-serial computation is expected to have higher latency per operation, it is also expected to have significantly larger throughput than bit-parallel arithmetic. Bit-serial computing in SRAM arrays can be realized by storing data elements in a transposed data layout. Transposing ensures that all bits of a data element are mapped to the same bitline, thus eliminating the need for communication between bitlines. \emph{Neural Cache} introduces a Transpose Memory Unit (TMU) to convert data between transposed and regular format before writing or reading the SRAM arrays. A TMU consists of an 8T SRAM array which can access data from both vertical and horizontal directions. Only a few TMUs are required to fulfill the bandwidth demand of \emph{Neural Cache}, and static data elements such as the weights can be transposed offline. In addition, \emph{Neural Cache} includes extra logic in the column peripherals of the SRAM arrays. First, an XOR operation is performed based on the AND/NOR operations from the SRAM bitlines. Then, a full-adder with a register to store the carry is implemented by adding the required logic gates and a latch. Furthermore, additional logic is included to perform a predicated multiplication/division. Fig.~\ref{fig:sram_bl_peripherals} shows an example of the SRAM array bitline peripherals.

\begin{figure*}[t!]
\centering
\includegraphics[width=\textwidth]{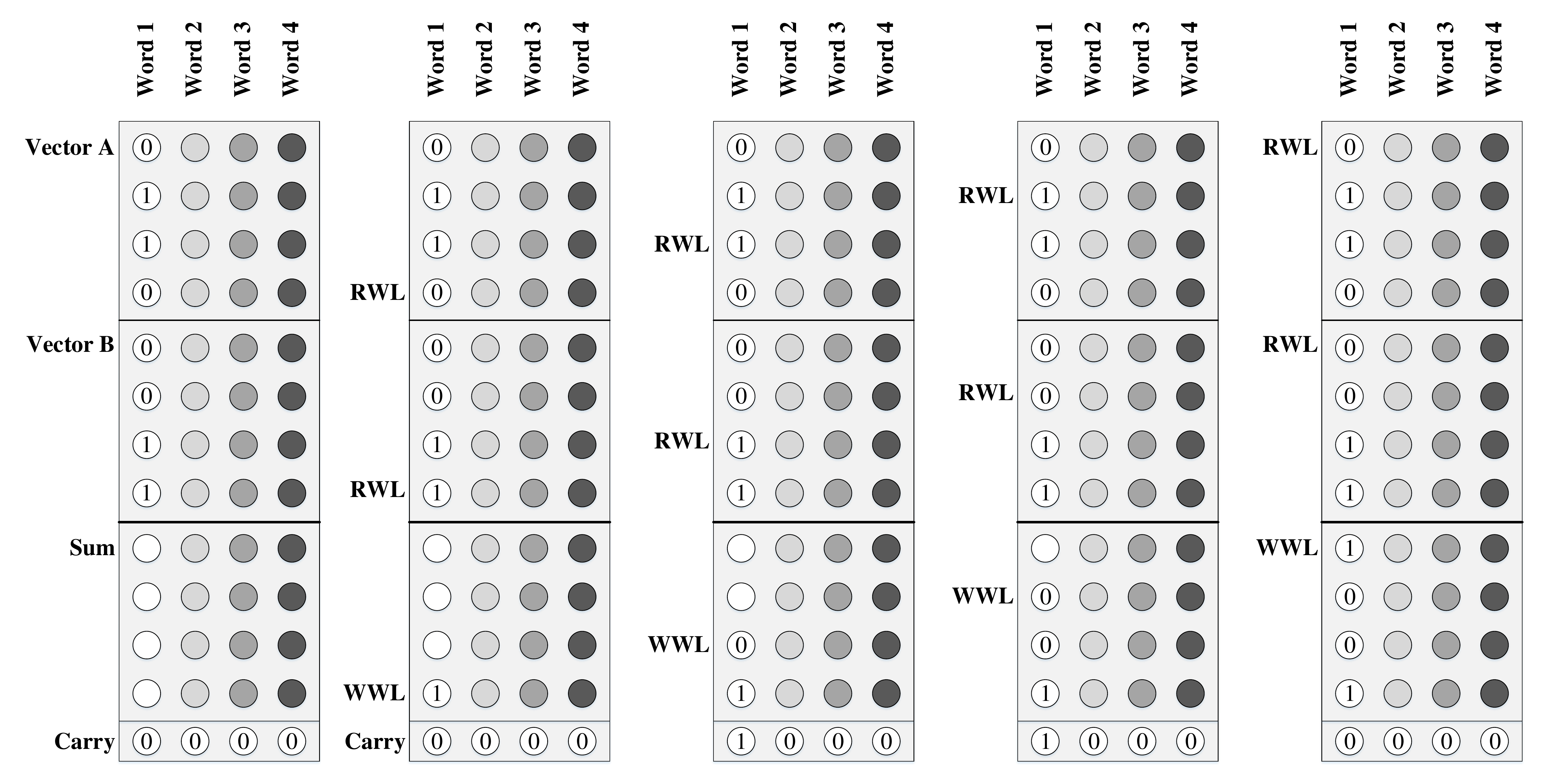}
\caption{Example of an addition operation in \emph{Neural Cache} (adapted from~\cite{eckert2018neuralcache}).}
\label{fig:neuralcache_addition_example}
\end{figure*}

Fig.~\ref{fig:neuralcache_addition_example} shows an example of an addition operation with a $12 \times 4$ SRAM array storing two vectors A and B, each with 4-bit elements, in transpose layout. The two vectors must be aligned so that the two words that are going to be added together share the same bitline. The addition algorithm of two $N$-bit elements is carried out bit-by-bit starting from the least significant bit (LSB) of the two words. $N$ empty rows are reserved to store the results, and there is an additional row of latches inside the column peripherals for the carry storage. In the first half of each cycle, two read wordlines (RWL) are activated simultaneously, generating the sum and carry-out bits through the sense amplifiers and logic gates in the column peripherals. In the second half of each cycle, a write wordline (WWL) is activated to store back the sum bit, and the carry-out bit overwrites the data in the carry latch becoming the carry-in of the next cycle. The addition takes $N+1$ cycles to complete. Multiplication and division are supported based on addition and a predication algorithm, and reduction is supported as a set of moves and additions.

The architecture of \emph{Neural Cache} is implemented on top of the LLC of an Intel Xeon processor, which consists of 14 cache slices with 20 ways in each slice. The last way (way-20) of each cache slice is reserved to enable normal processing for CPU cores, while the penultimate way (way-19) is reserved to store inputs and outputs. The remaining ways are utilized for storing filter weights and performing DNN computations. \emph{Neural Cache} computes DNN layers serially, which means that the inference is performed layer by layer with the operands stored in-place in memory. On the other hand, the convolutions within a CNN layer are computed in parallel. A single convolution consists of generating one element of the output feature map. Each LLC slice works on a uniformly distributed set of consecutive output elements from all the filters, and each of the SRAM arrays within a slice computes convolutions in parallel. Furthermore, \emph{Neural Cache} also exploits channel level parallelism in a single convolution. For each convolution, an SRAM array executes multiple MACs in parallel by mapping the channels to different bitlines, which is followed by a reduction step across channels.

To summarize, each column in an SRAM array performs a separate calculation and the thousands of memory arrays in the cache can operate concurrently. The inputs are streamed in from the reserved way-19, while the filter weights are loaded from DRAM only once per layer, and replicated in multiple SRAM arrays from different LLC ways and slices to improve data reuse, that is, filter weights are stationary in the compute arrays (way-1 to way-18) of each slice. In addition, \emph{Neural Cache} assumes 8-bit precision and quantized inputs and filter weights, as well as batching, to further increase the throughput of the accelerator.

\emph{Discussion}. \emph{Neural Cache} adds support for the most common DNN operations directly into a commodity LLC using SRAM memory cells. By employing a novel data layout and dataflow, the LLC structure and organization is exploited to efficiently execute CNNs with a low area overhead. \emph{Neural Cache} provides better performance and energy efficiency compared to a modern desktop CPU and GPU. However, the LLC may be limited by storage capacity to efficiently execute large DNNs, and some bit-serial operations such as the multiplication or the reduction may be very slow, limiting the performance compared to state-of-the-art DNN accelerators.

\subsubsection{Bit Prudent}
Wang et al. proposed Bit Prudent~\cite{wang2019bitprudent}, an in-cache accelerator that extends the in-SRAM architecture of \emph{Neural Cache}, to further improve the inference of CNNs by leveraging the neural network redundancy and massive parallelism.

Based on the observation that DNNs are normally oversized, and tend to have a significant amount of weights with a value close to zero~\cite{han2015learning}, static pruning is an effective technique to reduce the model size and the number of computations by removing unimportant connections and/or nodes depending on the weight values. After an iterative process of pruning and training, the pruned model retains accuracy while requiring significantly less memory storage and computations, resulting in large performance improvements and energy savings.

The authors propose to exploit the high degree of neural network redundancy in two ways. First, they prune and fine-tune the trained network model, and develop two distinct methods, coalescing and overlapping, to run inferences efficiently with sparse models. Second, they propose an architecture for network models with a reduced bit-width benefiting from the bit-serial computation.

Despite the numerous benefits of static DNN pruning, there are several challenges for exploiting the sparsity of pruned neural networks in \emph{Neural Cache}. In first place, the vector parallelism in SRAM arrays requires that computation cannot be skipped even if only one of the vector elements needs it. They solve this problem by developing techniques which create dense computation by coalescing non-zero filter channels. Filter channels are gathered into a dense format using a novel offline structured channel pruning and retraining process, while input channels are gathered dynamically at runtime using a new hardware coalescing unit. This specialized hardware unit consists of a crossbar switch, with a reconfiguring peripheral, that outputs the needed inputs for a given filter depending on a bit-mask of pruned channels. However, input coalescing can increase input loading time. Moreover, different filters may be heterogeneously pruned and the same input data can no longer be broadcast to all filters, further increasing the input loading time.

\begin{figure}[t!]
\centering
\subfloat[]{
    \includegraphics[width=0.95\columnwidth]{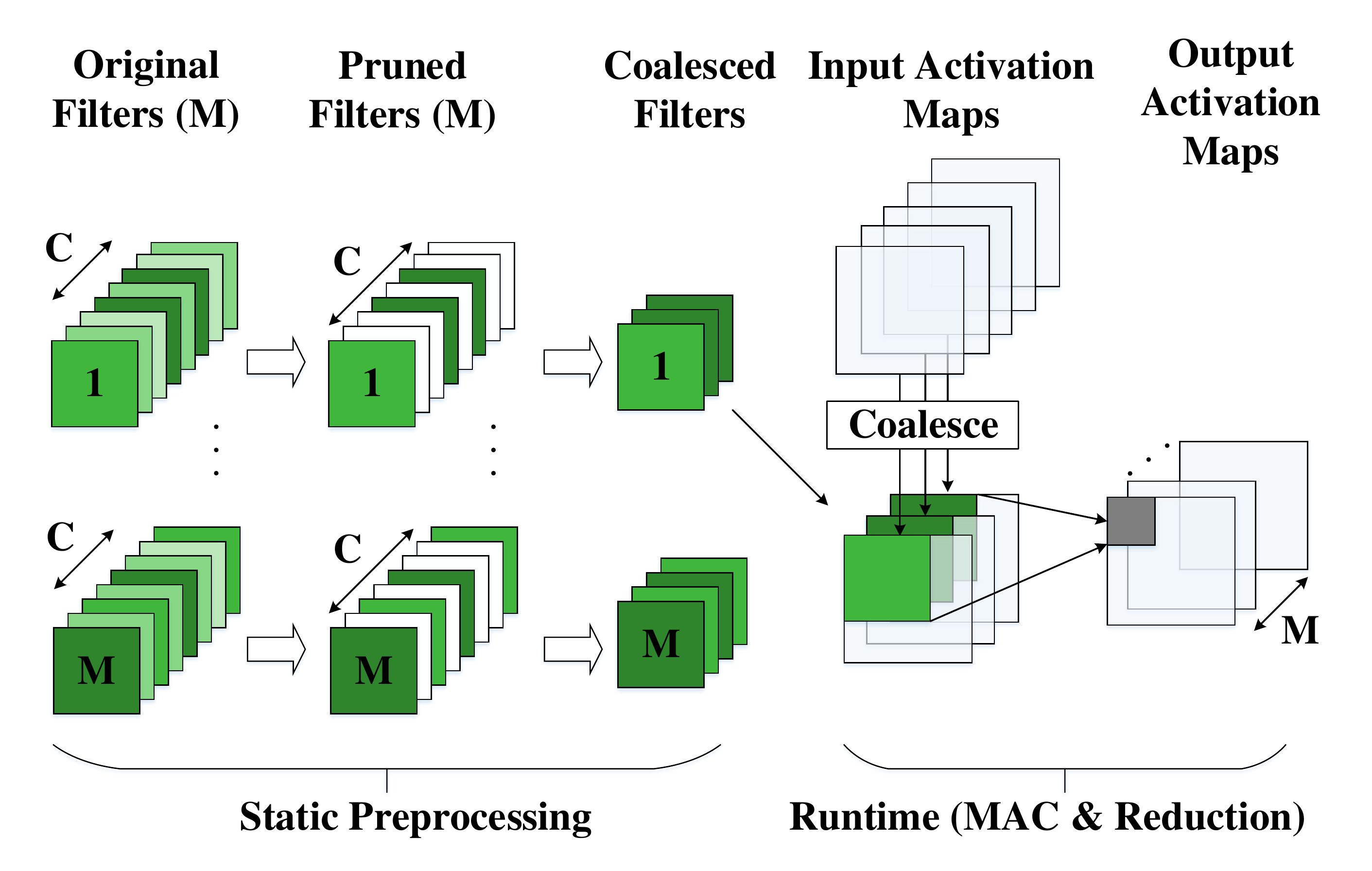}
    \label{fig:sparse_conv_coalescing}
} \\
\subfloat[]{
    \includegraphics[width=0.95\columnwidth]{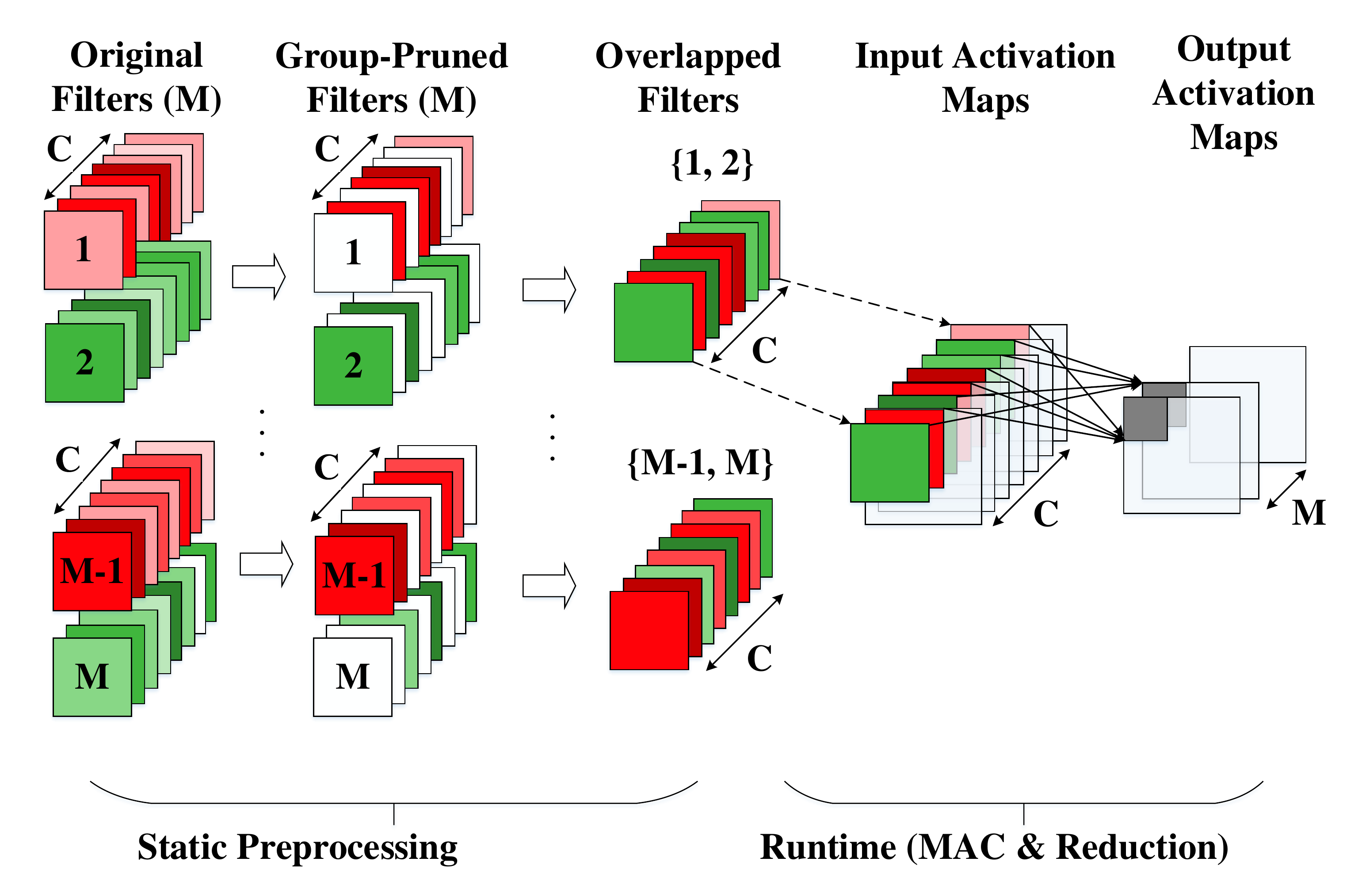}
    \label{fig:sparse_conv_overlapping}
}
\caption{Sparse Convolution with (a) Coalescing and (b) Overlapping (adapted from~\cite{wang2019bitprudent}).}
\label{fig:bit_prudent}
\end{figure}

They tackle this problem by proposing an input-loading aware pruning method that restricts the channels that can be pruned in each filter. Then, they explore a filter channel overlapping technique which does not change the input data mapping from the original unpruned network models, removing the need for input coalescing. The key idea is to overlap the filters which are sparse at different channels into one filter. The overlapping technique adds more constraints to the pruning method limiting the benefits of pruning, but its hardware overhead is lower and it is much simpler to implement. Fig.~\ref{fig:sparse_conv_coalescing} and Fig.~\ref{fig:sparse_conv_overlapping} show an example of a sparse convolution with coalescing and overlapping, respectively. In the sparse convolution with coalescing, filter channels are statically pruned and coalesced while generating a mask that keeps track of the pruned channels of each filter. Then, input activations are coalesced at runtime, according to the mask, avoiding useless computations. In contrast, in the sparse convolution with overlapping, filter channels are statically pruned by groups so that the non-pruned channels of different filters can be overlapped into a single filter and, hence, avoid the need for coalescing the input activations. Both methods require small changes to the dataflow and execution scheme of \emph{Neural Cache}, but they are implemented and evaluated independently since the two techniques are not orthogonal.

On the other hand, the authors also develop efficient compute techniques for low-precision neural networks. The bit-serial compute paradigm naturally takes advantage of low-precision input activations and weights, since compute cycles of bit-serial algorithms are proportional to bit-width. However, the bit-serial multiply-accumulate of \emph{Neural Cache} is inefficient when weights have ultra-low precision such as ternary or binary, due to the extra starting steps for multiplication. They redesign the process of multiply-accumulate for ternary/binary weights to combine the multiplication and accumulation in fewer cycles using logical operations and additions.

\emph{Discussion}.  The bit prudent architecture adds support for sparse and reduced precision DNN models in \emph{Neural Cache}, employing novel techniques such as input channel coalescing or filter overlapping. The evaluation of the different proposals shows that the reduced precision techniques achieve the best performance and energy efficiency at the cost of high accuracy loss. In contrast, the accelerator designs supporting sparse models improve the \emph{Neural Cache} efficiency with negligible accuracy loss, highly improving the performance over a CPU and GPU. In addition, they compare against state-of-the-art sparse DNN accelerators achieving much better throughput per area per energy. However, the support for sparse models is limited to CNNs with structured channel pruning, and they do no take into account the possibility to use dynamic precision techniques. Moreover, each proposal is implemented and evaluated independently, and not combined in a single accelerator design, losing opportunities for higher benefits in terms of performance and energy consumption.

\subsection{3D-Stacked Memory based Architectures}\label{s:ndp_sota_3d}
\subsubsection{Neurocube}
Kim et al. proposed Neurocube~\cite{neurocube/ISCA.2016.41}, a programmable digital neuromorphic architecture based on 3D high-density memory. The authors implement an accelerator for efficient neural network computing that is integrated in the logic layer of a 3D stack Hybrid Memory Cube (HMC) based on DRAM. Neurocube aims to provide a balance between the programmablity and scalability of GPGPUs, and the performance and energy efficiency of ASICs, by exploiting the advantages of 3D stack memory such as the high bandwidth and low latency of the TSV connections, or the high level of parallelism due to the independent vaults. Neurocube is highly optimized for the execution of CNNs, but its high degree of programmability and scalability allows to map and run different types of DNNs, including support for both the training and inference phases.

\begin{figure}[t!]
\centering
\includegraphics[width=\columnwidth]{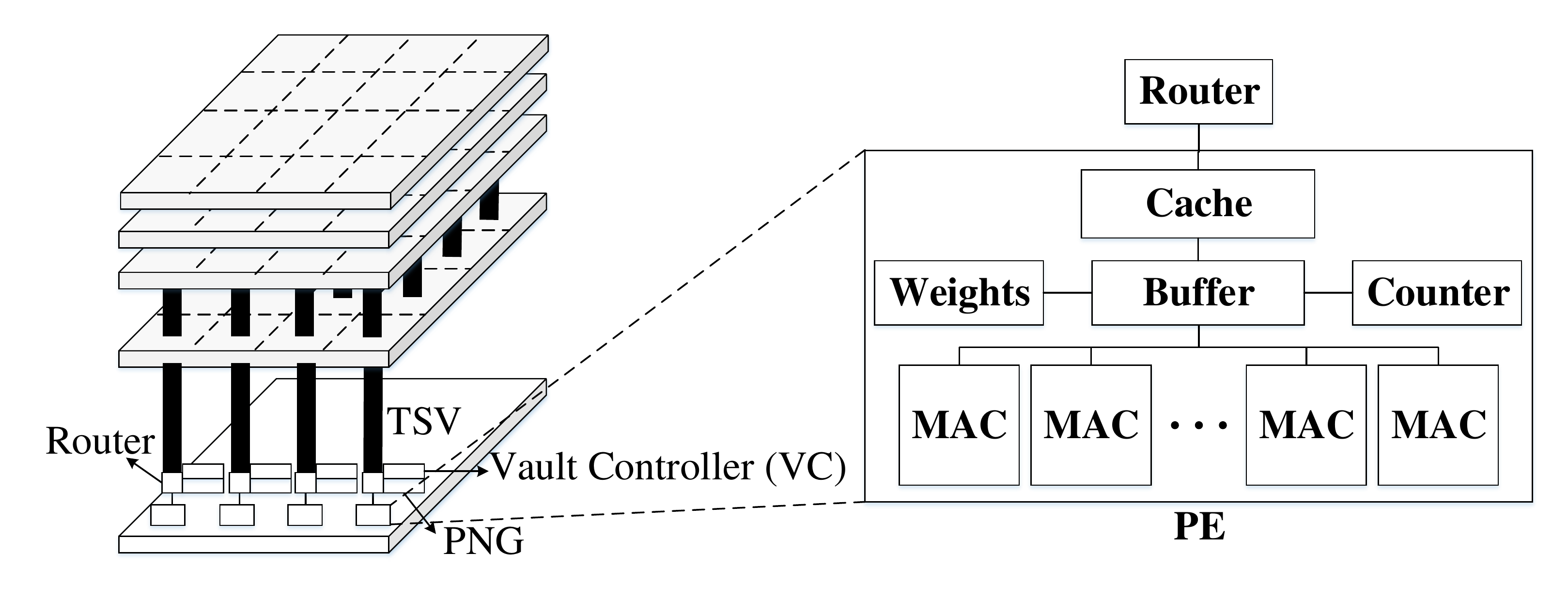}
\caption{Overview of the Neurocube architecture and organization of a PE (adapted from~\cite{neurocube/ISCA.2016.41}).}
\label{fig:neurocube_arch}
\end{figure}

As described in Section~\ref{s:background_3d}, a typical HMC is composed of multiple DRAM layers integrated in 3D with a logic layer at the bottom. The DRAM layers of a HMC are spatially divided into multiple memory channels or vaults, which can be accessed in parallel. The Neurocube architecture consists of clusters of processing engines (PE) connected by a 2D mesh NoC in the processing layer. Each PE of the logic layer is associated to a single memory vault, and can operate independently, and communicate through the TSVs and a vault controller (VC). The organization of each PE is similar to a tile of previous state-of-the-art DNN accelerators such as DaDianNao~\cite{DaDianNao}, and is composed of multiple memory buffers to store weights and inputs as well as some MACs to perform operations. Fig.~\ref{fig:neurocube_arch} shows a general overview of the Neurocube architecture and a PE.

The memory centric computing principle embeds specialized state-machines within the vault controllers of the HMC to drive data into the PE clusters. Each vault controller includes a Programmable Neurosequence Generator (PNG) unit that generates the commands to orchestrate the corresponding operations of the DNN layers. The PNGs employ a finite state machine (FSM) with counters that are initialized depending on the number of MACs in each PE and the DNN layer topology. Based on the fact that DNNs are data-driven and have statically known memory access patterns, the PNGs keep track of the data to be accessed at each moment, and produce the corresponding sequence of memory addresses required in the execution of each layer. Neurocube processes a DNN layer by layer, and for each layer the computations are translated to three nested loops that can be mapped to finite state machines using three counters, that is, a loop across all neurons in the layer, a loop across all connections for each single neuron in the layer, and a loop across all the MACs. Initially, the host programs the counters of all PNGs by loading the corresponding configuration parameters. Then, the PNGs compute the memory addresses required for each output neuron in the layer, and encapsulate the state of the input neurons and their associated weights in packets that are moved to the PEs.

On the other hand, Neurocube employs an output stationary dataflow for both convolutional and FC layers, meaning that each MAC from a PE is responsible for the computations of a different output neuron at a time, but with small optimizations depending on the layer type and its size. For example, if the size of the synaptic weights matrix of a convolutional layer is small, all weights are stored in a local PE memory for reuse. The PEs form different vaults that communicate among them using a 2D mesh NoC router. The work of a given layer is distributed between the vaults and its associated PEs, and to reduce the communication cost of the NoC, the weights of convolutional filters are replicated among the vaults while dividing the input images into overlapped segments. In this way each vault can operate on its set of output neurons without having to communicate with the rest during most of the execution. Similarly, the weight matrix of FC layers is divided among vaults and the input vector is replicated. In case that the input vector does not fit, it is also partitioned increasing the NoC traffic. In order to sequence through the correct number of input neurons for each output neuron, each PE has an operation counter (OP-counter) that keeps track of the input neuron currently being processed by the MAC units. The cache in each PE is used to store packets that arrive out of order, while the inputs and weights for the current iteration are stored in a temporal buffer. When all the operands for the current iteration are available, the MACs are triggered and the operands for the next iteration are searched from the cache.

\emph{Discussion}. Neurocube is one of the first proposals implementing a DNN accelerator in the logic layer of a HMC, tackling the typical thermal and area concerns of the 3D-stacked memory designs. Neurocube can be used for both DNN inference and training, providing better performance and energy efficiency compared to a GPU. The accelerator exploits the high bandwidth offered by the HMC, and implements an straightforward dataflow with a FSM to control the memory accesses and the execution of the different DNN layers. However, their dataflow is not optimized to exploit the intrinsic features of a 3D-stacked architecture, and no computations are performed withing the DRAM layers, leaving opportunities to further improve the design with PIM capabilities.

\subsubsection{TETRIS}
Continuing the same line of research, Gao et al. proposed TETRIS~\cite{gao2017tetris}, a scalable and efficient accelerator for DNN inference based on 3D-stacked HMC using DRAM layers. Like Neurocube, TETRIS also exploits NDP in the logic layer of a HMC. It presents an optimized hardware architecture coupled with software scheduling and partitioning techniques that exploit the intrinsic characteristics of 3D memory. First, the authors show that the high throughput and low energy characteristics of 3D memory allow to rebalance the NN accelerator design, using more area for processing elements and less area for SRAM buffers. Second, they move portions of the NN computations close to the DRAM banks to decrease the bandwidth pressure and increase performance and energy efficiency. Third, they demonstrate that despite the use of small SRAM buffers, the presence of 3D memory simplifies the dataflow scheduling for NN computations, and present an analytical scheduling scheme. Finally, they develop a hybrid partitioning scheme that parallelizes the NN computations across multiple vaults and stacks.

\begin{figure}[t!]
\centering
\subfloat[]{
    \includegraphics[width=0.95\columnwidth]{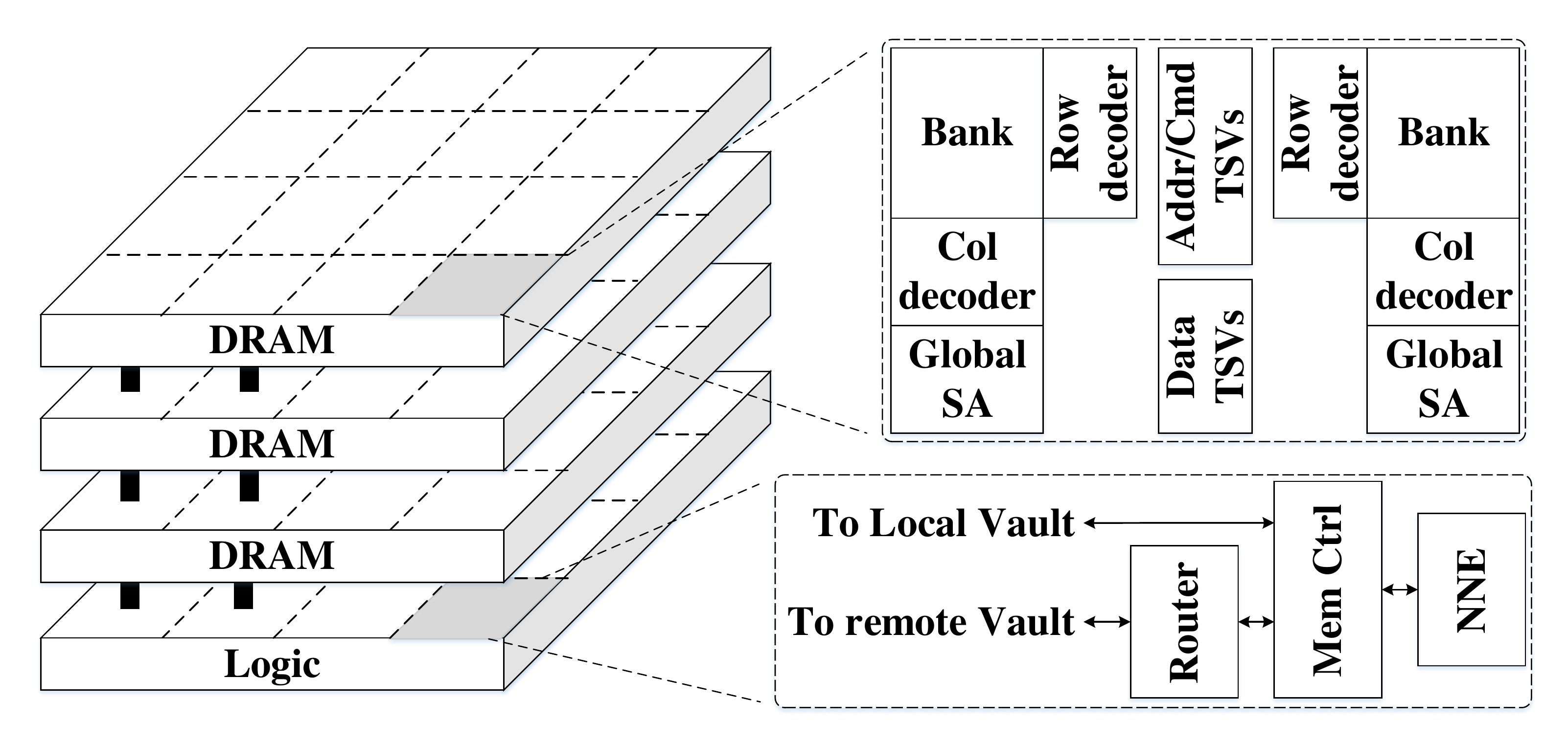}
} \\
\subfloat[]{
    \includegraphics[width=0.95\columnwidth]{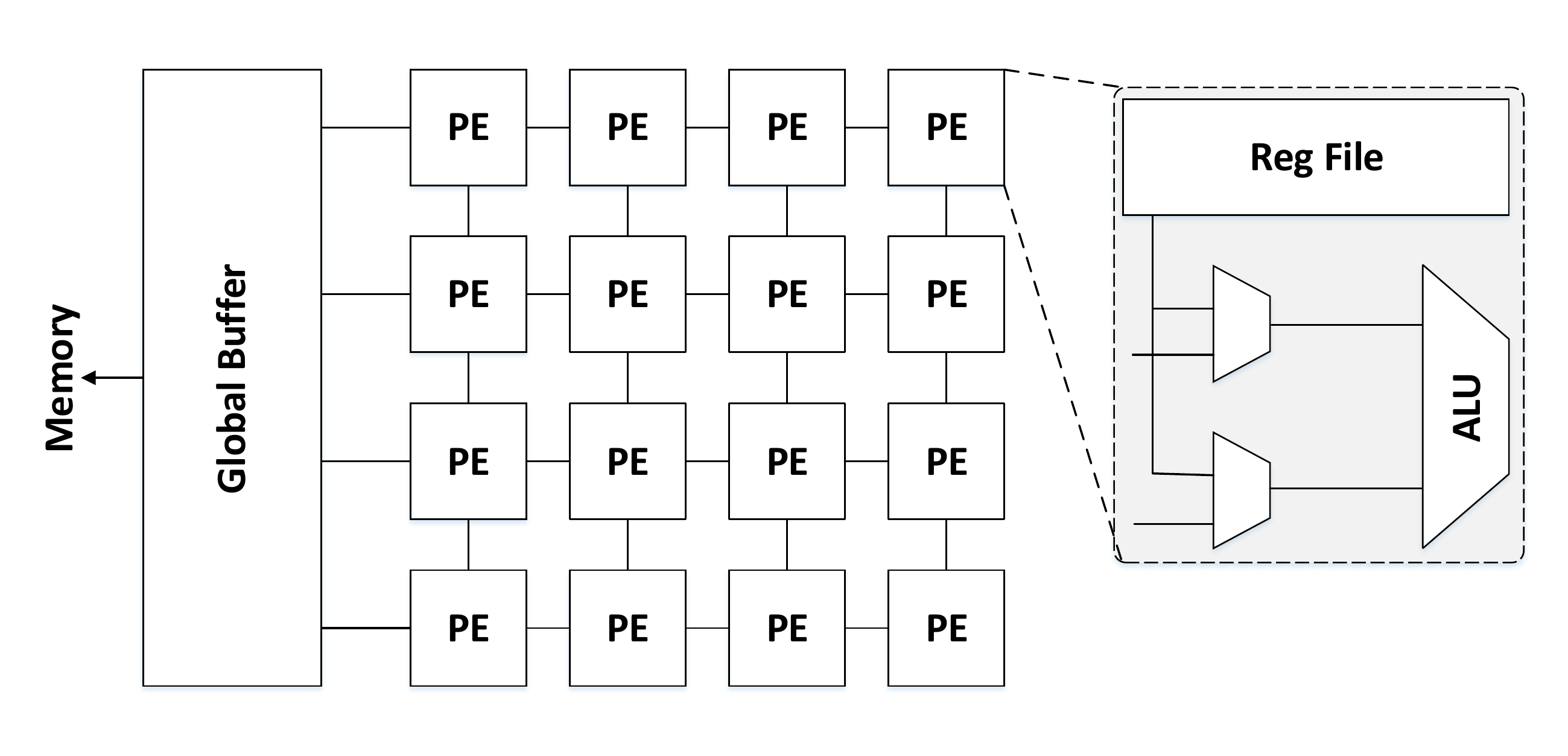}
}
\caption{(a) Overview of the TETRIS architecture and (b) organization of a Neural Network Engine (NNE) (adapted from~\cite{gao2017tetris}).}
\label{fig:tetris_architecture}
\end{figure}

Similar to Neurocube, the TETRIS architecture employs a HMC with a stack of multiple DRAM dies vertically divided into 16 vaults. Each vault is associated to one Neural Network Engine (NNE) in the logic layer connected through TSVs, and the different NNEs are interconnected with a 2D mesh NoC. Unlike Neurocube, the NNE of each vault consists of a systolic array of processing elements (PEs) with a global SRAM buffer shared among the PEs, and where each PE has a register file and a MAC to locally store the inputs/weights and perform computations. Multiple vault engines can be used to process a single NN layer in parallel, and the memory buffers in each NNE and PE make use of prefetching techniques to increase the data reuse from memory. The NNEs are based on conventional 2D accelerators such as Eyeriss~\cite{chen2016eyeriss}, which also implement a row stationary dataflow that improves data reuse while reducing the data movements. Row stationary (RS) maps 1D convolutions onto a single PE to utilize the PE register file for local data reuse, and orchestrates the 2D convolution dataflow on the 2D array interconnect so that the data propagation among PEs remains local. Fig.~\ref{fig:tetris_architecture} shows a high-level overview of the TETRIS architecture and a single NNE.

On the other hand, the area of the logic layer of a HMC is constrained by the size of the 3D package which, along the power constrains of 3D memory, limit the scaling of the PE arrays and its buffers. One of the key observations that motivates the TETRIS architecture is that the logic layer requires a better balance between the number of PEs and the size of the memory buffers. Compared to prior works that tend to use more area for the memory buffers, the authors evaluate this trade-off, and decide to spend the same amount of area for both (i.e. 50:50 ratio) to achieve higher performance and fully exploit the bandwidth of the HMC. As a side effect of reducing the amount of memory in the logic layer, there may be more data movements due to an increase of unfinished accumulations or partial sums, so the data has to be fetched again to finish the calculations. To alleviate this problem, they put extra adders in the DRAM dies close to the banks, and introduce a new atomic update that performs the read-add-write operation inside the memory layer, reducing the overall cost of the accumulations.

The efficiency of a DNN accelerator depends heavily on the dataflow scheduling and partitioning of the NN computations. The scheduling is particularly important for TETRIS since it uses small on-chip buffers, which could potentially increase the accesses to DRAM. Moreover, since TETRIS includes one NN accelerator per vault, it is crucial to effectively partition and parallelize the NN workloads across vaults. The scheduling problem can be decomposed into two subproblems: the \emph{mapping} and the \emph{ordering}. TETRIS leverages the row stationary dataflow from Eyeriss~\cite{chen2016eyeriss} for the mapping of the inputs, weights and outputs of a 2D convolution inside the systolic array of PEs from a given vault. Then, the authors propose a bypass ordering to schedule the 2D convolutions, storing just one of the three streams in the global shared buffer, and relying on the low cost accesses of 3D memory for the rest. For example, the Output Weight (OW) bypass avoids the global buffer for the outputs and weights and only stores the inputs to be reused.

The mapping exploits the PE array interconnect and register files, while the ordering focuses on how to buffer the data in the global shared buffer to maximize on-chip reuse. The ordering involves the blocking of the input/output channels and batch size, but the bypass ordering simplifies the problem compared to general loop blocking algorithms. They analytically derive the optimal scheduling by finding the best blocking parameters while minimizing the number of DRAM accesses given one of the three bypass ordering variants. Then, to partition the work of a NN layer and distribute it among the vaults, they consider different partitioning schemes, and propose a simple cost model that minimizes the overall memory access energy taking into account a penalty for remote vault accesses. Finally, TETRIS implements a hybrid approach that balances the benefits from feature map partitioning and output partitioning.

\emph{Discussion}. Overall, the results show that TETRIS significantly improves the performance and reduces the energy consumption over DNN accelerators with conventional, low-power DRAM memory systems such as Eyeriss as well as Neurocube. TETRIS provides an improved dataflow scheduling and NN partitioning compared to Neurocube, exploiting better the data reuse while reducing both local and remote DRAM accesses. In addition, the inclusion of accumulators inside the memory layers offers a nice solution to reduce the cost of memory accesses, and opens up opportunities to do further research in this line.

\subsection{ReRAM based Architectures}\label{s:ndp_sota_reram}
As described in Section~\ref{s:background_reram}, the introduction of memristor devices, along with the massively parallel analog MAC operation provided by ReRAM crossbar arrays, has brought a new set of opportunities in the area of NDP architectures, although they also present some new challenges, such as the high overhead of the digital-analog conversions. In this section we discuss several ReRAM based approaches for the acceleration of DNNs.

\subsubsection{PRIME}
Chi et al. proposed PRIME~\cite{PRIME}, a PIM architecture for accelerating DNN computations in ReRAM-based main memory. In PRIME, a portion of ReRAM crossbar arrays can be configured as accelerators for NN applications or as normal memory for a larger memory space. The authors propose a microarchitecture and circuit designs, as well as a software/hardware interface for software developers to implement various DNNs on PRIME.

\begin{figure}[t!]
\centering
\subfloat[]{
    \includegraphics[width=0.45\columnwidth]{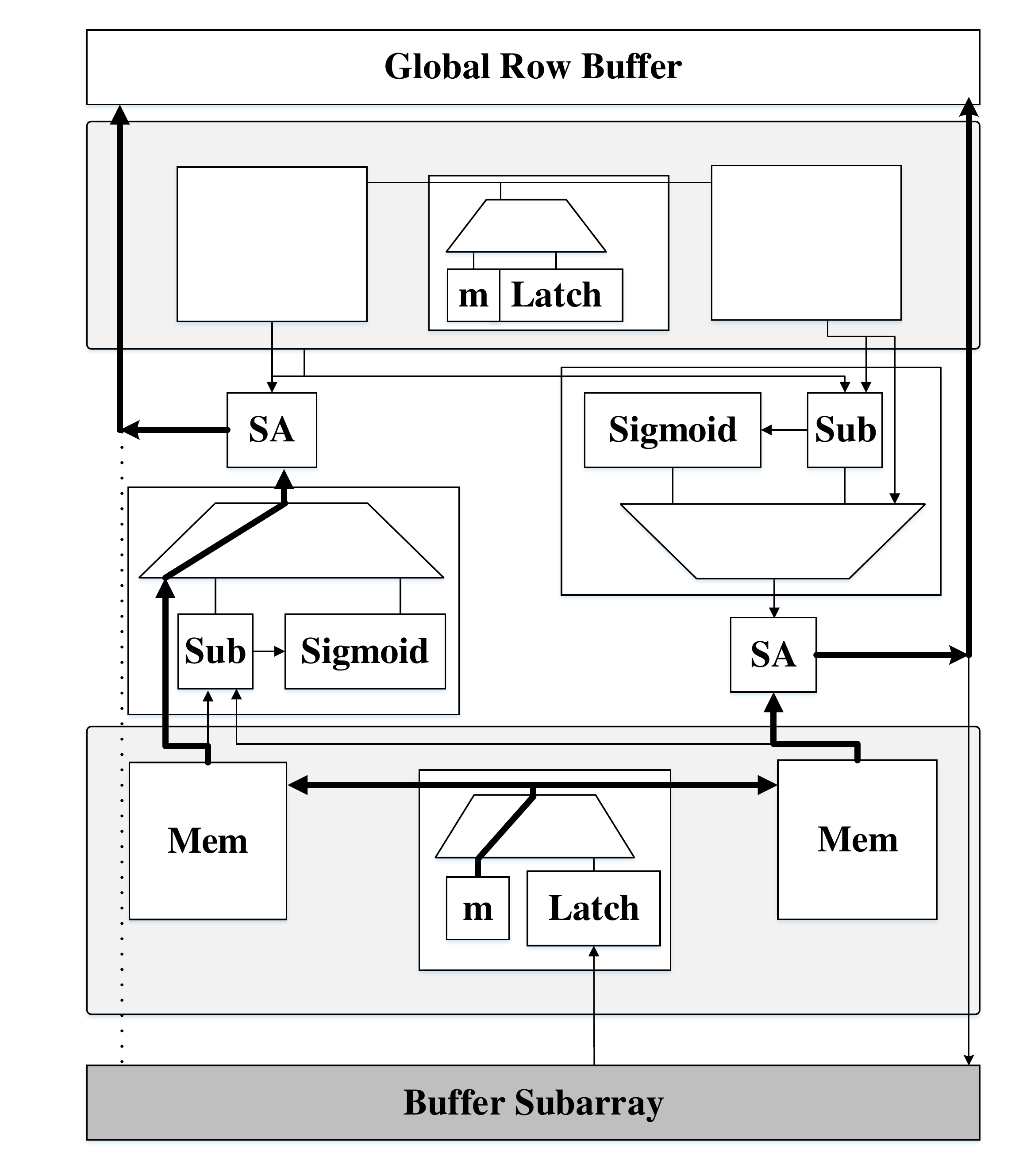}
}
\subfloat[]{
    \includegraphics[width=0.45\columnwidth]{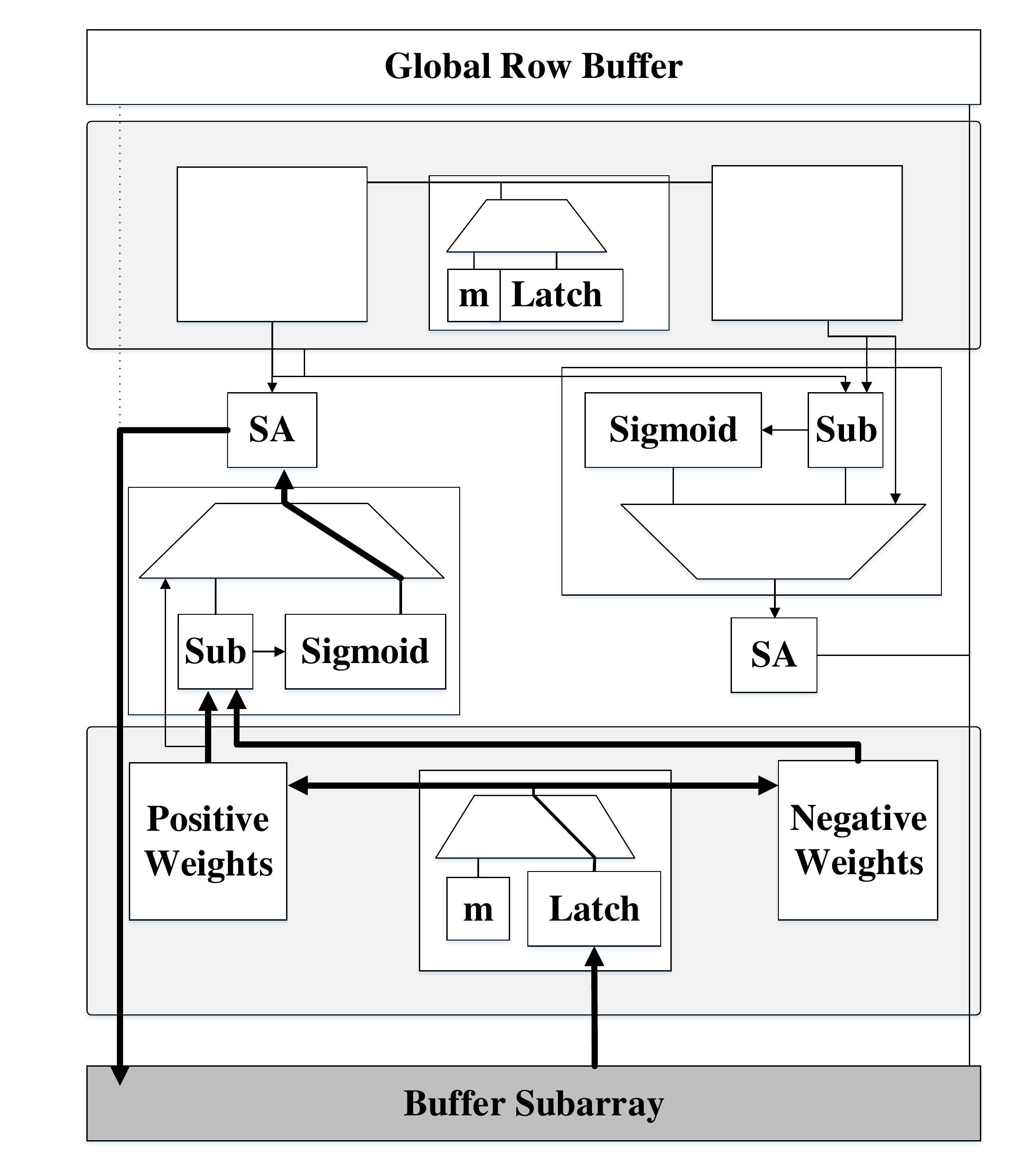}
}
\caption{PRIME FF subarray configurations (adapted from~\cite{PRIME}). (a) Memory mode; (b) Computation mode.}
\label{fig:prime_ff_modes}
\end{figure}

PRIME divides a ReRAM bank into three types of subarrays: memory (Mem), buffer and full function (FF). Mem subarrays only store data whereas FF subarrays can either store data or perform DNN computations as shown in Fig.~\ref{fig:prime_ff_modes}. To switch the FF subarrays from memory to compute mode, data stored in them are moved to the memory subarrays. Then, weights of the mapped DNNs are written to the FF subarrays, and the periphery is reconfigured by a controller. The opposite process happens to change from compute to memory mode. On the other hand, the Buffer subarrays store the data for FF subarrays without requiring the involvement of the CPU. SRAM buffers are not suitable for this task due to the lower storage capacity and high area overhead. FF subarrays are connected to their closest memory buffer through private data ports, benefiting from the high bandwidth of in-memory data movement and the ability to work in parallel to the CPU.

\begin{figure}[t!]
\centering
\includegraphics[width=0.7\columnwidth]{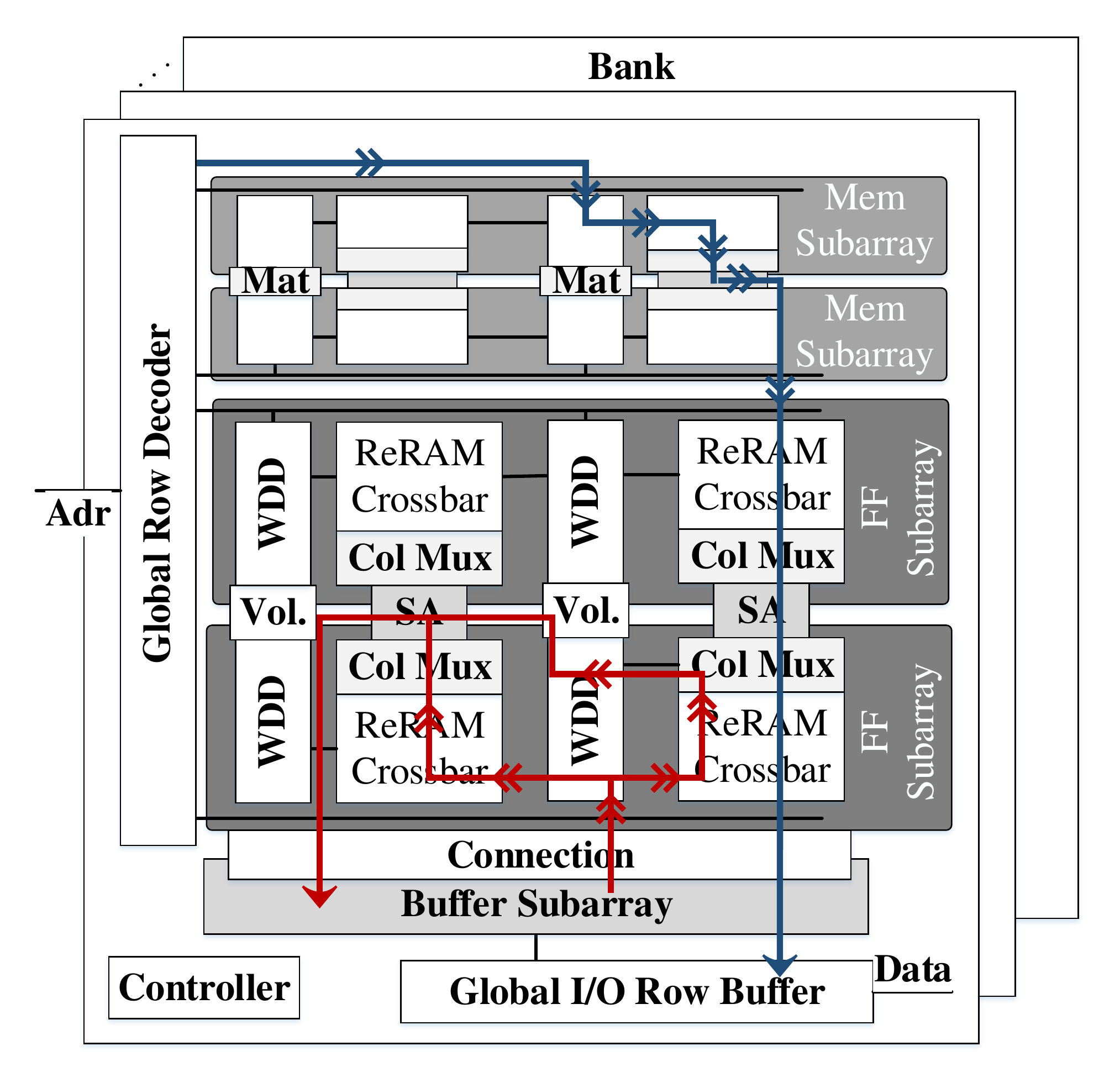}
\caption{Overview of the PRIME architecture (adapted from~\cite{PRIME}).}
\label{fig:prime_architecture}
\end{figure}

PRIME efficiently accelerates NN computation by leveraging ReRAM’s computation capability and the PIM architecture. To enable the NN computation function in FF subarrays, they modify the decoders, drivers, column multiplexers (MUX), and sense amplifiers (SA) of the periphery. Fig.~\ref{fig:prime_architecture} shows a high-level overview of the PRIME architecture. They noted that an SA performs similar function as an ADC and the same is also true for read/write drivers and DACs. Hence, with only small modifications, they reuse the read/write drivers and SAs to perform the function of DACs and ADCs, respectively. This sharing of periphery between computation and memory lowers the area overhead.

Precision is one of the most critical challenges for ReRAM based NN computations. Since NN applications can tolerate low precision of inputs and weights, they assume 3-bit and 4-bit precision of input voltage and synaptic weights, respectively, which implies 8 voltage and 16 resistance levels, respectively. The output precision required is 6-bit with truncation, and a dynamic fixed-point format is used. In order to reduce the accuracy loss, they employ a scheme whereby two 3-bit inputs are composed into one 6-bit input and two 4-bit cells are used for representing one 8-bit weight. To implement FC layers, they separate the synaptic weight matrix in two matrices for storing positive and negative weights and store them into two different crossbars. Then, inputs are read from the memory buffer and stored in a latch of the decoder/driver unit. MVM of positive and negative weights is implemented in ReRAM arrays, and the resulting currents are sent to an analog substractor unit and a sigmoid unit of the column multiplexer. Finally, the results are converted to digital in the SA, and written back to the memory buffer. They also discuss the implementation of convolution and max/mean pooling layers. The mapping of the DNN layers to ReRAM is optimized at compilation time depending on the size of the DNN (e.g., small, medium or large). PRIME allows to use multiple banks to implement large scale DNNs, while running in a pipelined fashion to improve throughput and avoid reprogramming the FF subarrays at every execution of a layer.

\emph{Discussion}. By leveraging both a PIM architecture and the efficiency of using ReRAM for DNN computation, PRIME distinguishes itself from prior works on DNN acceleration with significant performance improvements and energy savings on multiple ML workloads. The organization of PRIME in different types of subarrays, the schemes to encode/decode the inputs/outputs and weights, and the modifications of the memory peripherals, provide an interesting solution to some of the ReRAM based architecture challenges. However, PRIME supports positive and negative weights but only unsigned inputs, and the analog dot-product computations are lossy because the resolution of the ADCs does not always match the precision of the computed dot-product. Furthermore, the analog-digital conversions still represent an important overhead, and the modified SAs may be slow since multiple comparisons against different voltage levels are required to make each conversion.

\subsubsection{ISAAC}
Shafiee et al. presented ISAAC~\cite{ISAAC}, an accelerator for CNN inference with in-situ analog arithmetic in ReRAM crossbars. ISAAC takes a PIM approach similar to PRIME, where memristor crossbar arrays not only store weights, but are also used to perform dot-product operations in an analog manner. This requires the integration of several digital and analog components, as well as overcoming multiple challenges.

\begin{figure}[t!]
\centering
\includegraphics[width=\columnwidth]{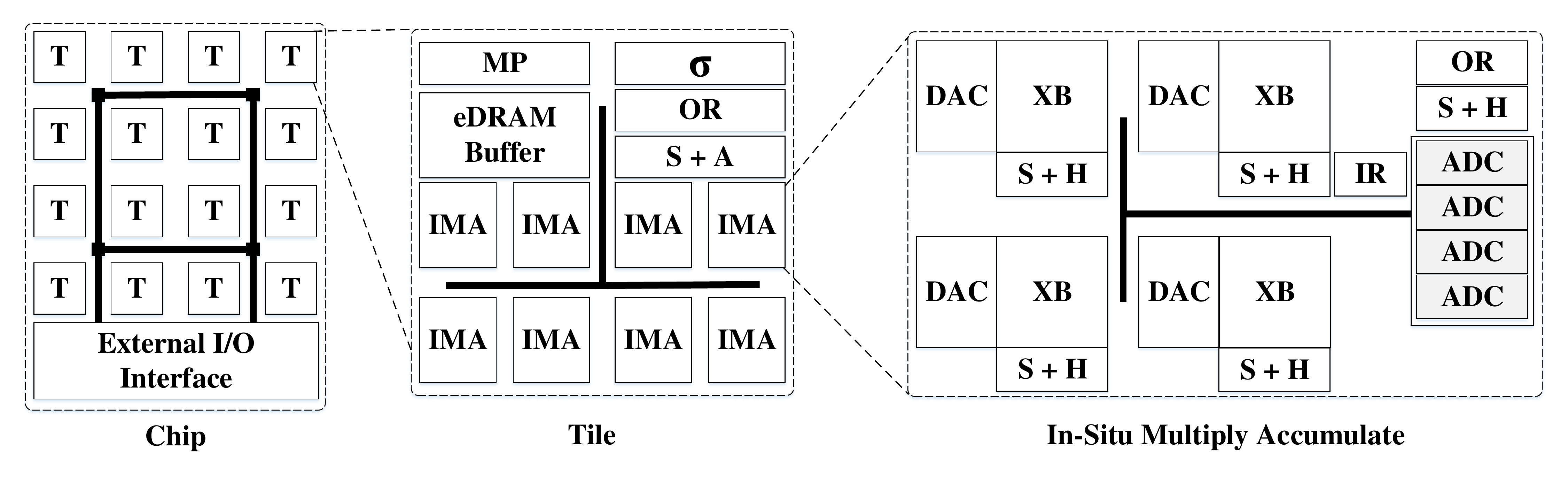}
\caption{Overview of the ISAAC architecture (adapted from~\cite{ISAAC}).}
\label{fig:isaac_architecture}
\end{figure}

First, the authors present a design with a pipelined architecture, employing dedicated crossbars for each DNN layer, and eDRAM buffers that aggregate data between pipeline stages. Fig.~\ref{fig:isaac_architecture} shows a high-level overview of the ISAAC architecture. Similar to DaDianNao~\cite{DaDianNao}, the system is organized into several tiles each designed with multiple In-situ Multiply-Accumulate (IMA) units to perform analog dot-products, shift-and-add (S+A), sigmoid and max pool (MP) unit, as well as some buffers and registers to store data. Within the core of each IMA, there are numerous ReRAM crossbar arrays, DACs/ADCs and Sample-and-Hold (S+H) units. ISAAC processes one layer at a time, distributing those computations across all tiles to maximize parallelism, and executing multiple layers and neurons on a single tile with time multiplexing. On the other hand, ReRAM crossbars are slow to re-program and, thus, each crossbar of ISAAC is dedicated to process a set of neurons in a given CNN layer. The outputs of that layer are fed to other crossbars that are dedicated to process the next CNN layer, and so on. As soon as enough outputs are generated by a layer and aggregated in an eDRAM buffer, the next layer can start its operations. By designing such a pipeline, the buffering requirements between layers is reduced, allowing to dedicate most of the chip for dot product engines. In order to improve the throughput of the initial layers and create a more balanced pipeline, additional crossbars can be employed for those layers by replicating the weights.

Second, the authors observed that the key overheads in a crossbar are the analog-to-digital (ADC) and digital-to-analog converters (DAC), which are required to communicate between units operating in different domains. Then, they propose strategies and define new data encoding techniques that are amenable to analog computations and can reduce the high overheads of DACs/ADCs. Each ADC is shared by multiple crossbar arrays. Weights are stored as 16-bit fixed-point values in $16/w$ $w$-bits cells of a single row (e.g., $w = 2$). A 16-bit digital input is represented using 16 consecutive voltage levels, each recording a 0/1 bit of the 16-bit input. Consequently, a pipelined execution is also employed within IMAs.

The analog product is computed as 16 sequential operations which requires only 1-bit DAC. A sample-and-hold circuit receives the bitline current, feeding it to a shared ADC unit when available, while a new set of inputs starts to be processed. Lastly, the partial sums are merged using shifts-and-adds. The crossbar inputs are provided as 2's complement, and the negative synaptic weights are represented as a bias. When the weights in a column are collectively large, they are stored in flipped form which ensures that the MSB of the sum-of-products is always 0, reducing the size of ADCs by one bit. Due to the nearly exponential relationship between the resolution and cost of ADC, and the large contribution of ADC in overall power consumption, this optimization has large impact on overall efficiency.

\emph{Discussion}. ISAAC design includes a balance of ReRAM storage/compute, ADCs, and eDRAM storage on a chip. ISAAC provides large improvements in throughput and energy efficiency compared to prior NDP accelerators. Along PRIME, ISAAC is one of the first PIM accelerators for CNNs based on ReRAM crossbars, exploiting the ability to perform analog dot-products within the crossbars. The pipelined architecture is an attractive solution to the long ReRAM writing latency, and the strategies to reduce the ADC/DAC overheads are also promising. However, ReRAM crossbars cannot be efficiently re-programmed on the fly, and the operations of some layers like the normalization cannot be easily adapted. In addition, their deep pipeline suffers from bubbles and stalls when not many inputs (e.g., images) can be successively fed into the accelerator. Moreover, the ADCs still represent an important energy overhead, and some digital components such as the shared data bus or the eDRAM buffers consume a large portion of area.

\subsubsection{Pipelayer}
Following the works of PRIME and ISAAC, Song et al. proposed PipeLayer~\cite{PipeLayer}, a pipelined ReRAM-based accelerator for both training and inference of CNNs. Since NN training involves weight updates and intricate data dependencies, most works only support inference, and assume that weights are written in ReRAM cells only at the beginning. By comparison, PipeLayer presents techniques for implementing NN training also in ReRAM.

\begin{figure}[t!]
\centering
\includegraphics[width=0.80\columnwidth]{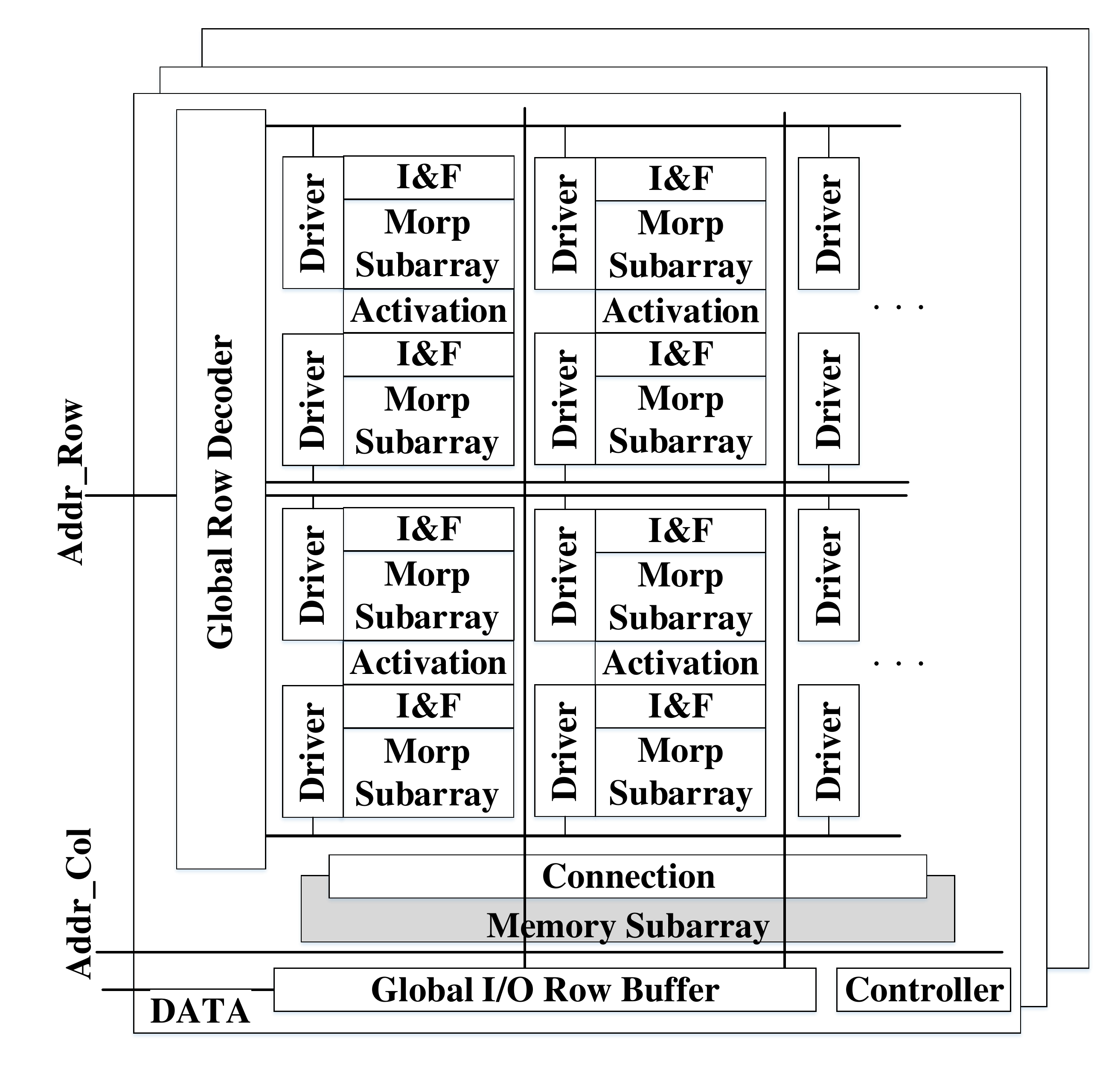}
\caption{Overview of the PipeLayer architecture. (adapted from~\cite{PipeLayer}).}
\label{fig:pipelayer_architecture}
\end{figure}

Fig.~\ref{fig:pipelayer_architecture} shows a high-level overview of the PipeLayer architecture. Similar to PRIME, they divide the ReRAM crossbar arrays into two types: memory and morphable subarrays. The morphable subarrays can perform both computation and data-storage, while the memory subarrays only store data such as results from the morphable subarrays. Both forward and backward computations have data dependencies. The results of forward computations are stored in memory subarrays, which are used in the backward computations for generating the errors and partial derivatives to update the weights.

\begin{figure}[t!]
\centering
\includegraphics[width=\columnwidth]{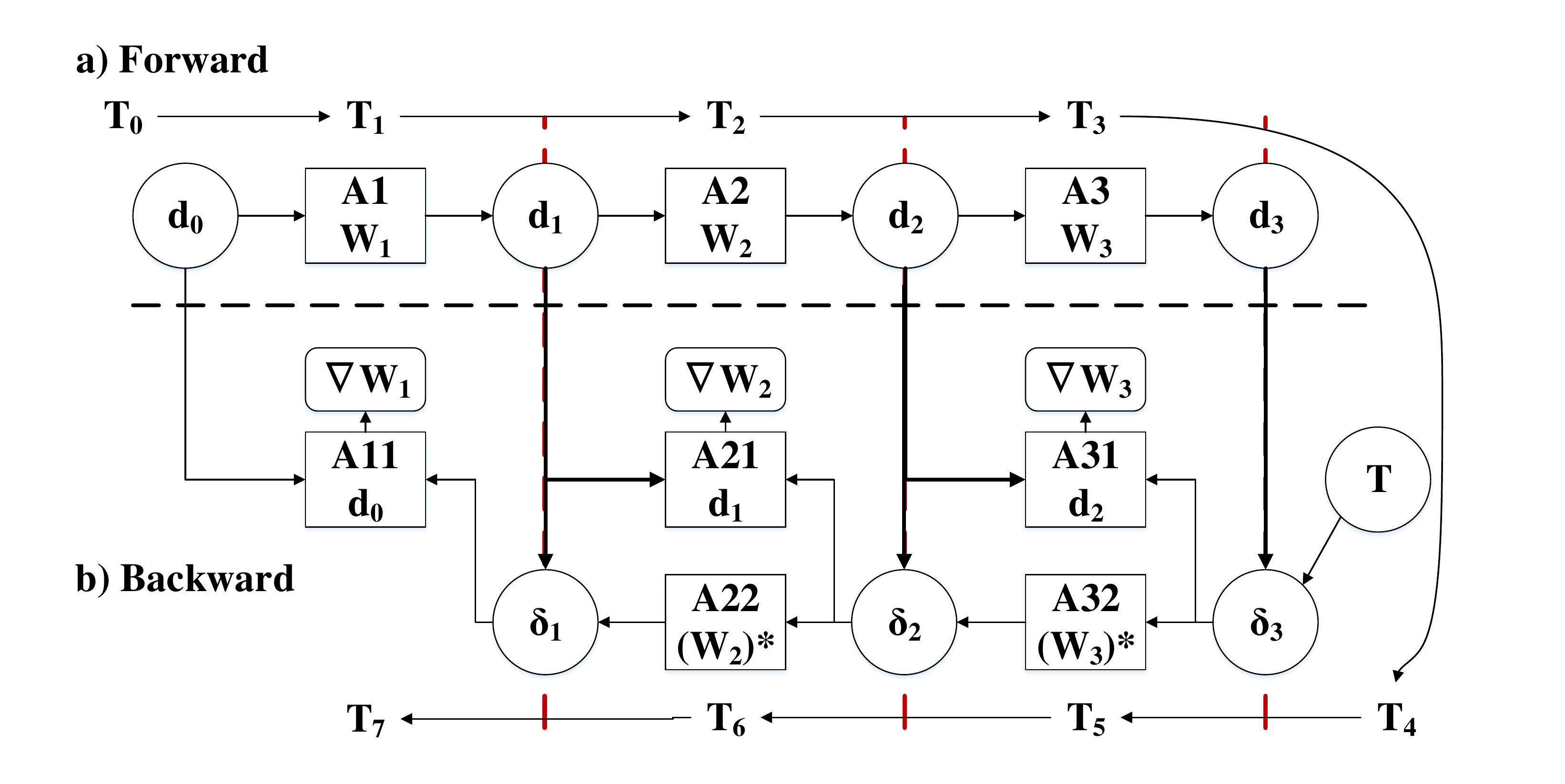}
\caption{Example of PipeLayer configured for training (adapted from~\cite{PipeLayer}).}
\label{fig:pipelayer_training}
\end{figure}

Fig.~\ref{fig:pipelayer_training} shows the training of a three-layer CNN with PipeLayer. In $T_1$, inputs stored in the memory subarray $d_0$ enter the morphable subarray A1 to perform a MVM operation. The results are written to the memory subarray $d_1$. This process continues until the results of the forward computation are stored in $d_3$. Then, backward computation starts in $T_4$, where errors $\delta_l$ (l denotes the layer) and partial derivatives ($\nabla W_l$) are generated. First, the error for the third layer ($\delta_3$) is calculated in $T_4$ and stored in a memory subarray. In $T_5$, two calculations happen concurrently depending on ($\delta_3$). First, the partial derivatives $\nabla W_3$ are calculated using the results in $d_2$ and $\delta_3$. Second, the error $\delta_2$ of the second layer is calculated from $\delta_3$, and based on $\nabla W_3$, weights in A31 and A32 are updated. Continuing in this manner, $\nabla W_1$ is calculated in $T_7$. Note that the cycles (e.g. $T_0$, $T_1$, ...) are logical cycles, meaning that each cycle may take several physical clock cycles depending on the implementation.

In training, the notion of batch limits the number of inputs that can be processed consecutively, because the inputs in the next batch need to be processed based on the updated weights. As previously discussed, the deep pipeline in ISAAC is vulnerable to pipeline bubbles and execution stalls, specially when the amount of consecutive inputs is small, resulting in an inefficient scheme for training. The authors analyze the data dependencies and weight update procedure in training algorithms and propose an efficient pipeline to exploit both intra/inter-layer parallelism. In order to exploit intra-layer parallelism, they noted that, since the number of inputs to a ReRAM crossbar array in each logical cycle is large, mapping all the kernels to a single crossbar leads to an inefficient design. Hence, they map and replicate them in multiple crossbars and, then, collect and add their outputs. The number of replicated copies of crossbars storing the same weights shows the granularity of parallelism, and by choosing a right value, a trade-off between hardware overhead and throughput can be achieved.

They further noted that, during training, the inputs processed before a weight-update (or batch) do not have any dependency. Since the batch size is usually much larger than one during the training (e.g. 64), they propose a pipelined training architecture where inputs inside a batch can be processed in a pipelined manner but inputs of next batch cannot enter the pipeline until the previous batch has been fully processed. To avoid the need of DACs, instead of voltage-level based input, they utilize a weighted spike-coding approach similar to ISAAC, and its higher latency is tolerated by the pipelined design of different layers. To avoid the need of ADCs, they use a “integration and fire” component (I\&F) which integrates analog currents and stores the generated output spikes in a counter.

\emph{Discussion.} PipeLayer enables a highly pipelined execution of both training and inference, without introducing the potential stalls in previous works, and opening up opportunities to do further research on ReRAM-based DNN training accelerators. Compared to a GPU implementation, their design achieves large improvements in performance and energy efficiency, but they do not compare against other state-of-the-art DNN accelerators for training. Similar to PRIME and ISAAC, PipeLayer requires a large amount of ReRAM crossbars due to the pipelined execution, and the throughput of the training may be limited by the slow writing latency and complex re-programming of ReRAM crossbars. In addition, their training dataflow is not much different from previous proposals, and one of the major drawbacks of a pipelined training is the increased requirement of memory subarrays, since the data stored during the forward computation must not be overwritten until consumed in the backward computation. Finally, similar to the modified SA of PRIME, the cost of the integration and fire circuit may be higher than a conventional ADC, limiting the performance of the accelerator.

\subsubsection{CASCADE}
Chou et al. proposed CASCADE~\cite{CASCADE}, a DNN accelerator connecting ReRAMs to extend the analog dataflow in an end-to-end PIM paradigm. A key limitation of ReRAM-based PIM architectures is the cost of the multi-bit analog-to-digital (A/D) conversions that are required to perform In-ReRAM analog dot product computations, which can defeat the efficiency and performance benefits of PIM. The digital inputs to the ReRAM crossbar need to be converted to voltage pulses using DACs, and the outputs of the crossbar in the form of analog currents need to be integrated and digitized using ADCs. The resolution requirement of analog computation is pushed to accommodate tens or hundreds of products of multi-bit WL pulses with multi-bit ReRAM conductances that are summed together. High resolution ADCs are required, adding a significant overhead that scales with the crossbar size and device resolution.

A second limitation of in-ReRAM computation is that even a single layer in a state-of-the-art DNN/RNN can be too large to fit on a practical ReRAM crossbar. Therefore, one kernel computation needs to be separated and mapped to multiple ReRAM crossbars. The resulting partial sums from multiple crossbars need to be digitized and accumulated in the digital domain. In addition, it is impractical to assume that an 8-bit weight value can be reliably stored in one ReRAM cell with current technologies. Multi-level cell (MLC) requires the use of more complex DACs/ADCs, and can be more easily affected by noise and process variations. In consequence, it is more practical to map a multi-bit weight value to multiple ReRAM cells. Similarly, it is also more practical to separate an input to units of 1 bit and apply them serially. All of these practical approaches lead to more partial sums that need to be digitized and digitally accumulated. To summarize, in-ReRAM computation consists of in-ReRAM dot products, A/D conversions, and digital accumulation of partial sums.

\begin{figure*}[t!]
\centering
\subfloat[]{
    \includegraphics[width=0.30\textwidth]{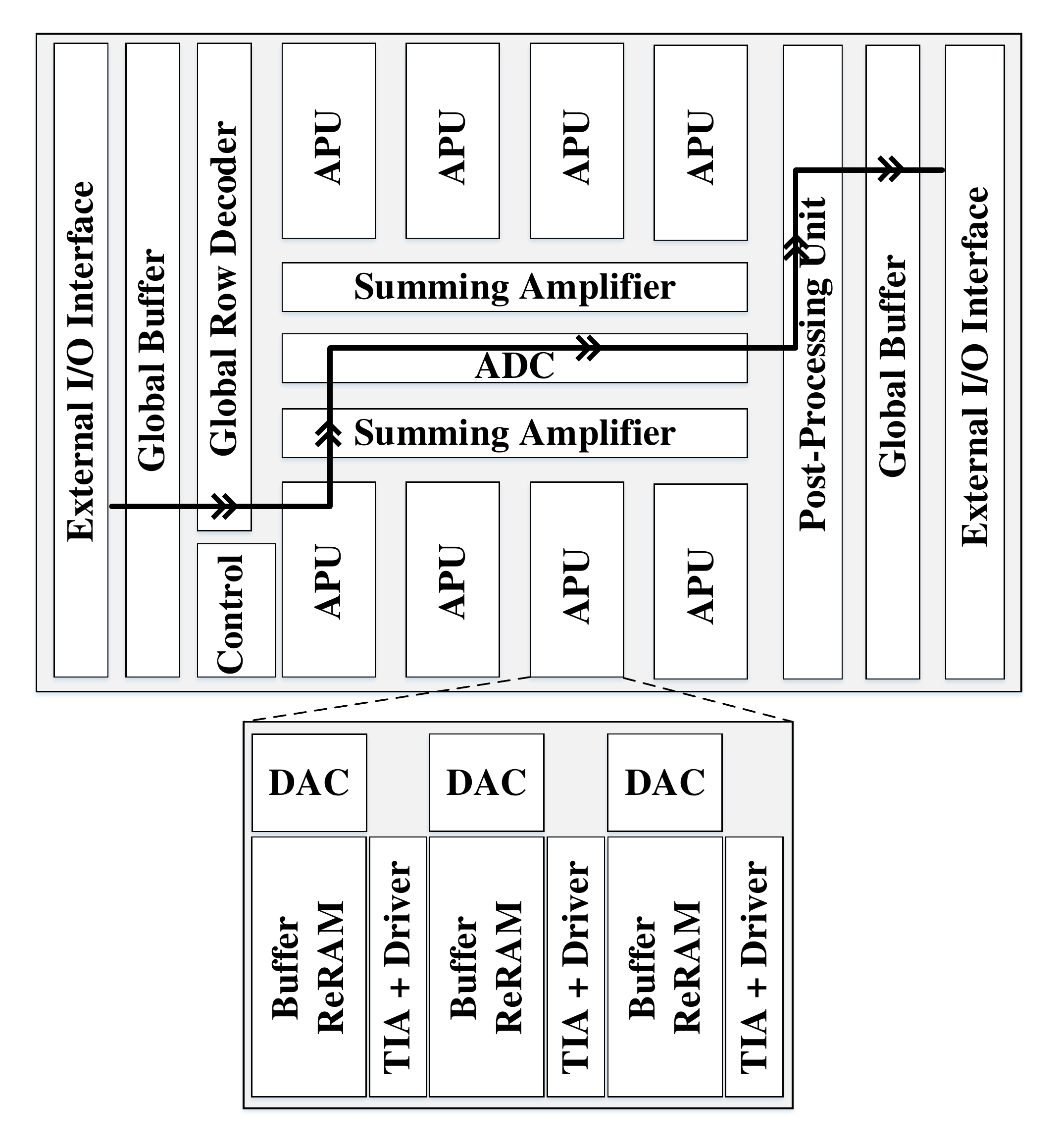}
    \label{fig:cascacde_architecture}
}
\subfloat[]{
    \includegraphics[width=0.65\textwidth]{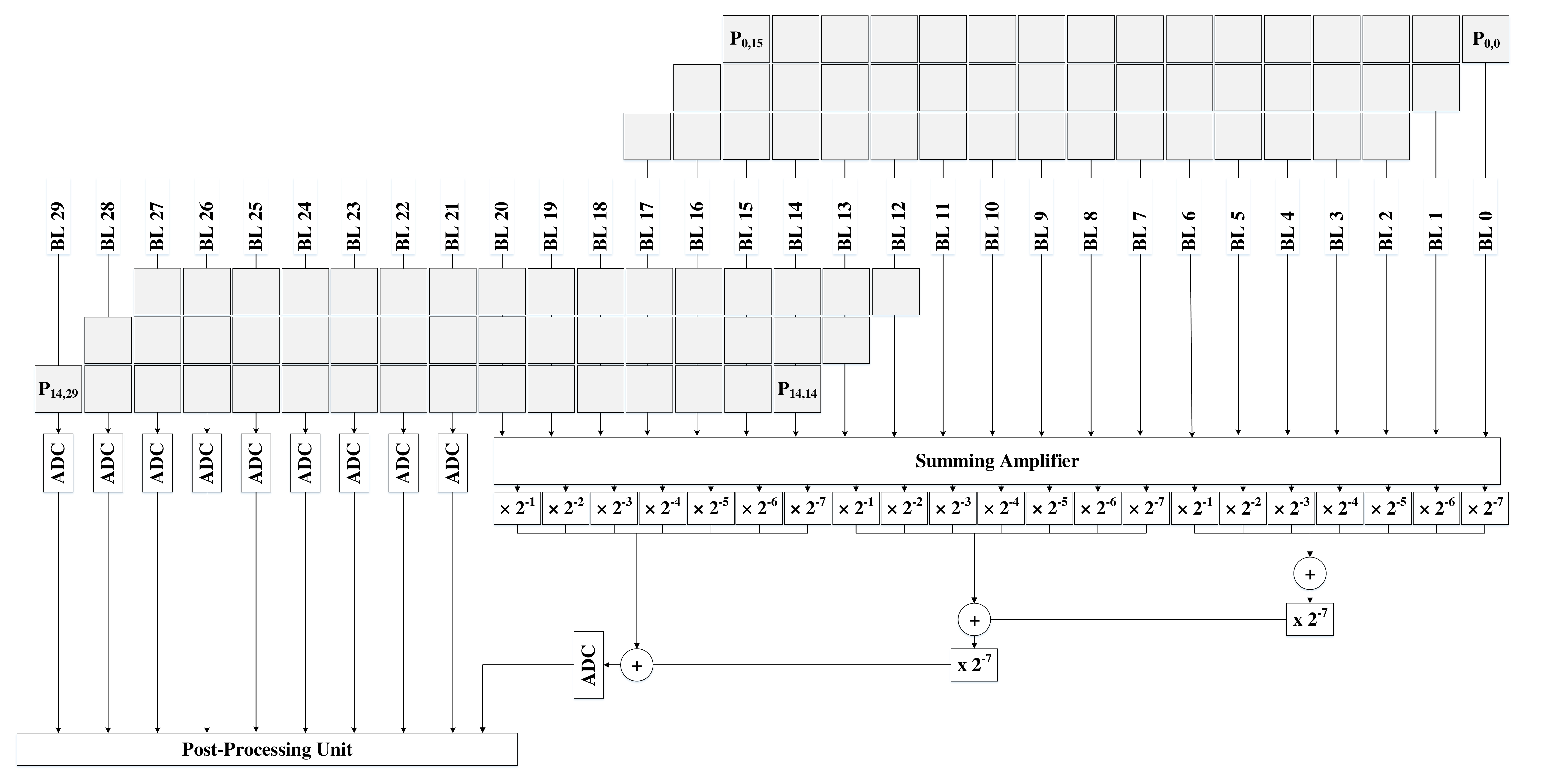}
    \label{fig:cascade_accumulations}
}
\caption{(a) Overview of the CASCADE architecture; and (b) final accumulation with a summing amplifier (adapted from~\cite{CASCADE}).}
\label{fig:cascade}
\end{figure*}

CASCADE is an in-ReRAM computation architecture for DNNs and RNNs, that specifically addresses the problems of high-cost A/D conversion and digital partial sum accumulation. Fig.~\ref{fig:cascacde_architecture} shows a high-level overview of the CASCADE architecture, which is made of multiple Analog Processing Units (APU). Each APU connects a multiply-accumulate (MAC) ReRAM crossbar array with a couple of buffer ReRAM arrays to extend the processing in analog domain and in-memory: dot products are followed by partial-sum buffering and accumulation to implement a complete DNN/RNN layer. The output currents of MAC ReRAMs are converted to proportional voltages using TIAs, and stored in the buffer ReRAMs. The transimpedance amplifier (TIA) interface is designed to enable a variation-tolerant, robust analog dataflow. Moreover, to limit the BL resolution to 6 bits, and reduce the impact of variation and noise, they use 1-bit weight mapping and moderate-sized MAC ReRAM arrays of $64 \times 64$. With bit-serial input streaming and binary weight mapping in a MAC ReRAM, only two voltage references are needed, one for read and one for write, simplifying routing and driver circuitry.

In addition, they propose to use a lower voltage to write to buffer ReRAMs, and 1T1R MLCs to control the write current, which leads to improved endurance and lower energy consumption. The authors also present a new memory mapping scheme named R-Mapping to enable the in-ReRAM accumulation of partial sums; and an analog summation scheme is used to reduce the number of A/D conversions required to obtain the final sum. Fig.~\ref{fig:cascade_accumulations} shows the final accumulation with a summing amplifier. The partial sums are divided into groups to efficiently obtain the output of a required resolution. The MSB group directly contributes to the required output, while the partial sums of the LSB group are connected to analog summing amplifiers that scale the current before summing. Finally, the digital values are added together to produce the final sum.

\emph{Discussion.} Overall, CASCADE is a DNN/RNN accelerator that executes a model layer by layer, and reduces the number of A/D conversions, compared to ISAAC and PRIME, by extending the analog dataflow with additional ReRAM buffers to perform the in-ReRAM accumulation of partial sums. Compared to recent in-ReRAM computation architectures, CASCADE provides large improvements in energy efficiency while maintaining a competitive throughput. However, the execution and dataflow of the accelerator is not described in detail, and the cost of writing in the ReRAM crossbars is not properly discussed, specially for the buffer ReRAMs that require multi-level cells (MLCs). Furthermore, the overhead of the TIAs is not compared against typical ADCs. Finally, activation, normalization and pooling layers are still performed in the digital domain, opening up opportunities to further extend the analog dataflow.

\subsubsection{RAPIDNN}
Imani et al. proposed RAPIDNN~\cite{RAPIDNN}, a DNN acceleration framework with neuron-to-memory transformation. As already discussed, one of the main challenges to extend the adoption of ReRAM crossbars to perform NN computations is the conversion between analog-digital domains. ADCs and DACs are slow and represent a large portion of area and power consumption in previous accelerators based on ReRAM crossbars. RAPIDNN addresses this issue by reinterpreting a DNN model and mapping it into a PIM accelerator, which is designed using NVM blocks that model four fundamental DNN operations, i.e., multiplication, addition, activation functions, and pooling, giving support to most of the typical DNN layers. Unlike ISAAC or PRIME, RAPIDNN supports all DNN functionalities in a digital-based memory design.

\begin{figure}[t!]
\centering
\includegraphics[width=1.0\columnwidth]{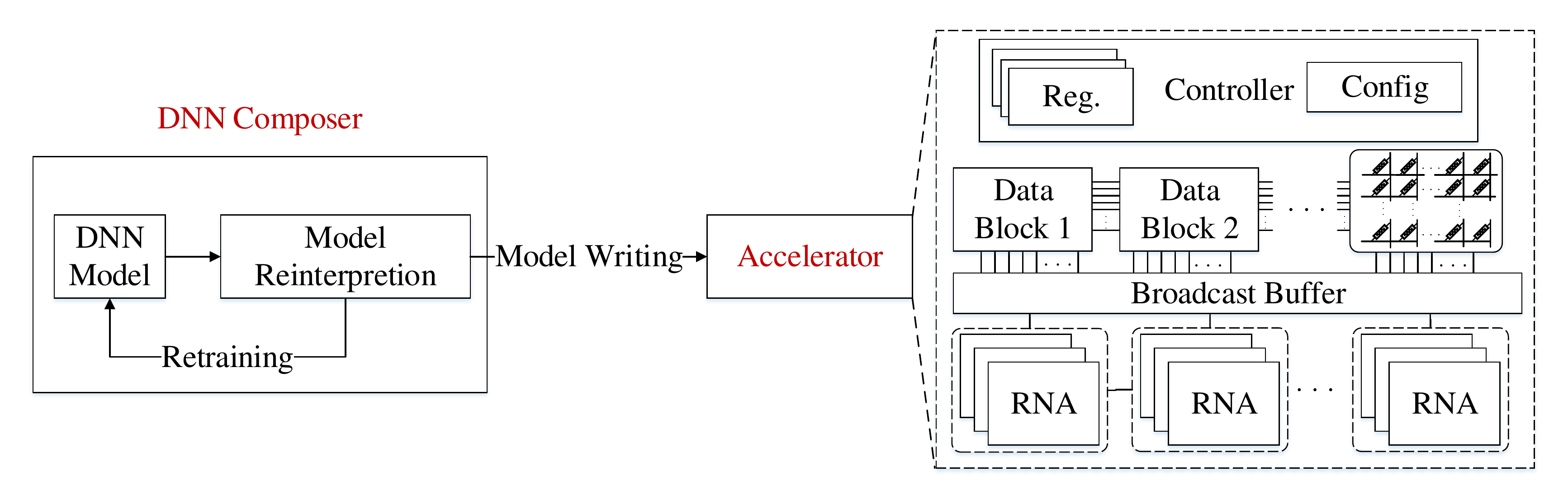}
\caption{Overview of the RAPIDNN framework (adapted from~\cite{RAPIDNN}).}
\label{fig:rapidnn_framework}
\end{figure}

The RAPIDNN accelerator supports multiplication and addition operations inside a ReRAM crossbar with single-level cells, and other operations, activation function and pooling, are modeled with associative memory (AM) blocks which are a form of lookup table. Their key observation is that, even though the operations of a DNN model are continuous functions, they can be approximated as step-wise functions without losing the quality of inference. Once a step-wise approximation is developed, they create computation tables which store the finite pre-computed values, and map them into specialized memory blocks capable of in-memory computations.

The RAPIDNN framework consists of the DNN Composer and the PIM accelerator as shown in Fig.~\ref{fig:rapidnn_framework}. The DNN Composer extracts representative operands of a DNN model, e.g., weights and input values, using clustering methods such as K-Means. The clustering of the weights is performed by layer in FC and by filter in CONV layers. For the inputs, since they change dynamically, they use a subset of the training set to generate the intermediate inputs of each layer and perform a clustering of those to obtain their quantization approach. The activation functions are also quantized into a subset of values. After performing the clustering, they test the network accuracy and if needed they perform retraining, and repeat the clustering process again to minimize the accuracy loss. Finally, the DNN Composer computes all the possible combinations of products between the set of clustered inputs and weights in an offline process, and stores the pre-computed results into the accelerator memory blocks.

\begin{figure*}[t!]
\centering
\includegraphics[width=1.0\textwidth]{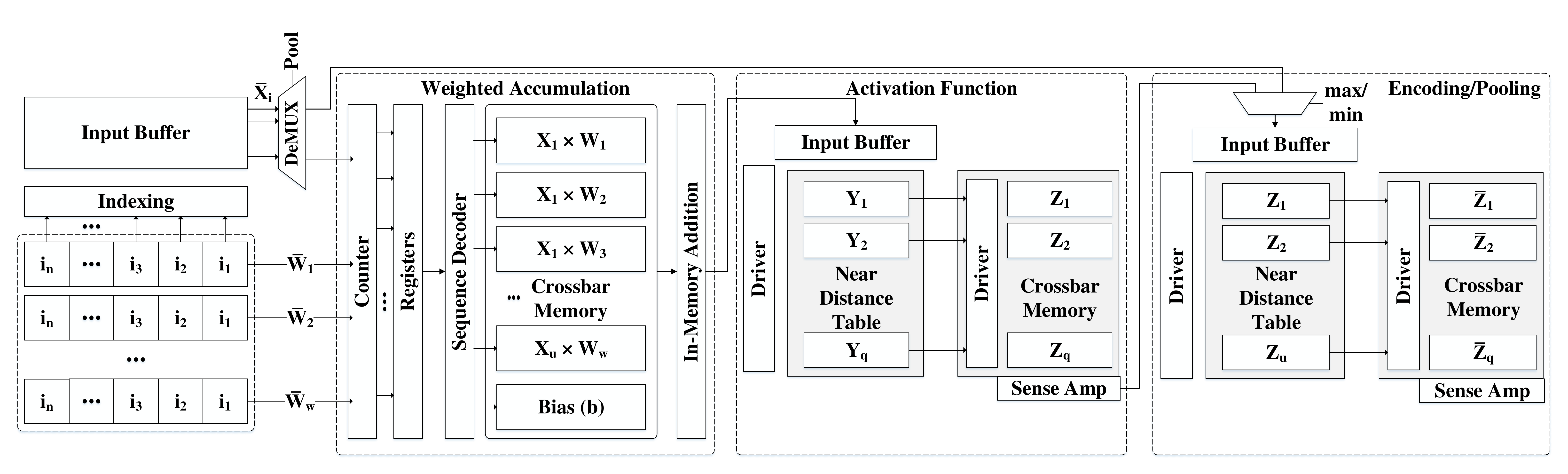}
\caption{Architecture of an RNA block in the RAPIDNN accelerator (adapted from~\cite{RAPIDNN}).}
\label{fig:rna_block}
\end{figure*}

On the other hand, the PIM accelerator is composed of two main blocks: Data blocks and Resistive Neural Acceleration (RNA) blocks. The Data Blocks are a type of NVM (e.g. ReRAM) storing all the parameters of a DNN model, while the RNAs are made of logic and AM tables consisting of a crossbar and a Nearest Distance CAM (ND-CAM). Fig.~\ref{fig:rna_block} shows the architecture of an RNA block. At runtime, the accelerator identifies computation results to be accessed based on an efficient in-memory search capability. Each RNA is dedicated to compute the output of a single neuron. The weighted accumulation is performed by counting the number of times each pair of encoded input/weight appears. Then, the counts will be reinterpreted as shifts and adds. The pre-computed products are stored in a crossbar memory which integrates the logic to perform shifts and additions. The additions are performed with NOR operations in the form of tree adders. Finally, both the activation function and the encoding are done with a ND-CAM and a lookup table. The ND-CAM finds the address of the closest value which is used to access the lookup table.

\emph{Discussion.} RAPIDNN is the first DNN accelerator which maps all functionalities inside the memory block using direct digital-based computations, and without any analog-to-digital conversion. The approximation of most operations, using crossbars that store pre-computed values and ND-CAMs that perform the searches, shows promising results, and leaves room for further improvement since the ND-CAM may not work correctly in some cases further increasing the accuracy loss. RAPIDNN achieves better performance and energy efficiency compared to DaDianNao, ISAAC and PipeLayer, while ensuring less than 1\% accuracy loss. However, RAPIDNN executes all the layers in a pipelined architecture which incurs in high power consumption. Moreover, RAPIDNN not only employs clustering but also approximates the weighted accumulations and the activations, which incurs in higher accuracy loss compared to linear quantization, and may require retraining. In addition, using clustering the accelerator still needs to perform some FP operations.

\section{Conclusions and Future Perspectives}\label{s:conclusions}
Recent advances in traditional memory systems, as well as new memory technologies, have renewed the interest in an old research topic (i.e. 1970~\cite{LiM}), currently known by the name of Near-Data Processing (NDP). In addition, many modern applications are data-intensive and demand a high level of parallelism and memory bandwidth. While deep learning techniques, and machine learning in general, are promising solutions to a broad range of applications, the characterists of conventional memories and processors limits their potential. Addressing these challenges requires fundamental breakthroughs in memory and computer architectures. In this paper, we present a survey of techniques for designing DNN accelerators with NDP architectures. We describe the key insights of different recent works, organized the works in several categories based on the memory technology employed to highlight their similarities and differences. Finally, in this section we conclude the paper with a discussion of future work.

Based on the observation that DRAM/SRAM arrays can simultaneously activate multiple rows to perform bitwsise operations, many prior works focus on adding PIM capabilities to commodity memories with minor changes, so that the main structure and interface to communicate memory and CPU/GPU systems remains the same. Future works can remove the need to keep the architectural structure of traditional memory systems with custom accelerators. ASICs are an interesting solution since they are more flexible and offer the opportunity to combine different memory technologies, conventional and emerging, with PIM capabilities to provide a computer system optimized for data-centric applications. For example, Ambit could be implemented in an HMC, and be further extended and improved to exploit this new memory technology. Moreover, most works based on traditional memory systems only include computation capabilities in a specific level of the memory hierarchy, such as the LLC or main memory, but multiple levels of the hierarchy could be used combined with novel dataflows.

In another line of research, Ambit and \emph{Neural Cache} effectively exploit the bandwidth of typical DRAM and SRAM arrays, respectively, and the parallelism of multiple arrays, but only perform computations between two rows of a given array. While two rows may already offer a high level of parallelism, it can be a major drawback compared to ReRAM crossbars that can operate using the entire array. Further research could be done to exploit the analog operation of SRAM/DRAM arrays to increase the level of parallelism, which may require the use of modified memory cells and peripherals. Following this direction, \emph{Neural Cache} includes extra logic in the sense amplifiers of SRAM arrays to perform complex operations, and Ambit employs a row of 2T1C cells for the DRAM bitwise NOT operation. Other new memory arrays with additional logic in the peripherals and/or a mix of different cells to perform faster operations could be investigated. On top of that, new data layouts could be explored to facilitate the interaction between bitlines. Although additional circuitry may be expensive and reduce the memory density, ASICs can be flexible enough in terms of storage capacity to overcome the overheads.

In addition, it is critical to design efficient and reliable PIM DRAM/SRAM arrays taking into account not only the negative effects of process variations but also aging and soft errors.

On the other hand, Bit-Prudent extends \emph{Neural Cache} with support for sparse and low-precision models, but they only consider a structured channel pruning for CNNs and binary/ternary networks. New pruning methods and dynamic precision approaches can be studied and evaluated in combination with novel hardware changes to support these models in NDP architectures based on commodity memory. Besides, given the limitations of CMOS technology, the use of new emerging memory technologies, such as HMC or ReRAM, could also be interesting for these designs.

Most works based on 3D-stacked memory with HMCs perform the majority of the computations in the logic die. In contrast, TETRIS introduced some adders in the DRAM dies to perform accumulations, and although the technology process to fabricate the different layers may differ and be slower, solutions that perform more computations in the memory layers could be further investigated. Following this approach, the level of parallelism in HMCs can be increased exploiting not only the independent vaults but also the different memory layers, which may require a better organization of the data as well as novel dataflows.

In the same line, Neurocube and TETRIS are mostly focused on CNNs, but recent DNN models, such as RNNs or Transformers, may also require new data mappings and execution schemes. In addition, these accelerators tend to work on a single NN layer at a time, but considering that there is usually enough storage capacity for the entire model, a pipelined execution of the layers could be more practical.

On top of that, since there is an increasing number of applications that require the execution of multiple DNNs, efforts could be made for exploring new schedulers to support the parallel execution of multiple models, as well as potential computation reuse opportunities.

Furthermore, little research has been done on DNN accelerators for both inference and training exploiting the intrinsic features of 3D-stacked memories. Similarly, sparse DNN accelerators based on HMCs have received little attention. Moreover, some of these solutions may incur in additional traffic to communicate between vaults, turning the NoC into a potential bottleneck of the system, and requiring additional improvements. On the other hand, some of the main challenges to extend the adoption of 3D-stacked memory and HMCs are the thermal and area constrains, which limit the scaling of the architectures based on this technology and, hence, proper solutions are required to relax theses restrictions.

Following a parallel line of research, most ReRAM based architectures have focused on accelerating a limited range of computations/algorithms, such as bitwise operations or MVM, which represent the majority of the DNN computations. However, not every DNN model/layer/operation can be easily integrated into every NDP accelerator. For example, the normalization layer requires operations that cannot be easily adapted and accelerated using ReRAM crossbars and, hence, it may be important to investigate novel algorithm-hardware co-designs.

In addition, the difficulty in re-programming the memristor arrays on the fly has led many prior works to propose architectures with several ReRAM crossbars to instantiate the full DNN model once, and then proceed with a pipelined execution to improve the utilization of the resources. This solution can be very effective in an HPC environment where batches of multiple inputs can be easily generated to achieve high throughput, and where the area and power constraints are not so tight. However, embedded and mobile devices have strict restrictions, so the number of crossbars must be relatively small and not able to store the full model. In consequence, solutions that increase the ReRAM write endurance or reduce the ReRAM write latency/energy can be highly attractive for power-constrained systems, enabling architectures that execute one layer at a time and perform efficient context switching. Similarly, techniques to reduce the number of write operations such as data-compression will be very effective in dealing with the write-agnostic nature of ReRAM.

Equally important, initial proposals of ReRAM based accelerators for DNNs, such as PRIME and ISAAC, incur in high energy and area overheads due to the numerous A/D conversions, whose cost is directly proportional to the ADCs/DACs resolution. The bit-serial input voltage execution reduces the DAC resolution to just one bit. However, lowering the ADC resolution may affect the final accuracy of the DNN since the outputs would be truncated. Consequently, the design of alternative methods to perform A/D conversions is crucial to reduce these overheads. Besides, each ReRAM based accelerator employs a different encoding/decoding scheme for inputs and weights with its corresponding hardware. However, it is not clear how they compare against each other. For example, PipeLayer employs an integrate and fire scheme that replaces the ADCs, but a comparative evaluation is not provided.

On the other hand, follow up works have tried to reduce the use of DACs/ADCs either by operating in analog domain as much as possible to reduce the number of A/D conversions (i.e. CASCADE), or by digitilizing the full design approximating most of the operations with ND-CAM searches (i.e. RAPIDNN). Still, these approaches do not solve the problem completely and/or add other problems to be solved. Therefore, further research in this direction to propose a better design may be interesting.

Compared to DNNs, less progress has been made towards implementing Spiking Neural Networks (SNNs) on ReRAM. Similarly, training acceleration has received less attention than inference acceleration because the training phase is more computationally demanding than the inference phase. In addition, ReRAM crossbars are not well suited for DNN training due to the long writing latency of memristor arrays and the difficulty for re-programming the ReRAM crossbar on the fly. In consequence, most previous works focus on optimizing the inference phase. PipeLayer is one of the few ReRAM based accelerators supporting training. However, their training dataflow is mostly based on state-of-the-art DNN accelerators for training. Similar to ISAAC, PipeLayer employs a pipelined execution that requires a large number of crossbars, which may not be suitable for low power devices. In addition, it is not clear how they deal with all the ReRAM writing issues.

As the works based on 3D-stacked memory, most state-of-the-art ReRAM based accelerators do not support the execution of sparse/binary/ternary models nor dynamic precision schemes. For example, pruning methods that are more suitable to perform analog computations with ReRAM crossbars, or schemes that dynamically change the precision of the inputs benefiting from the bit-serial input voltage execution may be interesting lines of research.

The vulnerabilities of ReRAM crossbars, such as process variations, hard faults or resistance drift, pose a serious threat to the accuracy of in-ReRAM computations. In addition, crossbars and DACs/ADCs can be exposed to analog signal degradation or external noise sources that can further reduce the DNN model quality. Tolerating these errors is feasible only in error-resilient applications which comprise a fraction of the total applications. Consequently, advanced manufacturing processes will be even more crucial to reduce the ReRAM vulnerabilities, and future research should focus on improving the design of the accelerators to be more resilient to possible errors while maintaining the performance and energy efficiency.

Finally, most existing NDP architectures for DNNs employ HMCs with DRAM dies or ReRAM crossbars. A comparative evaluation of other high density 3D stack memory technologies such as Wide I/O and HBM would be also intesting, as well as combining 3D stack architectures with NVMs such as PCM and STT. Ideally, one would like an accelerator not only for DNNs but for multiple data-centric applications such as graph processing and databases. Computer architects have the opportunity to extend the benefits of NDP architectures to the entire spectrum of applications, by employing an heterogeneous hardware approach that combines different emerging memory technologies with PIM capabilities, to accelerate not only DNN models with its corresponding layers and operations but also a variety of data intensive applications.


\ifCLASSOPTIONcaptionsoff
  \newpage
\fi

\bibliographystyle{IEEEtran}
\bibliography{IEEEabrv, ref}

\begin{thebibliography}{10}
\providecommand{\url}[1]{#1}
\csname url@samestyle\endcsname
\providecommand{\newblock}{\relax}
\providecommand{\bibinfo}[2]{#2}
\providecommand{\BIBentrySTDinterwordspacing}{\spaceskip=0pt\relax}
\providecommand{\BIBentryALTinterwordstretchfactor}{4}
\providecommand{\BIBentryALTinterwordspacing}{\spaceskip=\fontdimen2\font plus
\BIBentryALTinterwordstretchfactor\fontdimen3\font minus
  \fontdimen4\font\relax}
\providecommand{\BIBforeignlanguage}[2]{{%
\expandafter\ifx\csname l@#1\endcsname\relax
\typeout{** WARNING: IEEEtran.bst: No hyphenation pattern has been}%
\typeout{** loaded for the language `#1'. Using the pattern for}%
\typeout{** the default language instead.}%
\else
\language=\csname l@#1\endcsname
\fi
#2}}
\providecommand{\BIBdecl}{\relax}
\BIBdecl

\bibitem{vonneumann_arch}
J.~Von~Neumann, ``First draft of a report on the edvac,'' \emph{IEEE Annals of
  the History of Computing}, vol.~15, no.~4, pp. 27--75, 1993.

\bibitem{pandiyan2014quantifying}
D.~Pandiyan and C.-J. Wu, ``Quantifying the energy cost of data movement for
  emerging smart phone workloads on mobile platforms,'' in \emph{2014 IEEE
  International Symposium on Workload Characterization (IISWC)}.\hskip 1em plus
  0.5em minus 0.4em\relax IEEE, 2014, pp. 171--180.

\bibitem{kestor_energy_datamovement}
G.~Kestor, R.~Gioiosa, D.~J. Kerbyson, and A.~Hoisie, ``Quantifying the energy
  cost of data movement in scientific applications,'' in \emph{2013 IEEE
  international symposium on workload characterization (IISWC)}.\hskip 1em plus
  0.5em minus 0.4em\relax IEEE, 2013, pp. 56--65.

\bibitem{NDPBW}
R.~Balasubramonian, J.~Chang, T.~Manning, J.~H. Moreno, R.~Murphy, R.~Nair, and
  S.~Swanson, ``Near-data processing: Insights from a micro-46 workshop,''
  \emph{IEEE Micro}, vol.~34, no.~4, pp. 36--42, 2014.

\bibitem{speech_recognition_dnn}
L.~Deng, G.~Hinton, and B.~Kingsbury, ``New types of deep neural network
  learning for speech recognition and related applications: An overview,'' in
  \emph{2013 IEEE international conference on acoustics, speech and signal
  processing}.\hskip 1em plus 0.5em minus 0.4em\relax IEEE, 2013, pp.
  8599--8603.

\bibitem{image_processing_dnn}
M.~Egmont-Petersen, D.~de~Ridder, and H.~Handels, ``Image processing with
  neural networks—a review,'' \emph{Pattern recognition}, vol.~35, no.~10,
  pp. 2279--2301, 2002.

\bibitem{machine_translation_dnn}
D.~Bahdanau, K.~Cho, and Y.~Bengio, ``Neural machine translation by jointly
  learning to align and translate,'' \emph{arXiv preprint arXiv:1409.0473},
  2014.

\bibitem{AlexNet}
A.~Krizhevsky, I.~Sutskever, and G.~E. Hinton, ``Imagenet classification with
  deep convolutional neural networks,'' \emph{Advances in neural information
  processing systems}, vol.~25, pp. 1097--1105, 2012.

\bibitem{GoogleNet}
C.~Szegedy, W.~Liu, Y.~Jia, P.~Sermanet, S.~Reed, D.~Anguelov, D.~Erhan,
  V.~Vanhoucke, and A.~Rabinovich, ``Going deeper with convolutions,'' in
  \emph{Proceedings of the IEEE conference on computer vision and pattern
  recognition}, 2015, pp. 1--9.

\bibitem{NVIDIA28nmDallyLecture}
W.~J. Dally, ``Gpu computing: To exascale and beyond,'' \emph{Lecture
  slides--http://www. nvidia. com/content/PDF/sc}, 2010.

\bibitem{wulf1995memorywall}
W.~A. Wulf and S.~A. McKee, ``Hitting the memory wall: implications of the
  obvious,'' \emph{ACM SIGARCH computer architecture news}, vol.~23, no.~1, pp.
  20--24, 1995.

\bibitem{bandwidthwall}
A.~Kagi, J.~R. Goodman, and D.~Burger, ``Memory bandwidth limitations of future
  microprocessors,'' in \emph{23rd Annual International Symposium on Computer
  Architecture (ISCA'96)}.\hskip 1em plus 0.5em minus 0.4em\relax IEEE, 1996,
  pp. 78--78.

\bibitem{HennessyPatterson12}
J.~L. Hennessy and D.~A. Patterson, \emph{Computer Architecture: A Quantitative
  Approach}.\hskip 1em plus 0.5em minus 0.4em\relax Morgan Kaufmann, 2012.

\bibitem{siegl2016data}
P.~Siegl, R.~Buchty, and M.~Berekovic, ``Data-centric computing frontiers: A
  survey on processing-in-memory,'' in \emph{Proceedings of the Second
  International Symposium on Memory Systems}, 2016, pp. 295--308.

\bibitem{DaDianNao}
Y.~Chen, T.~Luo, S.~Liu, S.~Zhang, L.~He, J.~Wang, L.~Li, T.~Chen, Z.~Xu,
  N.~Sun \emph{et~al.}, ``Dadiannao: A machine-learning supercomputer,'' in
  \emph{MICRO}, 2014.

\bibitem{TPU}
N.~P. Jouppi, C.~Young, N.~Patil, D.~Patterson, G.~Agrawal, R.~Bajwa, S.~Bates,
  S.~Bhatia, N.~Boden, A.~Borchers \emph{et~al.}, ``In-datacenter performance
  analysis of a tensor processing unit,'' in \emph{Proceedings of the 44th
  annual international symposium on computer architecture}, 2017, pp. 1--12.

\bibitem{ComputeDRAM}
F.~Gao, G.~Tziantzioulis, and D.~Wentzlaff, ``Computedram: In-memory compute
  using off-the-shelf drams,'' in \emph{Proceedings of the 52nd annual IEEE/ACM
  international symposium on microarchitecture}, 2019, pp. 100--113.

\bibitem{DRACC}
Q.~Deng, L.~Jiang, Y.~Zhang, M.~Zhang, and J.~Yang, ``Dracc: a dram based
  accelerator for accurate cnn inference,'' in \emph{Proceedings of the 55th
  Annual Design Automation Conference}, 2018, pp. 1--6.

\bibitem{DRISA}
S.~Li, D.~Niu, K.~T. Malladi, H.~Zheng, B.~Brennan, and Y.~Xie, ``Drisa: A
  dram-based reconfigurable in-situ accelerator,'' in \emph{2017 50th Annual
  IEEE/ACM International Symposium on Microarchitecture (MICRO)}.\hskip 1em
  plus 0.5em minus 0.4em\relax IEEE, 2017, pp. 288--301.

\bibitem{[PIM_PCM]}
S.~Li, C.~Xu, Q.~Zou, J.~Zhao, Y.~Lu, and Y.~Xie, ``Pinatubo: A
  processing-in-memory architecture for bulk bitwise operations in emerging
  non-volatile memories,'' in \emph{Proceedings of the 53rd Annual Design
  Automation Conference}, 2016, pp. 1--6.

\bibitem{PIM_STT_MRAM}
Y.~Pan, P.~Ouyang, Y.~Zhao, W.~Kang, S.~Yin, Y.~Zhang, W.~Zhao, and S.~Wei, ``A
  multilevel cell stt-mram-based computing in-memory accelerator for binary
  convolutional neural network,'' \emph{IEEE Transactions on Magnetics},
  vol.~54, no.~11, pp. 1--5, 2018.

\bibitem{Lecun_DEEP_LEARNING}
Y.~LeCun, Y.~Bengio, and G.~Hinton, ``Deep learning,'' \emph{nature}, vol. 521,
  no. 7553, pp. 436--444, 2015.

\bibitem{Schmidhuber_DEEP_LEARNING}
J.~Schmidhuber, ``Deep learning in neural networks: An overview,'' \emph{Neural
  networks}, vol.~61, pp. 85--117, 2015.

\bibitem{MLP}
F.~Rosenblatt, ``Principles of neurodynamics. perceptrons and the theory of
  brain mechanisms,'' Cornell Aeronautical Lab Inc Buffalo NY, Tech. Rep.,
  1961.

\bibitem{CNN}
Y.~LeCun, L.~Bottou, Y.~Bengio, and P.~Haffner, ``Gradient-based learning
  applied to document recognition,'' \emph{Proceedings of the IEEE}, vol.~86,
  no.~11, pp. 2278--2324, 1998.

\bibitem{RNN}
H.~Jaeger, \emph{Tutorial on training recurrent neural networks, covering BPPT,
  RTRL, EKF and the" echo state network" approach}.\hskip 1em plus 0.5em minus
  0.4em\relax GMD-Forschungszentrum Informationstechnik Bonn, 2002, vol.~5,
  no.~01.

\bibitem{chua1971memristor}
L.~Chua, ``Memristor-the missing circuit element,'' \emph{IEEE Transactions on
  circuit theory}, vol.~18, no.~5, pp. 507--519, 1971.

\bibitem{strukov2008developingmemristor}
D.~B. Strukov, G.~S. Snider, D.~R. Stewart, and R.~S. Williams, ``The missing
  memristor found,'' \emph{nature}, vol. 453, no. 7191, pp. 80--83, 2008.

\bibitem{seal-lab-tech-report/reram-structure}
P.~Chi, S.~Li, C.~Xu, T.~Zhang, J.~Zhao, Y.~Wang, Y.~Liu, and Y.~Xie,
  ``Processing-in-memory in reram-based main memory,'' \emph{SEAL-lab Technical
  Report}, no. 2015-001, 2015.

\bibitem{Akinaga_ReRAM}
H.~Akinaga and H.~Shima, ``Resistive random access memory (reram) based on
  metal oxides,'' \emph{Proceedings of the IEEE}, vol.~98, no.~12, pp.
  2237--2251, 2010.

\bibitem{ReRAM_ACCELERATOR_SURVEY}
S.~Mittal, ``A survey of reram-based architectures for processing-in-memory and
  neural networks,'' \emph{Machine learning and knowledge extraction}, vol.~1,
  no.~1, pp. 75--114, 2019.

\bibitem{ISAAC}
A.~{Shafiee}, A.~{Nag}, N.~{Muralimanohar}, R.~{Balasubramonian}, J.~P.
  {Strachan}, M.~{Hu}, R.~S. {Williams}, and V.~{Srikumar}, ``Isaac: A
  convolutional neural network accelerator with in-situ analog arithmetic in
  crossbars,'' in \emph{2016 ACM/IEEE 43rd Annual International Symposium on
  Computer Architecture (ISCA)}, 2016, pp. 14--26.

\bibitem{davis2005prosandconsof3d}
W.~R. Davis, J.~Wilson, S.~Mick, J.~Xu, H.~Hua, C.~Mineo, A.~M. Sule, M.~Steer,
  and P.~D. Franzon, ``Demystifying 3d ics: The pros and cons of going
  vertical,'' \emph{IEEE Design \& Test of Computers}, vol.~22, no.~6, pp.
  498--510, 2005.

\bibitem{neurocube/ISCA.2016.41}
\BIBentryALTinterwordspacing
D.~Kim, J.~Kung, S.~Chai, S.~Yalamanchili, and S.~Mukhopadhyay, ``Neurocube: A
  programmable digital neuromorphic architecture with high-density 3d memory,''
  in \emph{Proceedings of the 43rd International Symposium on Computer
  Architecture}, ser. ISCA '16.\hskip 1em plus 0.5em minus 0.4em\relax IEEE
  Press, 2016, p. 380–392. [Online]. Available:
  \url{https://doi.org/10.1109/ISCA.2016.41}
\BIBentrySTDinterwordspacing

\bibitem{DEEPTRAIN}
D.~Kim, T.~Na, S.~Yalamanchili, and S.~Mukhopadhyay, ``Deeptrain: A
  programmable embedded platform for training deep neural networks,''
  \emph{IEEE Transactions on Computer-Aided Design of Integrated Circuits and
  Systems}, vol.~37, no.~11, pp. 2360--2370, 2018.

\bibitem{hybrid-memory-cube}
H.~M.~C. Consortium \emph{et~al.}, ``Hybrid memory cube specification 2.1,''
  \emph{hybridmemorycube. org}, 2014.

\bibitem{HMC_Analysis}
P.~Rosenfeld, E.~Cooper-Balis, T.~Farrell, D.~Resnick, and B.~Jacob, ``Peering
  over the memory wall: Design space and performance analysis of the hybrid
  memory cube,'' \emph{Univ. of Maryland Systems and Computer Architecture
  Group, Tech. Rep. UMD-SCA-2012-10-01}, 2012.

\bibitem{seshadri2017ambit}
V.~Seshadri, D.~Lee, T.~Mullins, H.~Hassan, A.~Boroumand, J.~Kim, M.~A. Kozuch,
  O.~Mutlu, P.~B. Gibbons, and T.~C. Mowry, ``Ambit: In-memory accelerator for
  bulk bitwise operations using commodity dram technology,'' in \emph{2017 50th
  Annual IEEE/ACM International Symposium on Microarchitecture (MICRO)}.\hskip
  1em plus 0.5em minus 0.4em\relax IEEE, 2017, pp. 273--287.

\bibitem{seshadri2013rowclone}
V.~Seshadri, Y.~Kim, C.~Fallin, D.~Lee, R.~Ausavarungnirun, G.~Pekhimenko,
  Y.~Luo, O.~Mutlu, P.~B. Gibbons, M.~A. Kozuch \emph{et~al.}, ``Rowclone: fast
  and energy-efficient in-dram bulk data copy and initialization,'' in
  \emph{Proceedings of the 46th Annual IEEE/ACM International Symposium on
  Microarchitecture}, 2013, pp. 185--197.

\bibitem{eckert2018neuralcache}
C.~Eckert, X.~Wang, J.~Wang, A.~Subramaniyan, R.~Iyer, D.~Sylvester,
  D.~Blaaauw, and R.~Das, ``Neural cache: Bit-serial in-cache acceleration of
  deep neural networks,'' in \emph{2018 ACM/IEEE 45th Annual International
  Symposium on Computer Architecture (ISCA)}.\hskip 1em plus 0.5em minus
  0.4em\relax IEEE, 2018, pp. 383--396.

\bibitem{aga2017computecache}
S.~Aga, S.~Jeloka, A.~Subramaniyan, S.~Narayanasamy, D.~Blaauw, and R.~Das,
  ``Compute caches,'' in \emph{2017 IEEE International Symposium on High
  Performance Computer Architecture (HPCA)}.\hskip 1em plus 0.5em minus
  0.4em\relax IEEE, 2017, pp. 481--492.

\bibitem{wang2019bitprudent}
X.~Wang, J.~Yu, C.~Augustine, R.~Iyer, and R.~Das, ``Bit prudent in-cache
  acceleration of deep convolutional neural networks,'' in \emph{2019 IEEE
  International Symposium on High Performance Computer Architecture
  (HPCA)}.\hskip 1em plus 0.5em minus 0.4em\relax IEEE, 2019, pp. 81--93.

\bibitem{han2015learning}
S.~Han, J.~Pool, J.~Tran, and W.~Dally, ``Learning both weights and connections
  for efficient neural network,'' in \emph{Advances in neural information
  processing systems}, 2015, pp. 1135--1143.

\bibitem{gao2017tetris}
M.~Gao, J.~Pu, X.~Yang, M.~Horowitz, and C.~Kozyrakis, ``Tetris: Scalable and
  efficient neural network acceleration with 3d memory,'' in \emph{Proceedings
  of the Twenty-Second International Conference on Architectural Support for
  Programming Languages and Operating Systems}, 2017, pp. 751--764.

\bibitem{chen2016eyeriss}
Y.-H. Chen, T.~Krishna, J.~S. Emer, and V.~Sze, ``Eyeriss: An energy-efficient
  reconfigurable accelerator for deep convolutional neural networks,''
  \emph{IEEE journal of solid-state circuits}, vol.~52, no.~1, pp. 127--138,
  2016.

\bibitem{PRIME}
P.~Chi, S.~Li, C.~Xu, T.~Zhang, J.~Zhao, Y.~Liu, Y.~Wang, and Y.~Xie, ``Prime:
  A novel processing-in-memory architecture for neural network computation in
  reram-based main memory,'' \emph{ACM SIGARCH Computer Architecture News},
  vol.~44, pp. 27--39, 06 2016.

\bibitem{PipeLayer}
L.~{Song}, X.~{Qian}, H.~{Li}, and Y.~{Chen}, ``Pipelayer: A pipelined
  reram-based accelerator for deep learning,'' in \emph{2017 IEEE International
  Symposium on High Performance Computer Architecture (HPCA)}, 2017, pp.
  541--552.

\bibitem{CASCADE}
T.~Chou, W.~Tang, J.~Botimer, and Z.~Zhang, ``Cascade: Connecting rrams to
  extend analog dataflow in an end-to-end in-memory processing paradigm,'' in
  \emph{Proceedings of the 52nd Annual IEEE/ACM International Symposium on
  Microarchitecture}, 2019, p. 114–125.

\bibitem{RAPIDNN}
M.~{Imani}, M.~{Samragh Razlighi}, Y.~{Kim}, S.~{Gupta}, F.~{Koushanfar}, and
  T.~{Rosing}, ``Deep learning acceleration with neuron-to-memory
  transformation,'' in \emph{2020 IEEE International Symposium on High
  Performance Computer Architecture (HPCA)}, 2020, pp. 1--14.

\bibitem{LiM}
H.~S. {Stone}, ``A logic-in-memory computer,'' \emph{IEEE Transactions on
  Computers}, vol. C-19, no.~1, pp. 73--78, 1970.

\end{thebibliography}

\begin{IEEEbiography}[{\includegraphics[width=1in,height=1.25in,clip,keepaspectratio]{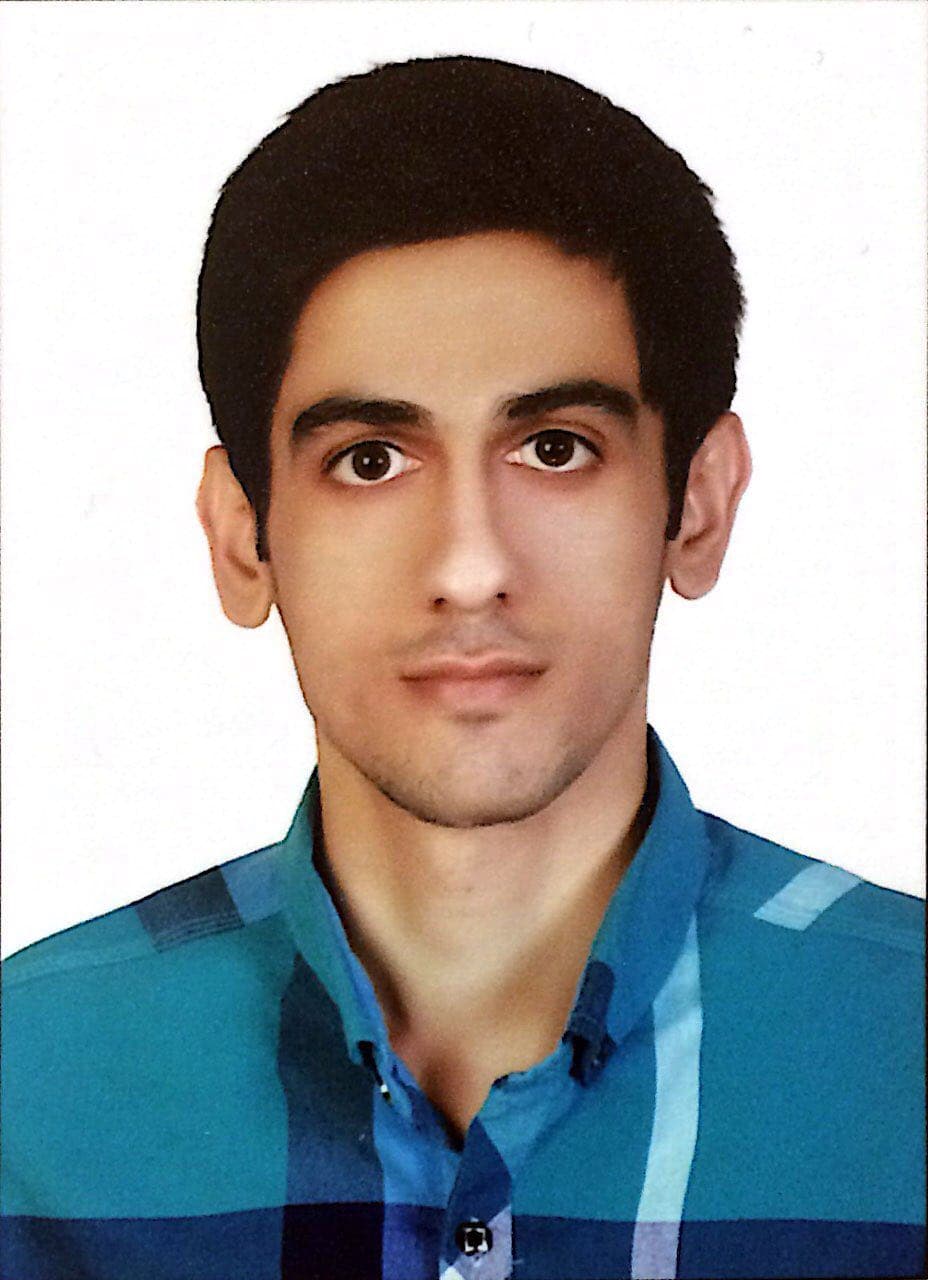}}]{Mehdi Hassanpour}
received his B.S degree in Computer Engineering in 2019 from Sharif University of Technology, Tehran (Iran). He joined ARCO research group in February 2020 and is currently pursuing his PhD. His research focuses on Near-Data Processing (NDP).
\end{IEEEbiography}

\begin{IEEEbiography}[{\includegraphics[width=1in,height=1.25in,clip,keepaspectratio]{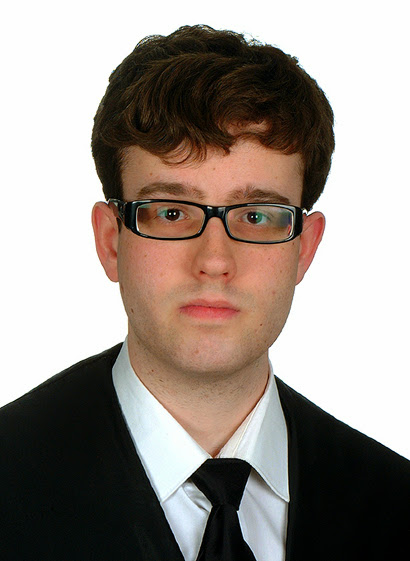}}]{Marc Riera}
received his B.S degree in Computer Engineering in 2013, his MS degree in MIRI: High Performance Computing in 2015, and his PhD in Computer Architecture in 2020, all from Universitat Politècnica de Catalunya (UPC - BarcelonaTech). He joined the ARCO research group in 2014, and is currently a postdoctoral researcher at ARCO. His research interests focus on the area of Accelerator Architectures, Machine Learning and Near-Data Processing (NDP).
\end{IEEEbiography}

\begin{IEEEbiography}[{\includegraphics[width=1in,height=1.25in,clip,keepaspectratio]{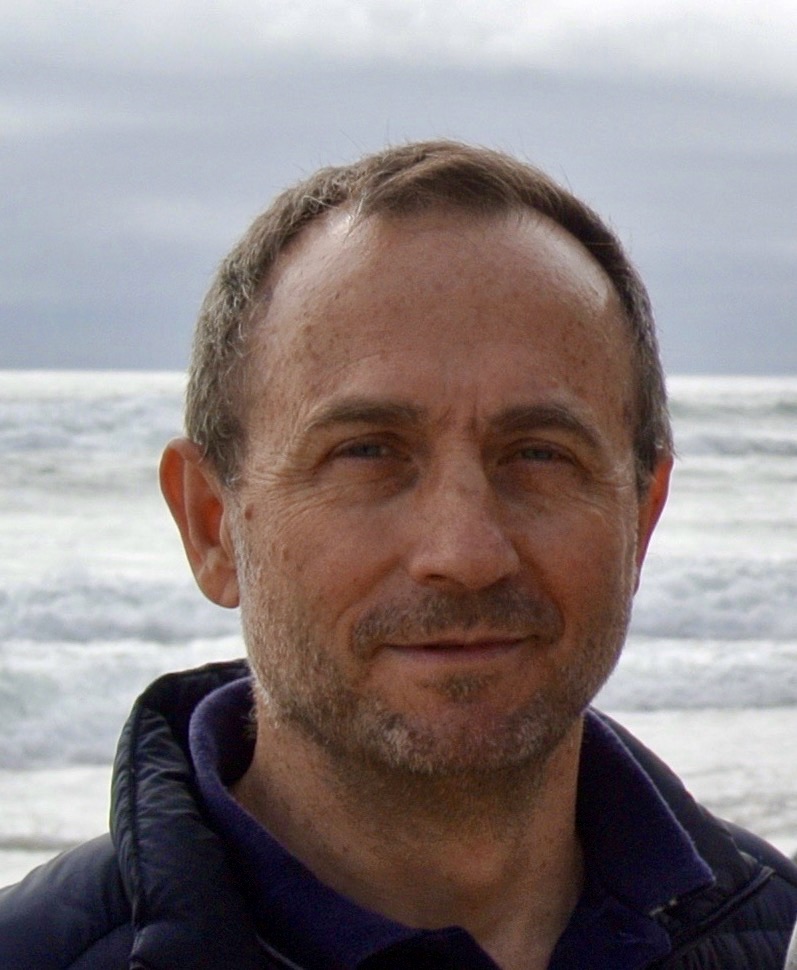}}]{Antonio Gonz\'{a}lez}
(PhD 1989) is a Full Professor at the Computer Architecture Department of the Universitat Politècnica de Catalunya, Barcelona (Spain), and the director of the Architecture and Compilers research group. He was the founding director of the Intel Barcelona Research Center from 2002 to 2014. His research has focused on computer architecture and compilers, with a special emphasis on cognitive computing systems and graphics processors in recent years. He has published over 380 papers, and has served as associate editor of five IEEE and ACM journals, program chair for ISCA, MICRO, HPCA, ICS and ISPASS, and general chair for MICRO and HPCA. He is a Fellow of IEEE and ACM.

\end{IEEEbiography}

\end{document}